\documentclass[aps,twocolumn,showpacs,showkeys,nofootinbib,superscriptaddress]{revtex4-1}
\pdfoutput=1
\usepackage{graphicx}     % Include figure files
\usepackage{amsmath}
\usepackage{amssymb, amsfonts}
\usepackage{array}
\usepackage{braket}
\usepackage{cases}
\usepackage[utf8]{inputenc}
\usepackage[overload]{empheq}
\usepackage{subcaption}
\usepackage{relsize}  %%%  LRR added this
\usepackage{multirow}
\usepackage{dcolumn}  % For aligning decimals in tables
\usepackage{hyperref}%hyperlink figures, tables, references
\usepackage{xcolor}
\usepackage{relsize}%
\usepackage{cleveref}
\hypersetup{
	colorlinks   = true, %Colours links instead of ugly boxes
	urlcolor     = blue, %Colour for external hyperlinks
	linkcolor    = blue, %Colour of internal links
	citecolor    = blue %Colour of citations
}
\newcommand{\RN}[1]{%
	\textup{\uppercase\expandafter{\romannumeral#1}}%
}

\newcolumntype{d}[1]{D{.}{.}{#1}}
\newcolumntype{K}[1]{>{\centering\arraybackslash}p{#1}}
\newcommand{\DS}{\displaystyle}
\usepackage{floatflt}     % floating figures and tables
\usepackage{wrapfig}
\usepackage{color}
\usepackage{pstricks}

\makeatletter
\newcommand*{\rom}[1]{\expandafter\@slowromancap\romannumeral #1@}
\usepackage{color}
\definecolor{DarkRed}{rgb}{0.35,0.01,0.01}
\definecolor{Linen}{rgb}{0.98,0.98,0.94}
\definecolor{Blue}{rgb}{0.,0.,1.0}
\definecolor{DarkBlue}{rgb}{0.099,0.099,0.44}
\definecolor{DarkGreen}{rgb}{0.0,0.4,0.0}
\definecolor{Turquoise}{rgb}{0.0,0.9,0.7}
\begin{document}
%%%%%%%%%%%%%%%%%%%%%%%%%%%%%%

%\title{Spatial entanglement in two-electron artificial atoms}
%\title{Spatial entanglement in interacting few-electron artificial atoms}
\title{Tuning spatial entanglement in interacting few-electron quantum dots}
%%%%%%%%%%%%%%%%%%%%%%%%%%%%%%

\author{Dung N. Pham}
\affiliation{Department  of   Physics,  Worcester
	Polytechnic Institute, Worcester, Massachusetts 01609.}
\affiliation{Center for Computational NanoScience,  Worcester
	Polytechnic Institute, Worcester, Massachusetts 01609, USA.}
\author{Sathwik Bharadwaj}
\email{sathwik@wpi.edu}
\affiliation{Department  of   Physics,  Worcester
	Polytechnic Institute, Worcester, Massachusetts 01609.}
\affiliation{Center for Computational NanoScience,  Worcester
	Polytechnic Institute, Worcester, Massachusetts 01609, USA.}
\author{L. R. Ram-Mohan}
\email{LRRAM@wpi.edu}
\affiliation{Center for Computational NanoScience,  Worcester
	Polytechnic Institute, Worcester, Massachusetts 01609, USA.}
\affiliation{Departments  of   Physics,  Electrical and Computer
	Engineering, and Mechanical Engineering, Worcester
	Polytechnic Institute, Worcester, Massachusetts 01609, USA.}


%%%%%%%%%%%%%%%%%%%%%%%%%%%%%%
%%%%%%%%%%%%%%%%%%%%%%%%%%%%%%
%%%%%%%%%%%%%%%%%%%%%%%%%%%%%%

%%%%%%%%%%%%%%%%%%%%%%%%%%%%%%
\begin{abstract}
	Confined geometries such as  semiconductor quantum dots 
	are promising  candidates  for  fabricating quantum  computing
	devices. When several quantum dots  are in proximity, {\it spatial}
	correlation between electrons in the system becomes significant.  In
	this  article,  we  develop  a  fully  variational  action  integral
	formulation for  calculating accurate few-electron  wavefunctions in
	configuration space, irrespective of  potential geometry.
	To  evaluate  the Coulomb  integrals  with  high accuracy,  a  novel
	numerical  integration method  using multiple  Gauss quadratures  is
	proposed.  Using  this approach,  we investigate the  confinement of
	two electrons  in double  quantum  dots, and  evaluate the  spatial
	entanglement. We investigate the  dependence of spatial entanglement
	on  various  geometrical  parameters.  We  derive the
	two-particle  wavefunctions   in  the   asymptotic  limit  of  the
	separation  distance between  quantum  dots,  and obtain  universal
	saturation values  for the  spatial entanglement. Resonances  in the
	entanglement  values due  to avoided  level-crossings of  states are
	observed.  We  also demonstrate the formation  of electron clusters,
	and show  that the entanglement  value is  a good indicator  for the
	formation  of  such  clusters.   Further, we  show  that  a  precise
	tuning  of the  entanglement values  is feasible  with applied
	external electric fields.
\end{abstract}
%%%%%%%%%%%%%%%%%%%%%%%%%%%%%%

\maketitle

\begin{center}
	\today
\end{center}

%%%%%%%%%%%%%%%%%%%%%%%%%%%%%%
\section{Introduction}\label{sec:Intro}%Sec 2
%%%%%%%%%%%%%%%%%%%%%%%%%%%%%%
Similarities  can be drawn between the electronic properties of a single quantum dot and
a hydrogenic atom, which implies that we can also develop an analogous
Hund's multiplicity rule \cite{hund,levin}  for these artificial atoms
\cite{steffens,sivan, sanjeev}.   When two or more  quantum dots (QDs)
are in  the vicinity of  one another, the  system can be  thought of  as a
covalent   molecule  of   QDs~\cite{sdsarma}.   Tunability   in  their
inter-electron  interaction  with  the separation  distance  and  with
external  fields facilitates  an enhanced  level of  control in  their
electronic   properties~\cite{chuang,wallis,haug,michler_QD}.    Advances   in the  precise   fabrication   of
QDs~\cite{Heiss,Eng,Hendrickx} have  led  to
substantial  improvements in  the prospects  of developing  integrated
devices  for  applications  in  quantum  computing~\cite{Loss,Burkard,
	Hayashi,  Shulman,   Shi,  samarth,  flatte},   quantum  information
~\cite{Jones,Delbecq},      and      quantum      memory      circuits
~\cite{Watson,Li,Berger}.

Evaluation  and measurement  of the  entanglement in  a multi-particle
system has attracted extensive  theoretical and experimental interest
~\cite{Nielen_n_Chuang,Schliemann1,Barnum,Eckert,Zanardi,Wiseman}.   A
complete description  of a quantum  system is given by  a wavefunction
that has  both spin  and spatial  components. Hence  the entanglement
will also have  contributions from  spin  and spatial correlations
of  the wavefunction.   
When QDs are in  proximity, the
spatial   correlation  between   electrons  in   the  system   becomes
significant and leads to the  {\it spatial entanglement}. Such spatial
entanglement will  alter considerably with changes  in the geometrical
parameters of QDs and with external perturbations.

The spatial entanglement properties are of interest to realize solid state all-electronic
quantum computing devices. To this end, there have been several proposals to define 
deterministic teleportation protocols for quantum information processing in semiconducting nanowires \cite{kais1}, double QDs \cite{visser, adepoju}, quantum dot
arrays \cite{pasquale}, and coupled quantum wires \cite{buscemi}. Applications  of
spatial  entanglement   as  an  indicator  for   bound  and  unbound
states~\cite{Ferron}, as well  as the effects of a  magnetic field on
entanglement have been discussed earlier~\cite{Simonovic}. The spatial  entanglement properties in Helium  and Helium-like atoms
have   been   studied   in   the   literature~\cite{Dehesa,KamHo_He}.
Entanglement calculations for the  ground state in simplistic models,
such   as    the   Hooke's-atom   model~\cite{Coe}    and   symmetric
one-dimensional  quantum wells with  point-contact  modeling  for the  Coulomb
potential~\cite{Abdullah},  have  been considered. Recently, it has  also been  
shown  that the  spatial overlapping  of indistinguishable  particles can  
be employed  in quantum  information processes through local measurements~\cite{Franco}. For all such applications
it is important to fabricate the devices operating at the resonant entanglement values. A detailed
study to obtain the ``spectroscopy'' of quantum entanglement in QDs has not been considered so far. 

A thorough understanding of the spatial correlation of particles is also crucial to the fabrication and manipulation of charge qubits, which are usually coupled through the Coulomb interaction \cite{Petersson}. Although the literature on using charge qubits for quantum information purposes has been prolific \cite{Bonderson, Cao}, there are surprisingly sparse theoretical studies on the entanglement that is intrinsic in the spatial wavefunctions describing the qubits. More specifically, how the spatial coupling between the particles depends on the system, and on external parameters, is of great interest. 

In this article, we consider the confinement of two electrons in QDs formed through nanowire heterostructures. Such structures have been considered as an ideal candidate for quantum electronic devices \cite{Minot} and quantum information processes \cite{Szumniak, Perge, Perge2}, since particles can be very effectively localized, either through external fields \cite{Perge2,Fasth} or by superlattice nanowires \cite{gudiksen,Bjork,Wu,Bjork2,Bjork3}. In either case the lateral width has negligible contributions to the entanglement properties. It is crucial to note that although in this study we consider specifically one-dimensional (1D) confinement for the particles, the methodology and conclusions are generalizable to any high-dimensional bipartite systems. In this article, we consider only the 1D potential confinements, and the generalization to higher dimensions will be presented elsewhere. 

One well-studied approach for few-electron problems is the full configuration interaction (FCI) method, which uses a  basis set constructed from single-particle orbitals. Due to its success as the benchmark for few-particle atmoic and molecular calculations \cite{Abe, Savukov}, the FCI has been used in condensed matter setups \cite{Pujari, Blundell, Shepherd, Abolfath}. In this study, we   develop  an alternative  variational  formalism   for  calculating  the
coordinate-space representation of the few-particle wavefunctions in semiconductor confinements. Accurate energy values and wavefunctions can be obtained through our  geometry  discretization  scheme  based  on  the  principle  of stationary action. We note that full scale FCI calculations with Schrodinger-Poisson self consistent single particle basis \cite{sandiaFCI1, sandiaFCI2}, or with atomistic tight-binding basis \cite{rahman1, rahman2} are desirable for system with several electrons in quantum dots. Our method provides a useful pre-step before such large scale FCI for simulating new devices. It may also be used to benchmark FCI results \cite{nielsen_muller}. We emphasis that our method is best suited to solve few-body confinement problems.   

New results presented in this article are summarized below:
\begin{enumerate}
	\item We have studied the case of two electrons trapped in different variations of a double QDs. We obtain the eigen-spectrum, and measure  the  spatial entanglement  by
	computing the  linear entropy of  the system. We show  that through
	geometrical manipulations  of the  confining potential  and/or using
	external electric fields, one can tune the level of entanglement.
	
	\item We  derive exact representations for the wavefunctions in
	the   asymptotic limit, and obtain universal saturation values for
	the   entanglement.
	
	\item  Resonances  in  the  {\it   entanglement}  are  observed  as  a
	consequence of avoided  level-crossings (also called anti-crossings)
	in  the  energy spectrum  in  an  applied  electric field,  or  with
	variations of the width in  asymmetric double QDs.  Electron  cluster formations  are
	also detected at  the exited states, which lead  to additional local
	maxima.
	
\end{enumerate}

This paper is organized as follows: In Sec.~\ref{sec:fem}, we describe
the variational  scheme for solving the  few-particle action integral.
In  Sec.~\ref{sec:energy}, the  energy spectrum  for two  electrons in
$\rm{GaAs/Ga_{x}Al_{1-x}As}$  symmetric double  QDs are  discussed. We
also derive  the asymptotic representations to  explain the degeneracy
spectrum.  Spatial  entanglement properties  for the  symmetric double
QDs are discussed  in Sec.~\ref{sec:Ent_QD}.  Entanglement resonances,
and the two  electron-cluster formation in a system  of asymmetric QDs
are studied  in Sec.~\ref{sec:non_sym}.  We  analyze the effect  of an
applied electric field on the entanglement in Sec.~\ref{sec:Efield}. We  obtain the  entanglement  values  in the  asymptotic
separation    distance     in    Appendix~\ref{subsec:asymptotic_ent}.
Calculation  of  the  entanglement   properties  with  parabolic  dot
confinement  is  given  in  Appendix~\ref{sec:parabolic}.   Concluding
remarks are given in Sec.~\ref{sec:conclusions}.

%%%%%%%%%%%%%%%%%%%%%%%%%%%%%%
\section{Variational formulation for few-particle
	wavefunctions}%Sec 3
\label{sec:fem}
%%%%%%%%%%%%%%%%%%%%%%%%%%%%%%
Let us consider a two-electron wavefunction in  configuration space
written as
\begin{equation}
\Phi({ x_1},\sigma_1,{x_2},\sigma_2) = \psi({x_1},{x_2}) S(\sigma_1,\sigma_2), 
\end{equation}
where      $\psi(x_1,x_2)$     and
$S(\sigma_1,\sigma_2)$ are the  spatial and the spin  components of the
wavefunction,    respectively.    Here,    ${x_1}(\sigma_1)$    and
${x_2} (\sigma_2)$ represent the  position (spin) coordinates of the
electron 1 and  2, respectively. Within the envelope-function approximation \cite{bastardEPA}, the spatial    part of  
the charge carrier's envelope function $\psi({x_1,x_2})$ satisfies  the    time-independent
Schr\"odinger's equation of the form
\begin{align}\label{eq:SE}
\hspace*{-0.15in}\Bigg\{\!\!-\!\frac{\hbar^2}{2}\bigg[\frac{d}{dx_1}\Big(\frac{1}{m^*_1}\frac{d}{dx_1}\Big)  
\!+ \!      \frac{d}{dx_2}\Big(\frac{1}{m^*_2}\frac{d}{dx_2}\Big)\!\bigg]         +
& V({x_1},{x_2})\!\Bigg\}\psi \nonumber\\ 
 &= E\psi,
\end{align}
where $m^*_1\equiv m^*_1({x_1})$, $m^*_2\equiv m^*_2({x_2})$ are
the effective masses of  the two electrons, and $V({x_1},{x_2})$
is  the  effective  potential. $V({x_1},{x_2})$  contains  terms
arising from the geometrical confinement of the system, $V_0$, and the
Coulomb interaction between the electrons. It is given by
\begin{equation}
V(x_1,x_2) = V_0({x_1}) +
V_0({x_2}) + \frac{e^2}{4\pi\epsilon_o|{x_1}-{x_2}|}. 
\end{equation}
We multiply Eq.~(\ref{eq:SE}) with $\delta\psi^{*}$, a small variation in the function $\psi^{*}$
and integrate over all space $\Omega$ to obtain
\begin{eqnarray}
\hspace*{-0.2in}\int_{\Omega}\!\! dx_1
dx_2\,\delta\psi^{*}\Big[-\!\frac{\hbar^2}{2}\Big\{\frac{d}{dx_1}\!\left(\frac{1}{m^*_1}\frac{d}{dx_1}\right)  
+ \frac{d}{dx_2}\!\left(\frac{1}{m^*_2}\frac{d}{dx_2}\right)\Big\}\nonumber \\+
V({x}_1,{x}_2)-E\Big]\psi = 0.\nonumber\\ 
\end{eqnarray}
Using the integration by parts on the first two terms in the above integral we obtain 
\begin{eqnarray}
%\hspace*{-0.2in}
\delta\int_{\Omega}\!\! dx_1
dx_2
\bigg[\frac{\hbar^2}{2}\Big(\frac{d\psi^*}{dx_1}\cdot\frac{1}{m^*_1}\frac{d\psi}{dx_1}+
\frac{d\psi^*}{dx_2}\cdot\frac{1}{m^*_2}\frac{d\psi}{dx_2}\Big)\nonumber\\\hspace*{0.2in}+\,\psi^*\Big(V(x_1,x_2)-E\Big)\psi\bigg]\nonumber \\ 
\hspace*{0.2in} =\delta\int\!\! dx_1
dx_2\,\mathcal{L}(x_1,x_2) = 0,
\end{eqnarray}
where $\mathcal{L}(x_1,x_2)$ is identified as the two-particle Lagrangian density of the system. Hence,
the action integral corresponding to Eq.~(\ref{eq:SE}) is given by
\begin{eqnarray}\label{eq:action1}
A &=&\int dt\int_{\Omega}\!\! dx_1
dx_2\ \mathcal{L}(x_1,x_2),\nonumber\\&=& T\!\!\int_{\Omega}\!\! dx_1
dx_2
\bigg[\frac{\hbar^2}{2}\Big(\frac{d\psi^*}{dx_1}\!\cdot\!\frac{1}{m^*_1}\frac{d\psi}{dx_1}+
\frac{d\psi^*}{dx_2}\!\cdot\!\frac{1}{m^*_2}\frac{d\psi}{dx_2}\Big)\nonumber \\ 
&&\hspace*{1in} +\,\psi^*\Big(V(x_1,x_2)-E\Big)\psi\bigg]. 
\end{eqnarray}
Since we are considering a  steady-state problem, the time integral is
simply a  constant $T$. We are  interested in finding bound  states of
the  system, hence  Dirichlet  boundary conditions  are  imposed
along the periphery of the truncated finite domain.

\par
\begin{figure}[t!] %Fig 1
	\includegraphics[width=3in]{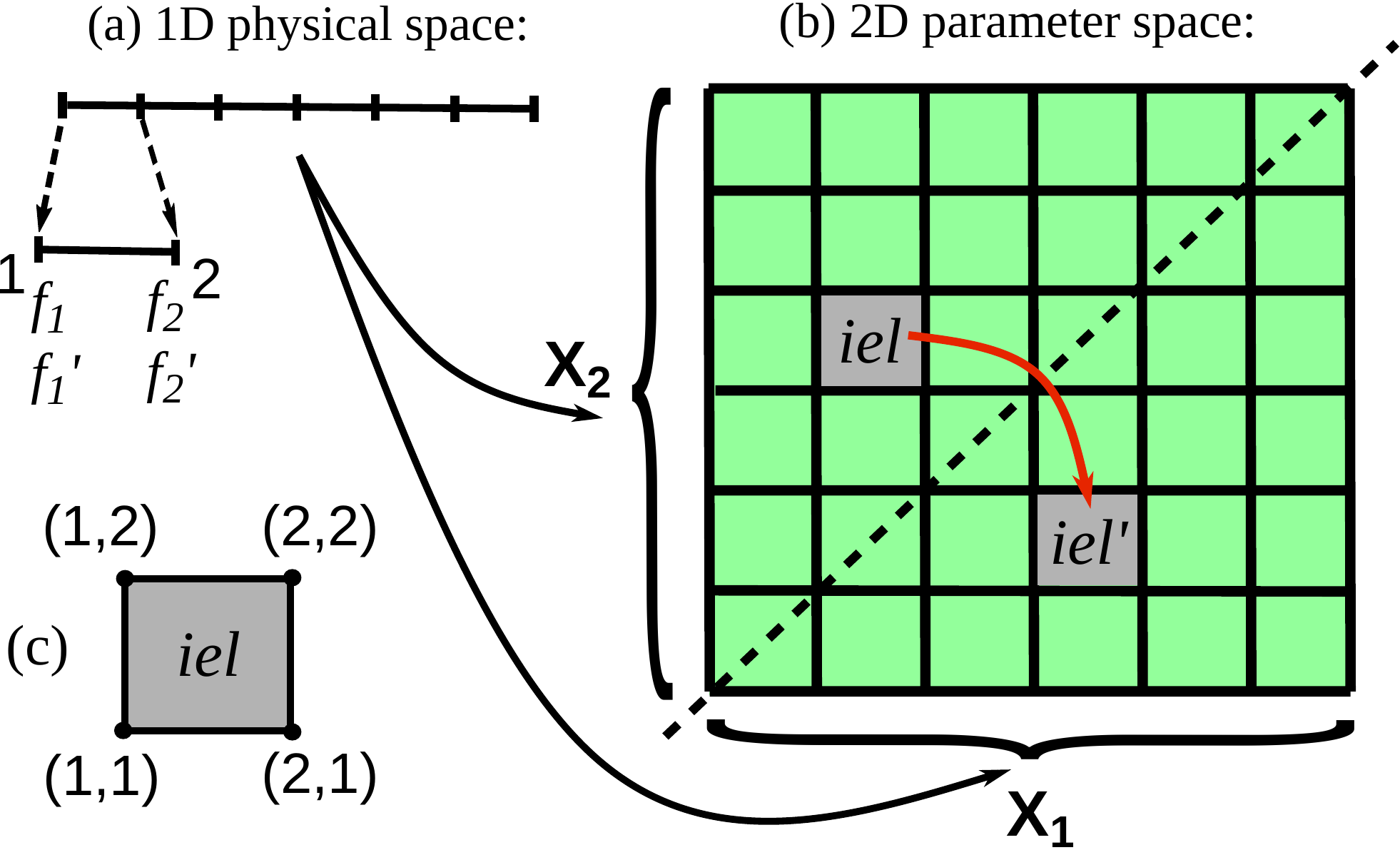}
	\caption{\label{fig:mapping}  (a)  A schematic representation of the discretization in one dimension is shown. For Hermite interpolations, each node has coefficients corresponding to the wavefunction and its first derivative values. (b) A mapping of the 1D discretization into a 2D-parameter space $(x_1,x_2)$ for a two-particle problem is shown. Elements $iel$ and $iel'$ are related through  the
		mirror reflection across the mutual axis (represented by the dashed line). (c) A sample 2D element within which the wavefunction is represented using Eq.~(\ref{eq:Interps}) is displayed.}
\end{figure}

 The domain $\Omega$ is  discretized into a refined finite element
mesh. A schematic for the discretization of a one-dimensional physical domain is shown in Fig.~\ref{fig:mapping}(a). In  a bipartite  problem, this discretization is applied to each particle's space ${x_1}$ and ${x_2}$. This results in a 2N-dimensional finite element mesh, with N being the dimension of the particle confinement of the problem. For example, two electrons confined in a 1D physical space results in a 2D parameter space as shown in Fig.~\ref{fig:mapping}(b). Here $x_1$ and $x_2$ represents the coordinate of the first and second particle, respectively. Although in this paper we consider specifically electrons trapped in 1D geometry, the method is generalizable to any number of dimensions.

Within  each 2N-dimensional parameter space element,  the two-particle spatial  wavefunction is
represented as  a linear combination of  the interpolation polynomials
$N_i({x}_k)$ with as-yet undetermined coefficients $\psi_{ij}$, given by
\begin{equation}\label{eq:Interps}
\psi^{iel}(x_1,x_2) =
\sum_{ij}\psi_{ij}N_i(x_1)
N_j(x_2), 
\end{equation}
where $iel$  is the  element index,  and $\psi_{ij}$  are undetermined
wavefunction values at nodes (vertices) of the element. 
In contrast to employing global basis functions, using a local element representation
 offers  exceptional   flexibility   in  obtaining   accurate
wavefunctions, since discretization can be done to systems of arbitrary shapes.  In our  calculation, we  employ Hermite  interpolation
polynomials~\cite{LRR_book},  in which the variational parameters are the wavefunction, and its first derivative values at each node. Using the Hermite interpolation polynomials guarantee the function  and the first derivative  continuity throughout the domain $\Omega$.    

To impose the
antisymmetric property of fermions, the  following scheme is used. Let
$iel$ be  an element  in the discretized  parameter space.  Since both
particles share  the same  physical space, for  a given  element $iel$
there  exists a  distinct element  $iel'$  related to  it through  the
mirror reflection across the mutual axis of the parameter space (see Fig.\ref{fig:mapping}(b)). For a
spatial  wavefunction to  be symmetric  (or antisymmetric),  we demand
that within $iel'$, the local representation satisfies the relation
\begin{equation}
\!\!\psi^{iel'}({x_1},{x_2})\! =\!
\pm\psi^{iel}({x_1},{x_2}) \!=\!\pm\!\!\sum_{ji}\psi_{ji}N_{j}({x_1})
N_{i}({x_2}), 
\end{equation}
where the  $\pm$ sign  corresponds to a symmetric  or antisymmetric
spatial  wavefunction, respectively.  Note  that  this procedure  also
reduces the number of variational  parameters by half, since the nodal
values associated with $iel'$ are now related to those of $iel$.

The  action  integral in Eq.~(\ref{eq:action1}) is then   evaluated  within   each element using Eq.~(\ref{eq:Interps}), and the total action is found by summing over all elements. In Fig.~\ref{fig:mapping}(c) we show an example of element in the 2D parameter space. Here the variational parameters are the wavefunction and their first derivative values that are corresponding to the vertices of the square element. A more detailed discussion of the finite element analysis and the Hermite interpolation polynomials can be found in Ref.~\cite{LRR_book,Zienkiewicz1977,Jin2002,LRR_TunnelingBC,gouriDhatt}.  

Using the principle of stationary action, a variation
of  Eq.~(\ref{eq:action1})  with  respect  to $\Psi^{\dagger}$ (a row vector containing all variational parameters corresponding to the function $\psi^{*}$)  results  in  a
generalized eigenvalue problem of the form
\begin{equation}
({\bf K} - E {\bf U})\Psi = 0,\label{eq:matrixeq}
\end{equation}
where ${\bf K}$  is the coefficient matrix corresponding  to the first 3 terms in the two-particle Lagrangian density defined in Eq.~(\ref{eq:action1}), ${\bf  U}$ is the  coefficient matrix
associated with  overlap integrals corresponding to the coefficient $E$ in Eq.~(\ref{eq:action1}),  and $\Psi$  is the  column vector
containing  all the  nodal values  to  be determined.  Once the  nodal
values  are obtained  by solving  the generalized  eigenvalue problem
in 
Eq.~(\ref{eq:matrixeq}),  the  wavefunction  at any  location  can  be
calculated through reinterpolation using  Eq.~(\ref{eq:Interps}).

Being  a variational  method based on  geometry
discretization,  the  accuracy  of our  scheme  in
computing wavefunctions is geometry-independent and can be applied to
QDs of any arbitrary shapes,
with any desired level of accuracy achievable by suitable mesh refinement. In contrast to the traditional
Raleigh-Ritz method, our scheme provides the necessary flexibility in the element-based trial wavefunctions, and 
more variational parameters in the form of nodal values of the wavefunction and its derivatives.  

Within the scope of this paper, we consider strictly one-dimensional confinement, since the lateral effects are negligible for QDs formed through the nanowire heterostructures. The justification for the reduction of Coulomb integrals to one dimension is given in the supplementary material.    
We note that special attention  is needed to evaluate the Coulomb
contribution  to the  action integral  since it  has singularities  at
${x_1}  ={x_2}$.   We  propose  an  efficient  way  to  evaluate
numerically  such integrals  using multiple  orders for Gauss quadratures, and this is discussed in the supplementary material. 

% A few words about the  diagonalization and computational time required
% are in  order. A  typical calculation employing  Hermite interpolation
% polynomials for the desired  accuracy of $\simeq10^{-13}-10^{-14}$ has matrices
% of dimensions 750,000,
% %x750,000
%  and takes about 2.4 minutes to construct, and 5.2 minutes to
% diagonalize and determine the lowest 10-12 eigenvalues using  48
% processors. The level of accuracy sought reflects the need to verify
% the level degeneracy in Table~\ref{table:eig_distance}.

%%%%%%%%%%%%%%%%%%%%%%%%%%%%%%
\section{Energy calculations for two-particle
  systems}\label{sec:energy}%Sec 3 

\begin{figure}[ht!] %Fig 2
	\includegraphics[scale=0.22]{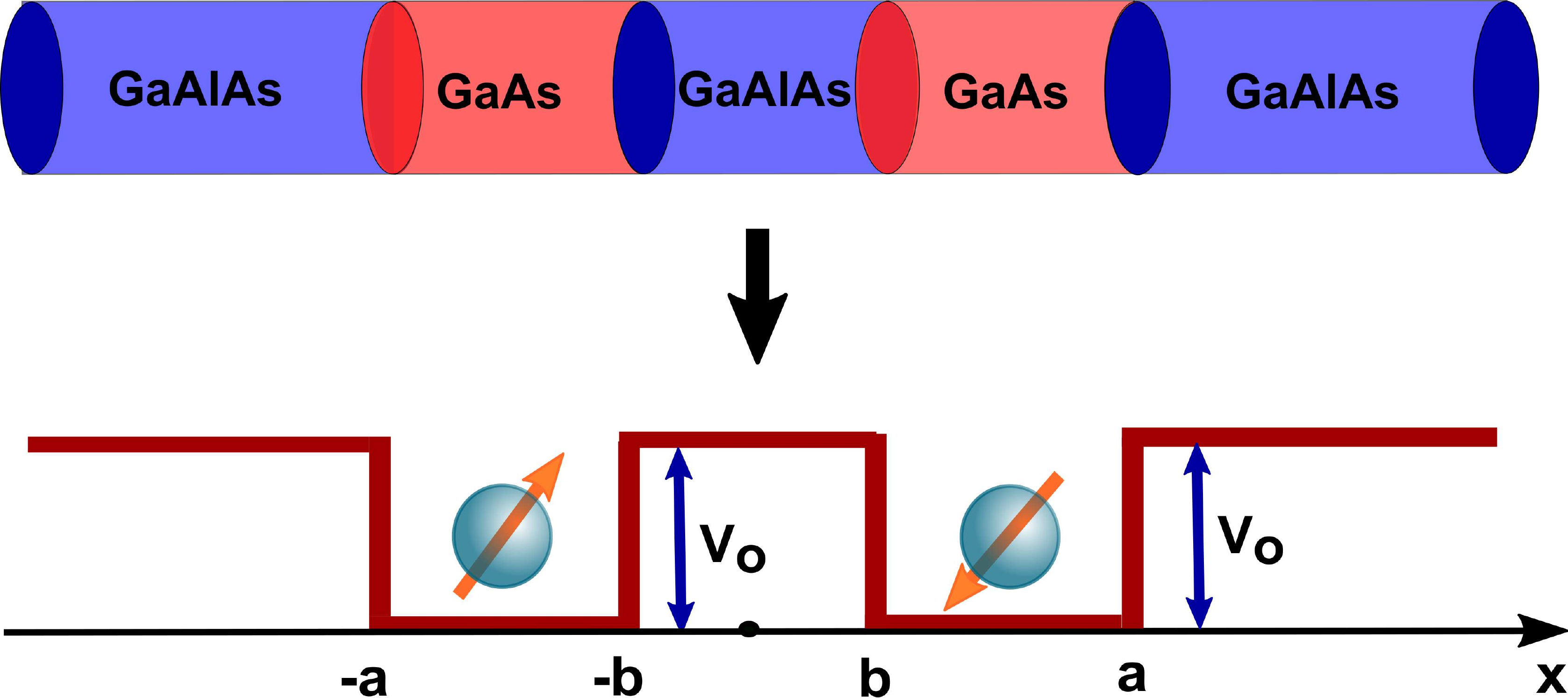}
	\caption{\label{fig:geometry} Schematic of a one-dimensional double quantum dot system.}
\end{figure}

\begin{table*}[ht]
	\centering
	\renewcommand{\arraystretch}{1.5}
	\begin{tabular}{c|c|c|c|c|c|c}
          \toprule
          ($\alpha, \beta$)  & $d \rightarrow \infty$ & $d$=30\,{\rm
 nm} &
       $d$= 10\,{\rm nm} & $d$= 6\,{\rm nm} & $d$=4\,{\rm nm} &$d$= 0\,{\rm nm}\\ 
          \hline
 \multirow{2}{*}{$(1,1)$}  
 & 0.03381551453605 & 0.06597445297395 &
 0.09117827187267 & 0.10132363931689 & 0.10766511883166 & 0.12308265436399 \\ 
 & 0.03381551453605 & 0.06597445297393 & 0.09117827187810 &
 0.10132363932279 & 0.10766511891643 & 0.12306768828638 \\			 
 \hline
 \multirow{4}{*}{$(1,2)$} 
 & 0.08413370110643 & 0.11633416555783 & 0.14217840394755 &
 0.15312310251314 & 0.16020213212699 & 0.17196886060204 \\
  & 0.08413370110643 & 0.11633416555779 & 0.14217840403115 &
 0.15312310253815 & 0.16020213239190 & 0.17190045732112 \\ 
 & 0.08413370110643 & 0.11704336845825 & 0.14608683015824 &
 0.15918731650000 & 0.16777544459370 & 0.18441352939932 \\ 
 & 0.08413370110643 & 0.11704336845822 & 0.14608683024372 &
 0.15918731661314 & 0.16777544608388 & 0.18438741290837 \\
 
 \hline
 \multirow{2}{*}{$(2,2)$}
 & 0.13445188767681 & 0.16741965485019 & 0.19756820520027 &
 0.21209234809771 & 0.22187638627095 & 0.22786101607874 \\ 
 & 0.13445188767681 & 0.16741965485014 & 0.19756820550444 &
 0.21209234852245 & 0.22187639053093 & 0.22790743753463 \\ 
 \botrule
 \end{tabular}
 \caption{Eigenvalues (measured in eV) of the first eight  states of a symmetric double
   QD are  tabulated for several  separation distances $d$  between the
   QDs.  The width of each quantum dot is w$_1=$w$_2=15\,$nm.  States
   are labeled with a pair of quantum numbers $(\alpha,\beta)$. The
   Coulomb energy increases with decreasing distance as seen in the table.}
	\label{table:eig_distance}
\end{table*}
\begin{figure*}[ht] %Fig 3
	%\captionsetup[subfigure]{labelformat=empty}
	\begin{subfigure}[h!]{0.3\textwidth}
          \includegraphics[scale=0.25]{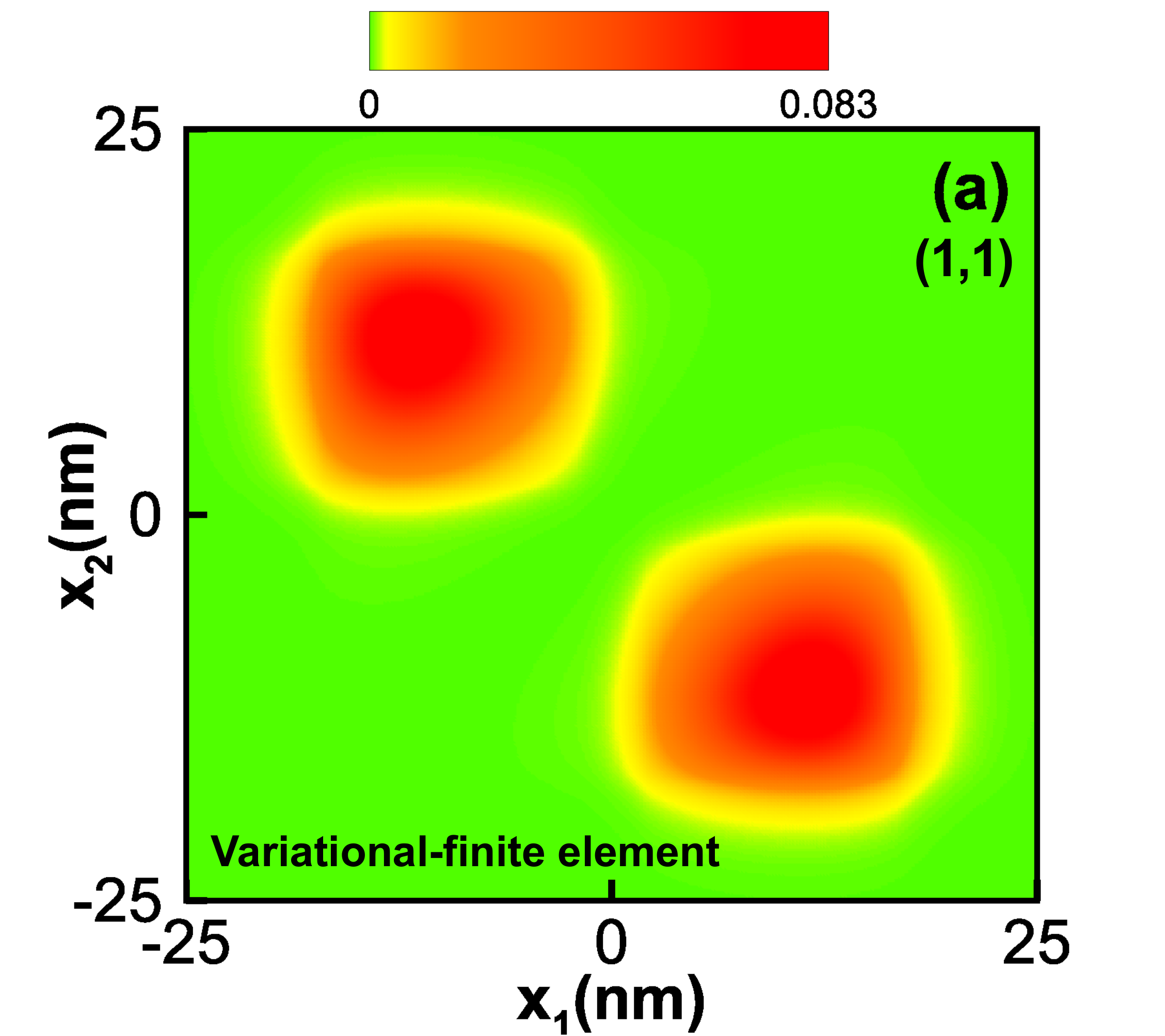}
	\end{subfigure}%
	\begin{subfigure}[h!]{0.3\textwidth}
		\includegraphics[scale=0.25]{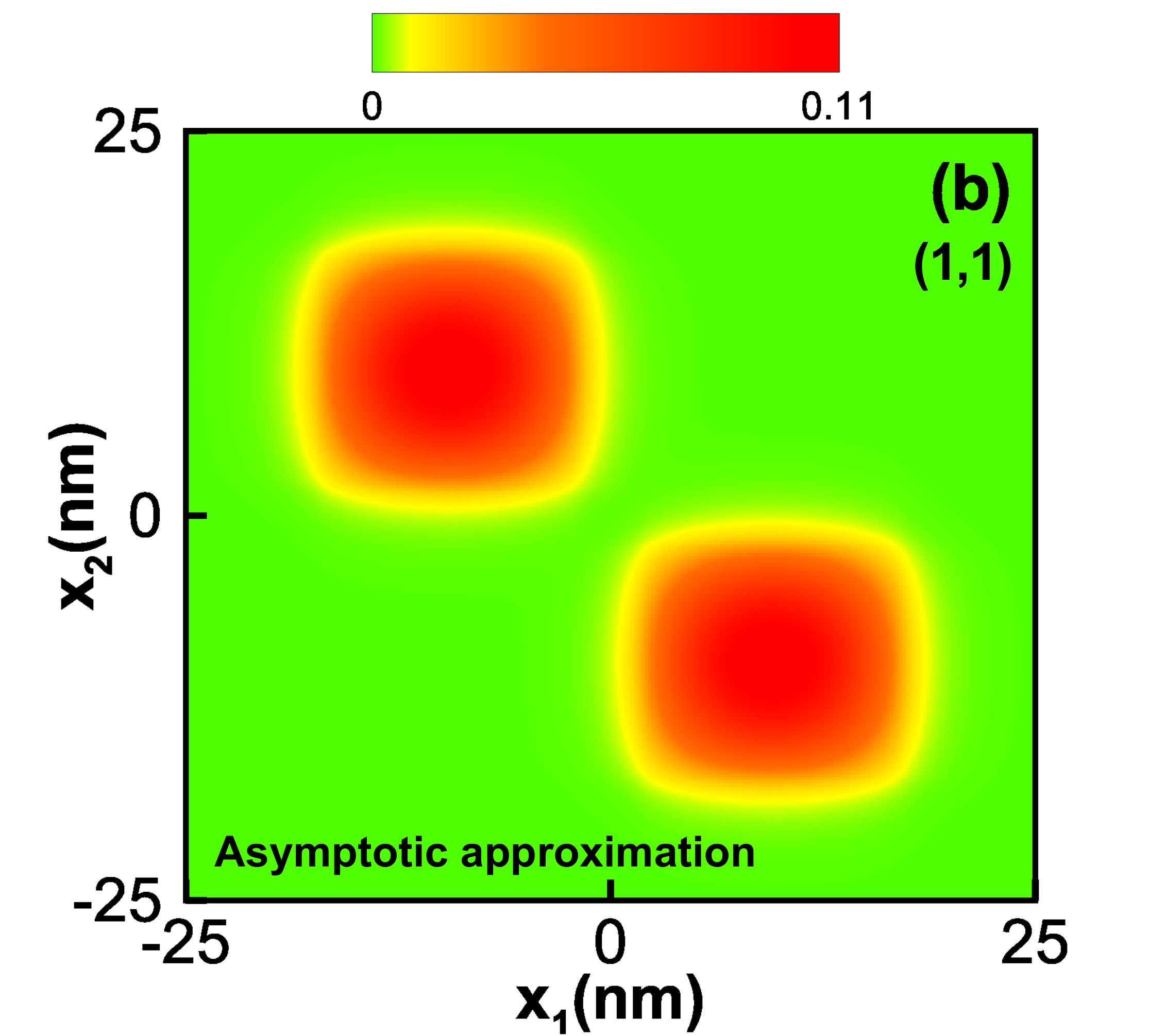}
	\end{subfigure}%
	\begin{subfigure}[h!]{0.3\textwidth}
		\includegraphics[scale=0.25]{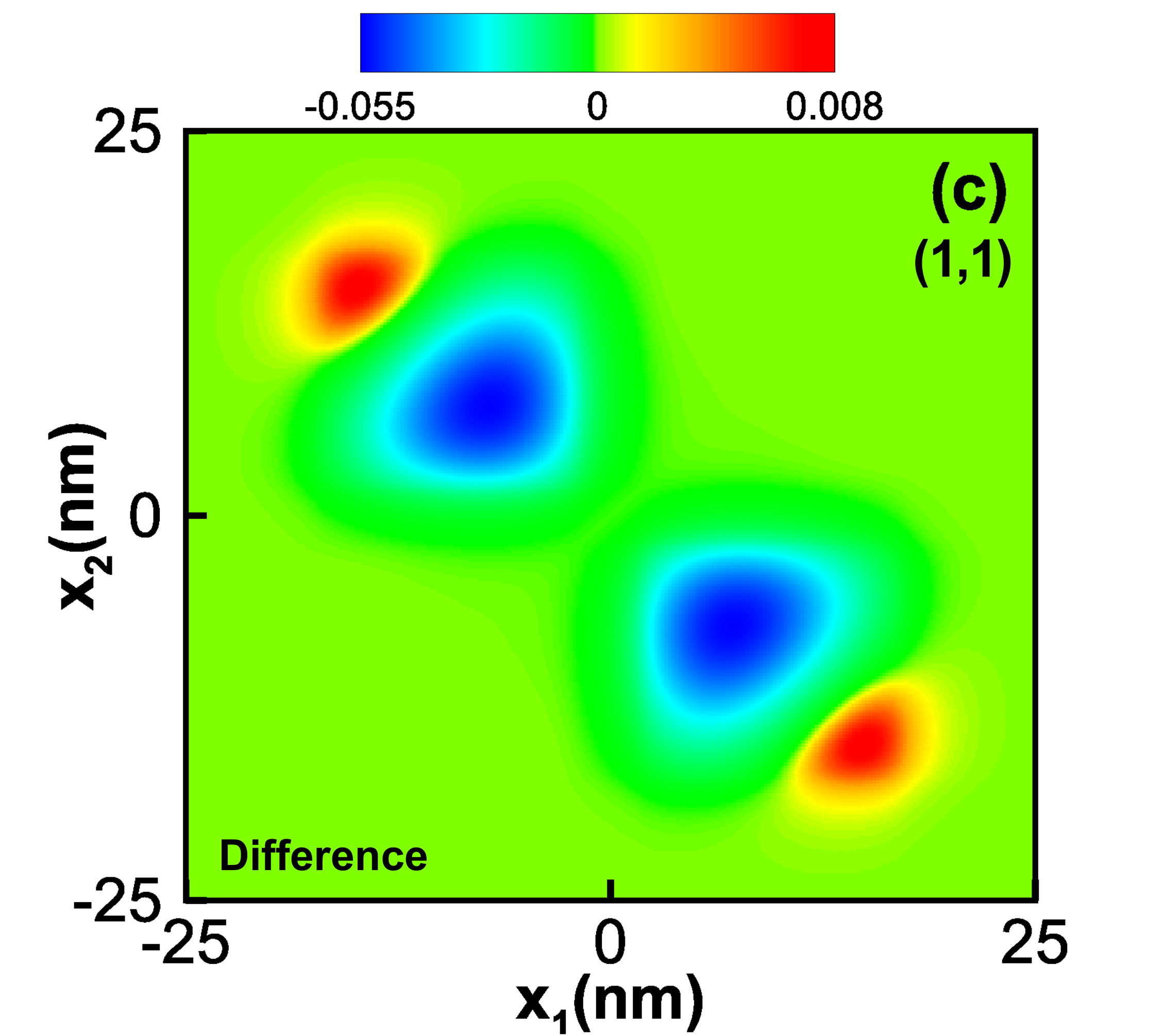}
		%	\caption{\label{fig:wave0_diff}}
        \end{subfigure}\\
        \vspace{0.2in}
	%\captionsetup[subfigure]{labelformat=empty}
	\begin{subfigure}[h!]{0.3\textwidth}
		\includegraphics[scale=0.25]{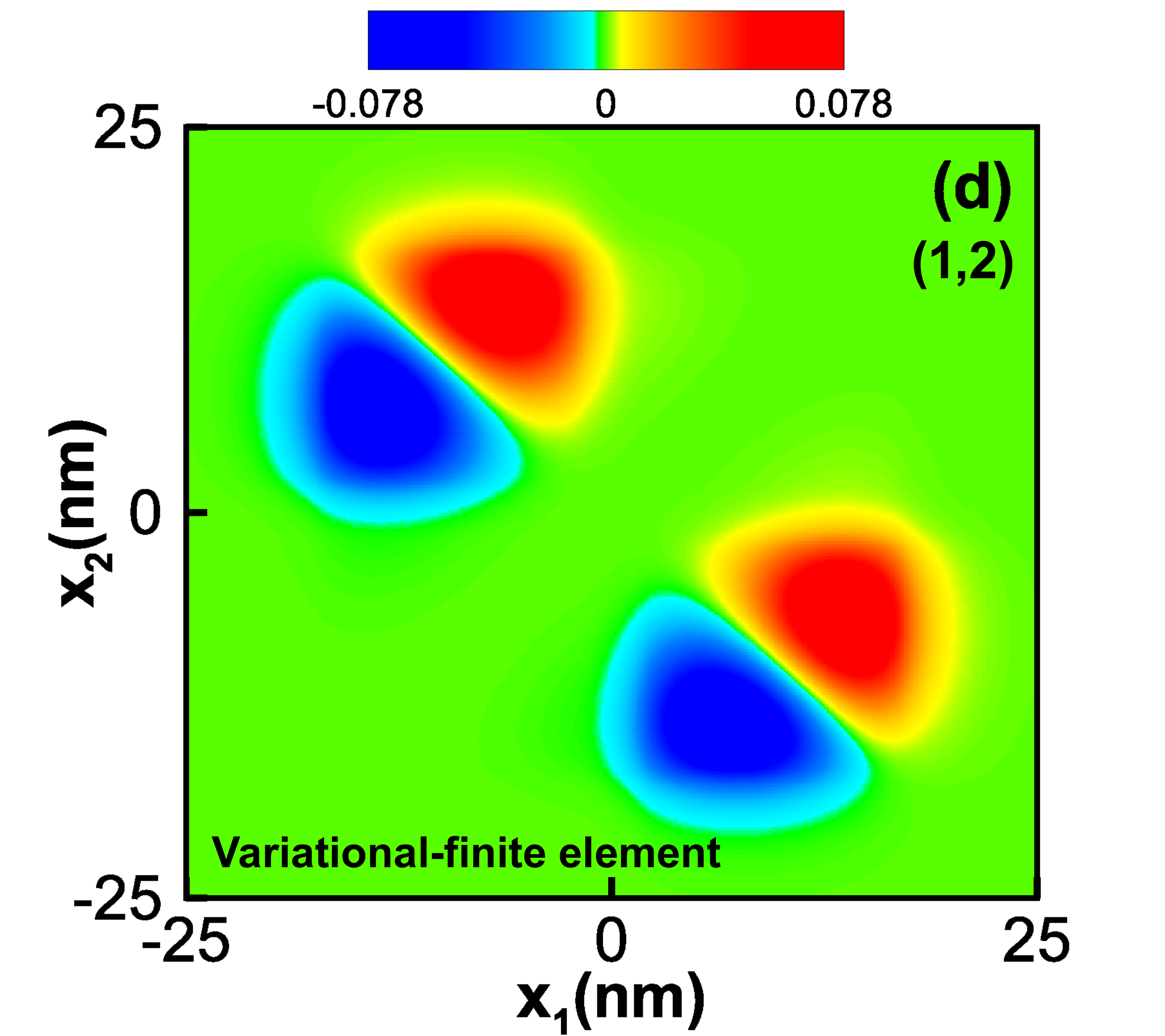}
		%	\caption{\label{fig:wave1_actual}}
	\end{subfigure}%
	%\captionsetup[subfigure]{labelformat=empty}
	\begin{subfigure}[h!]{0.3\textwidth}
		%	\centering
		\includegraphics[scale=0.25]{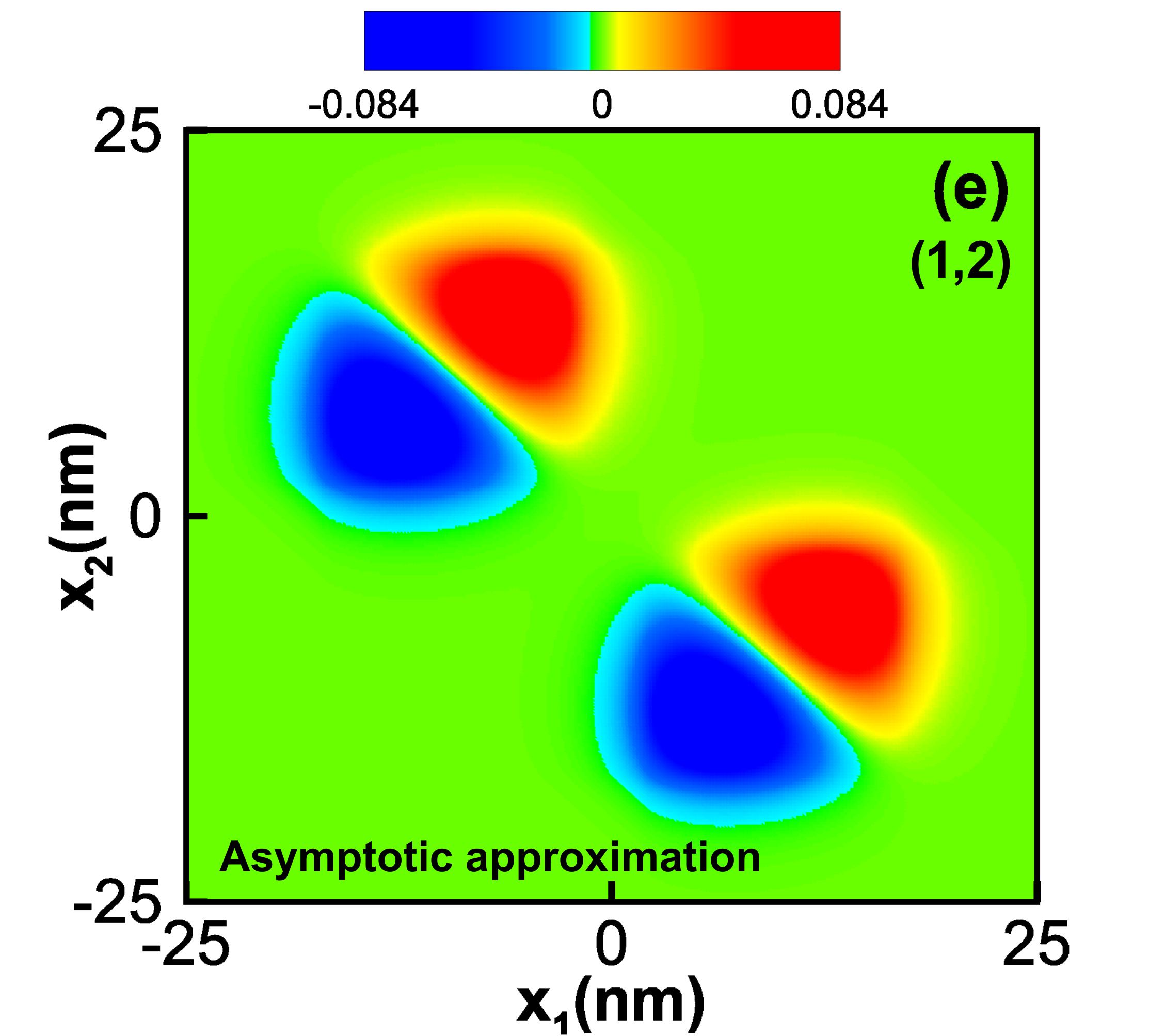}
		%	\includegraphics[scale=0.3]{./figures/png/symwell_150Awidth_40Asep_wave1_slater.png}
		%	\caption{\label{fig:wave1_slater}}
	\end{subfigure}%
	%\captionsetup[subfigure]{labelformat=empty}
	\begin{subfigure}[h!]{0.3\textwidth}
		%	\centering
		%	\includegraphics[scale=0.3]{./figures/png/symwell_150Awidth_40Asep_wave1_diff.png}
		\includegraphics[scale=0.25]{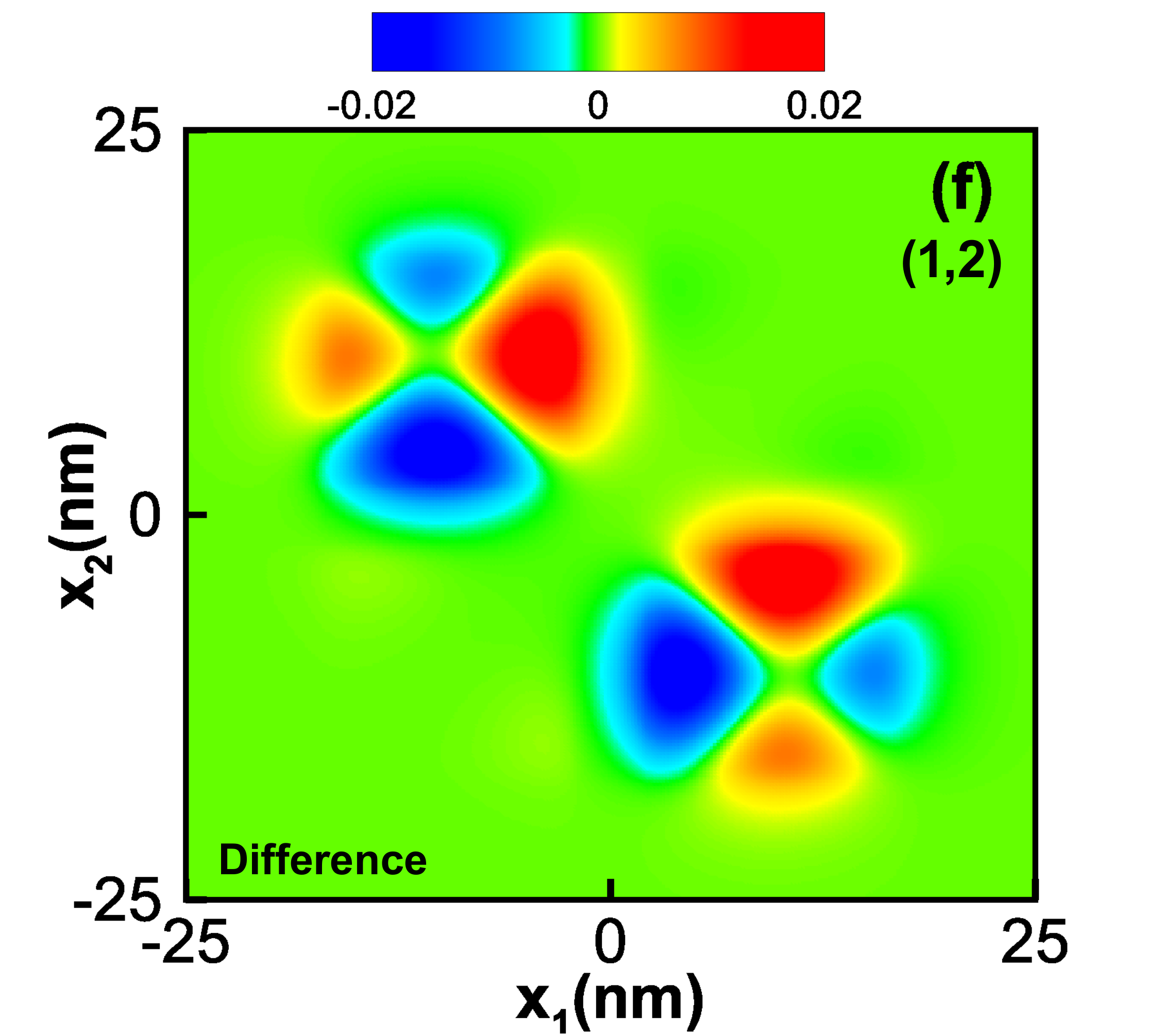}
		%	\caption{\label{fig:wave1_diff}}
	\end{subfigure}
	\caption{\label{fig:actual_vs_slater_compare}    Wavefunctions
          obtained through our  method for the state  (a) $(1,1)$, and
          (d) $(1,2)$ are shown  for the separation distance $d=4\,$nm
          between   the  QDs.   Wavefunctions  obtained   through  the
          asymptotic representation  in Eq.~(\ref{eq:asymptoticWave1})
          for the  state $(1,1)$  and $(1,2)$ are  shown in (b) and
          (e), respectively. The
          difference between  the two  representations are  plotted in
          (c)   and  (f).   The  width   of  each   quantum  dot   is
          w$_1=$w$_2=15\,$nm.}
\end{figure*}

Consider  a system  of  two  electrons in  the  conduction  band of  a
symmetric    $\rm{GaAs/Ga_{x}Al_{1-x}As}$    double   QDs. A realistic implementation of such a system can be a heterostructure nanowire composed of alternative layers of $\rm{GaAs}$ and $\rm{GaAlAs}$ stacked onto each other, as shown in Fig.~\ref{fig:geometry}.  For    a
stoichiometric ratio $x=0.37$ the potential depth is $V=0.276$ eV, the
effective  mass of  the  electron  $m^*_{in}=0.0665\,m_e$ inside,  and
$m^*_{out}=0.0858\,m_e$   outside  the   dots~\cite{band_parameters},
where  $m_e$  is  the  rest  mass  of  an  electron.   These  material
properties  are  considered for  all  calculations  presented in  this
paper. 

In  Table \ref{table:eig_distance},  eigenvalues  of  the first  eight
states in a symmetric double QD are presented for different separation
distances $d$.
We observe that as $d\rightarrow\,\infty$, we obtain 2- and 4-fold
degeneracies. This can be explained as follows.  

For  two-electron systems,  the  simplest approximation  of the
spatial   wavefunction   is   the   traditional   Slater   determinant
representation with single-particle wavefunctions, given by
\begin{align}\label{eq:slater}
\ket{\psi_{\rm spatial}}
 = \left\{\begin{array}{cc}
\DS\frac{1}{\sqrt{2}}\Big[\ket{a}_1\!\ket{b}_2 \pm \ket{b}_1\!\ket{a}_2 \Big], & a\neq b; \\
\ket{a}_1\!\ket{a}_2, & a=b,
\end{array}\right.
\end{align}
where  $\ket{a},\,\ket{b}$ are  the single-particle  wavefunctions for
the symmetric double  QD potential, and the subscript  is the particle
index.      The      total      wavefunction     is      given      by
\mbox{$\ket{\Psi}=\ket{\psi_{\rm      spatial}}\otimes\ket{S_{\rm
      spin}}$}.  Spatial  wavefunctions in  Eq.~(\ref{eq:slater})  are
associated    with   the  symmetric    or   antisymmetric    spin    parts
$\ket{S_{\rm spin}}$ so that the total wavefunction $\ket{\Psi}$
is antisymmetric under 
the  exchange  operator  $\hat{\mathcal{P}}$.   Note  that  the  above
representation leads to only singlets (for $a=b$) and 2-fold degenerate
states   (for   $a\neq   b$).   Moreover,   the   Slater   determinant
representation neglects  the Coulomb interaction that  is particularly
important for QDs in
proximity.

Here,  we first give  a  representation which  is  accurate at  the
asymptotic  limit  of large  separation distance $d$  between the
dots,  and then
explain       the      degeneracy       pattern      observed       in
Table.~\ref{table:eig_distance}.                                   Let
\mbox{$\mathbb{S}(n,i)        =        \left\{\ket{n,\alpha}_i\big{|}\
    \alpha\in\mathbb{N}\right\},$} be the basis set for an electron in
a  single  QD, where  $n=1,2$  is  the  dot  index, $\alpha$  is  the
single-particle quantum number, and $i=1,2$  is the particle index. In
the asymptotic limit, the Hamiltonian commutes ($i$) with the exchange
operator $\hat{\mathcal{P}}$, which permutes the particle indices, and
($ii$) with $\hat{\mathcal{W}}$, which permutes  the dot indices. Also, as
$d\rightarrow  \infty$,  the  Coulomb  interaction  vanishes  and  the
electrons are bound to one of  the QDs with zero tunneling probability
to the  other dot.   Hence we derive the spatial  part of the
wavefunction   as   linear   combinations  of   the   eigenstates   of
$\mathbb{S}(n,i)$ given by
\begin{widetext}
\begin{align}
 \hspace{-1in}\ket{\psi_{\rm s}^+} =
 & \frac{1}{2\sqrt{1+\delta_{\alpha\beta}}}\bigg[\Big(
   \ket{1,\alpha}_1\!\ket{2,\beta}_2 +
   \ket{1,\beta}_1\!\ket{2,\alpha}_2 \Big) +
   \Big(\ket{2,\alpha}_1\!\ket{1,\beta}_2 +
   \ket{2,\beta}_1\!\ket{1,\alpha}_2
   \Big)\bigg];\label{eq:asymptoticWave1}\\ 
		\hspace{-1in}\ket{\psi_{\rm s}^-} =
		& \frac{1}{2}\bigg[\Big(
           \ket{1,\alpha}_1\!\ket{2,\beta}_2 -
           \ket{1,\beta}_1\!\ket{2,\alpha}_2 \Big) -
           \Big(\ket{2,\alpha}_1\!\ket{1,\beta}_2 -
           \ket{2,\beta}_1\!\ket{1,\alpha}_2
           \Big)\bigg]\label{eq:asymptoticWave2};  \\
%\end{align}
%\end{widetext}
%\begin{widetext}
%\begin{align}
		\hspace{-1in}\ket{\psi_{\rm a}^+} =
		& \frac{1}{2}\bigg[\Big(
           \ket{1,\alpha}_1\!\ket{2,\beta}_2 -
           \ket{1,\beta}_1\!\ket{2,\alpha}_2 \Big) +
           \Big(\ket{2,\alpha}_1\!\ket{1,\beta}_2 -
           \ket{2,\beta}_1\!\ket{1,\alpha}_2
           \Big)\bigg];\label{eq:asymptoticWave3}\\ 
		\hspace{-1in}\ket{\psi_{\rm a}^-} =
		& \frac{1}{2\sqrt{1+\delta_{\alpha\beta}}}\bigg[\Big(
           \ket{1,\alpha}_1\!\ket{2,\beta}_2 +
           \ket{1,\beta}_1\!\ket{2,\alpha}_2 \Big) -
           \Big(\ket{2,\alpha}_1\!\ket{1,\beta}_2 +
           \ket{2,\beta}_1\!\ket{1,\alpha}_2
           \Big)\bigg],\label{eq:asymptoticWave4}	 	
\end{align}
\end{widetext}
where \mbox{$\hat{\mathcal{P}}\ket{\psi_{\rm
      s}^\pm}\!=\!\ket{\psi_{\rm s}^\pm}$},
\mbox{$\hat{\mathcal{P}}\ket{\psi_{\rm a}^\pm}\!=\!-\ket{\psi_{\rm
      a}^\pm}$}, \mbox{$\hat{\mathcal{W}}\ket{\psi_{\rm
      s}^\pm}\!=\!\pm \ket{\psi_{\rm s}^\pm}$}, and
\mbox{$\hat{\mathcal{W}}\ket{\psi_{\rm a}^\pm}\!=\!\pm \ket{\psi_{\rm
      a}^\pm}$}. As before, spatial wavefunctions in
Eqs.~(\ref{eq:asymptoticWave1})-(\ref{eq:asymptoticWave4}) are
associated with the symmetric or anti-symmetric spin parts
$\ket{S_{\rm spin}}$ so that $\ket{\Psi}$ is antisymmetric under the
exchange operator $\hat{\mathcal{P}}$. If $\alpha\!\neq\!\beta$, then
the four wavefunctions in
Eqs.\,(\ref{eq:asymptoticWave1})-(\ref{eq:asymptoticWave4}) form a set
of four-fold degenerate states. If, on the other hand,  $\alpha\! =\!\beta$ then
Eqs.~(\ref{eq:asymptoticWave2}) and (\ref{eq:asymptoticWave3})
vanish, resulting in a doublet. 
 
\par  For  a  finite  distance   $d$  between  the  QDs,  the  Coulomb
interaction  splits these  degeneracies,  and  the representations  in
Eqs.\,(\ref{eq:asymptoticWave1})-(\ref{eq:asymptoticWave4})   are   no
longer exact.  In  Table.~\ref{table:eig_distance}, we have classified
the eigenvalues in terms of the quantum numbers $(\alpha,\beta)$.  The
representation  $\ket{\psi_{\rm  s}^+}$,  the  accurate  wavefunctions
obtained from the technique described in Sec.~\ref{sec:fem}, and their
difference for the quantum numbers  $(1,1)$ and $(1,2)$ are plotted in
Figs.~\ref{fig:actual_vs_slater_compare}(a)-\ref{fig:actual_vs_slater_compare}(f). From
Figs.~\ref{fig:actual_vs_slater_compare}(c)                        and
\ref{fig:actual_vs_slater_compare}(f), we  see that  the disagreements
are significant  as a  consequence of  the Coulomb  interaction, which
impact  on  the amount  of  spatial  entanglement.  In  the  following
section we calculate the entanglement  in such double QDs, and discuss
the observed resonances.

%%%%%%%%%%%%%%%%%%%%%%%%%%%%%%
\section{Spatial entanglement in quantum dots}\label{sec:Ent_QD} %Sec 4
%%%%%%%%%%%%%%%%%%%%%%%%%%%%%% 
\subsection{Formalism} %Sec 4a
Any state  that describes a system  of identical fermions has  to obey
the antisymmetry  under the exchange  operator.  This makes  the state
necessarily nonseparable,  and hence  entangled.  The lowest  level of
entanglement is  provided by  the use of  single-particle states  in a
Slater   determinant   form    of   the   wavefunction~\cite{Ichikawa,
  Schliemann2,Ghirardi,Plastino,Killoran,Buscemi,Naudts}.   This level
of entanglement  may be  thought of  as a baseline  with which  a more
complete  calculation of  the entanglement  can be  compared.  In  the
following,  we  quantify entanglement  by  measuring  the {\it  linear
  entropy} of  the system.   It is  known in  the literature  that the
linear    entropy    is    a   good    indicator    of    entanglement
~\cite{Zanardi,Buscemi,Coe,Plastino,KamHo_He},  and  can  be  computed
efficiently even for a very large Hilbert space. The linear entropy is
defined as
\begin{equation}
{\cal E}_\ell = 1 - \rm{Tr}(\rho_1^2),
\end{equation}
where  $\rho_1$  is  the  reduced density  matrix.  Here  $\rho_1$  is
obtained by taking the trace over the second particle
\begin{equation}
\rho_1 = \rm{Tr}_2(\rho),
\end{equation}
where $\rho$ is the density matrix of the whole system. The spatial
and spin components are separable, and the density matrix can be
written as a tensor product of each contribution 
\begin{equation}
\rho = \rho_{{\rm spatial}}\otimes\rho_{{\rm spin}},
\end{equation} 
and 
\begin{equation}
\rm{Tr}(\rho_1^2) = \rm{Tr}(\rho_{spatial\,1}^2)\rm{Tr}(\rho_{spin\,1}^2).
\end{equation}
Since the trace of the spin and spatial parts are entirely separable, we
can consider them separately. For a two-electron system, the spin
contributions are readily evaluated as  
\begin{align}[left = {\rm{Tr}(\rho^2_{spin1}) = \empheqlbrace}]
& 1,\quad \ket{\uparrow\uparrow} \text{or} \ket{\downarrow\downarrow} \nonumber\\
& 0.5,\quad  \frac{1}{\sqrt{2}}\big[\ket{\uparrow\downarrow} \pm
 \ket{\downarrow\uparrow}\big]. 
\end{align} 
Thus  the  contribution  of  the  spin part  to  the  total  trace  of
$\rho^2_1$ is a constant. Hence, we lay it aside from our consideration of
the spatial  entanglement. In  the following the  spatial part  of the
density matrix is  referred to as $\rho$, and we  consider the measure
of spatial entanglement to be
\begin{eqnarray}
  {\cal E}_\ell       &    =  &        1-           \rm{Tr}(\rho^2_{{\rm
      spatial}\,1}) \nonumber \\
  &=&
  1-\int_{-\infty}^{+\infty}\!\!\int_{-\infty}^{+\infty}\!\!|
  \bra{\mathbf{r'_1}}\rho_1\ket{\mathbf{r_1}}|^2d\mathbf{r'_1}\rm{d}\mathbf{r_1}. 
\end{eqnarray}

Electron distributions  and the evolution of  the spatial entanglement
with varying parameters  are found to be very  analogous for symmetric
and  antisymmetric   partners,  and  a  detailed   comparison  of  the
entanglement  properties of  symmetric  and  antisymmetric states  are
given  in  the supplementary materials.  Therefore,  for
further analysis  in this paper, we  only discuss the results  for the
symmetric  states;  the   same  conclusions  can  be   drawn  for  the
antisymmetric case.

%%%%%%%%%%%%%%%%%%%%%%%%%%%%%%
\subsection{Symmetric double quantum dot}\label{subsec:sym} %Sec 4b
%%%%%%%%%%%%%%%%%%%%%%%%%%%%%%

\begin{figure*}[t!] %Fig 4
	\begin{subfigure}[h!]{0.28\textwidth}
		%\centering
		\includegraphics[scale=0.3]{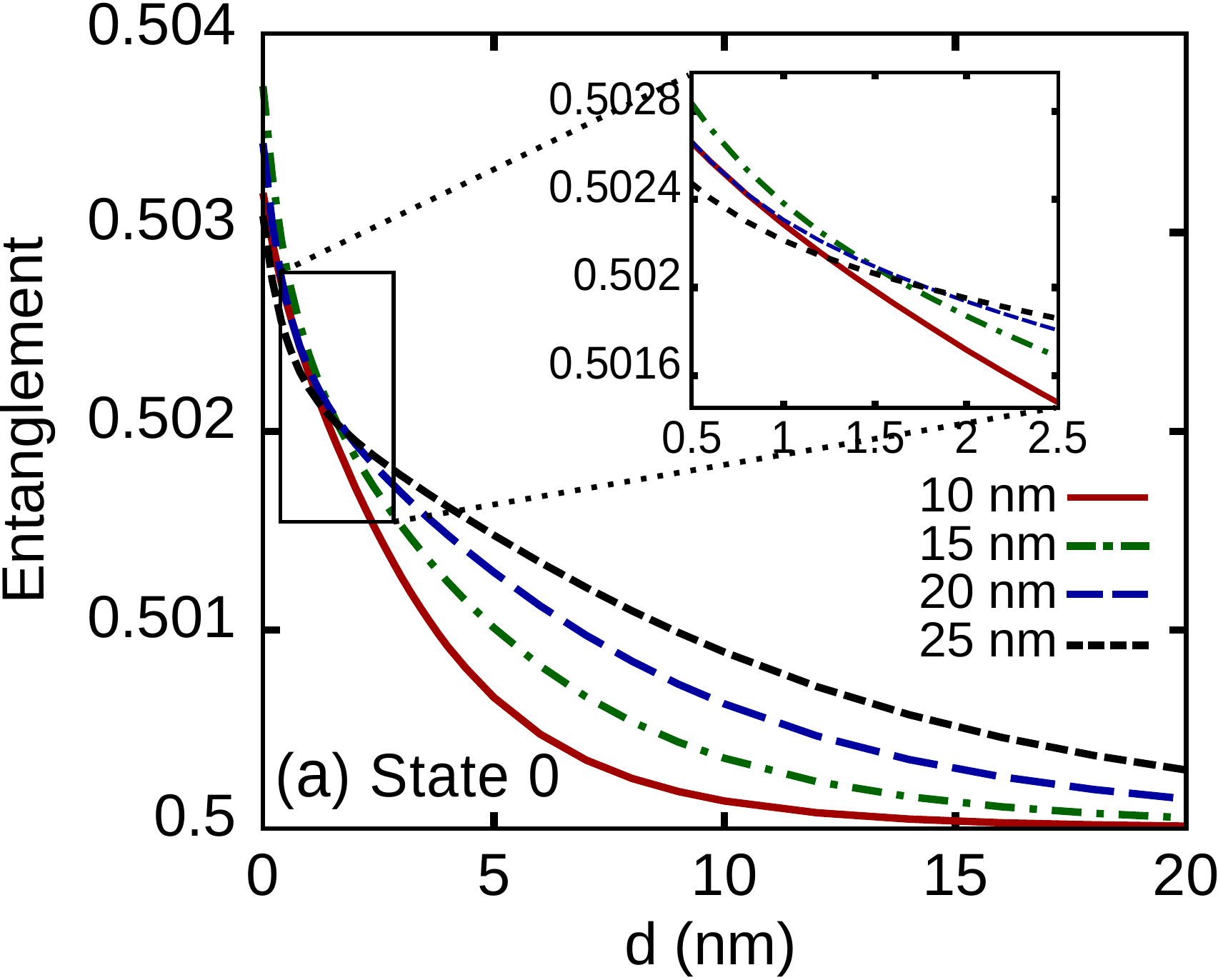}
		%\caption{State 0}
		%\label{fig:sym_ent_vs_distance0}
	\end{subfigure}%
	\begin{subfigure}[h!]{0.3\textwidth}
		%\centering
 \includegraphics[scale=0.3]{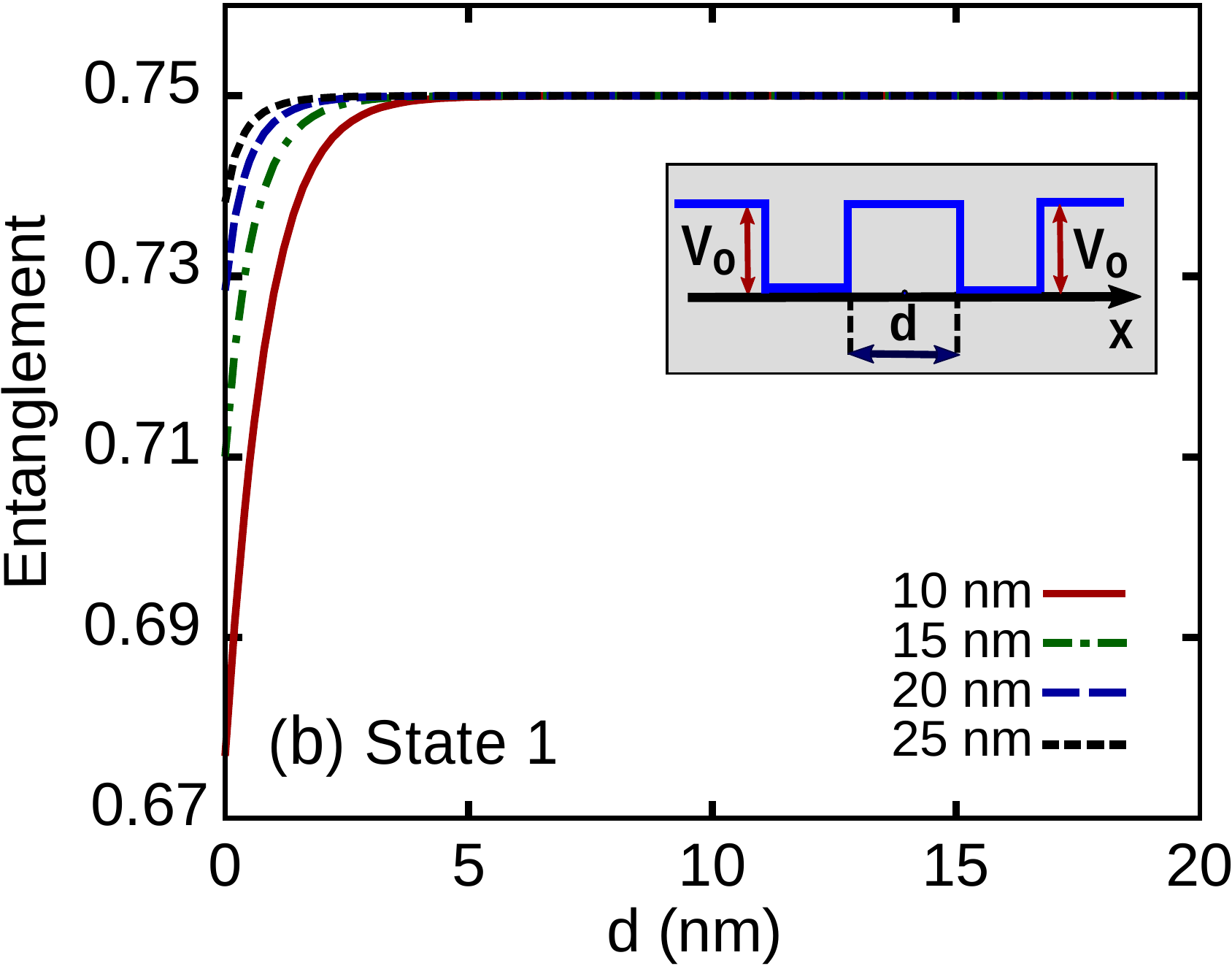}
		%\caption{State 1}
		%\label{fig:sym_ent_vs_distance1}
	\end{subfigure}%
	\begin{subfigure}[h!]{0.3\textwidth}
		%\centering
		\includegraphics[scale=0.3]{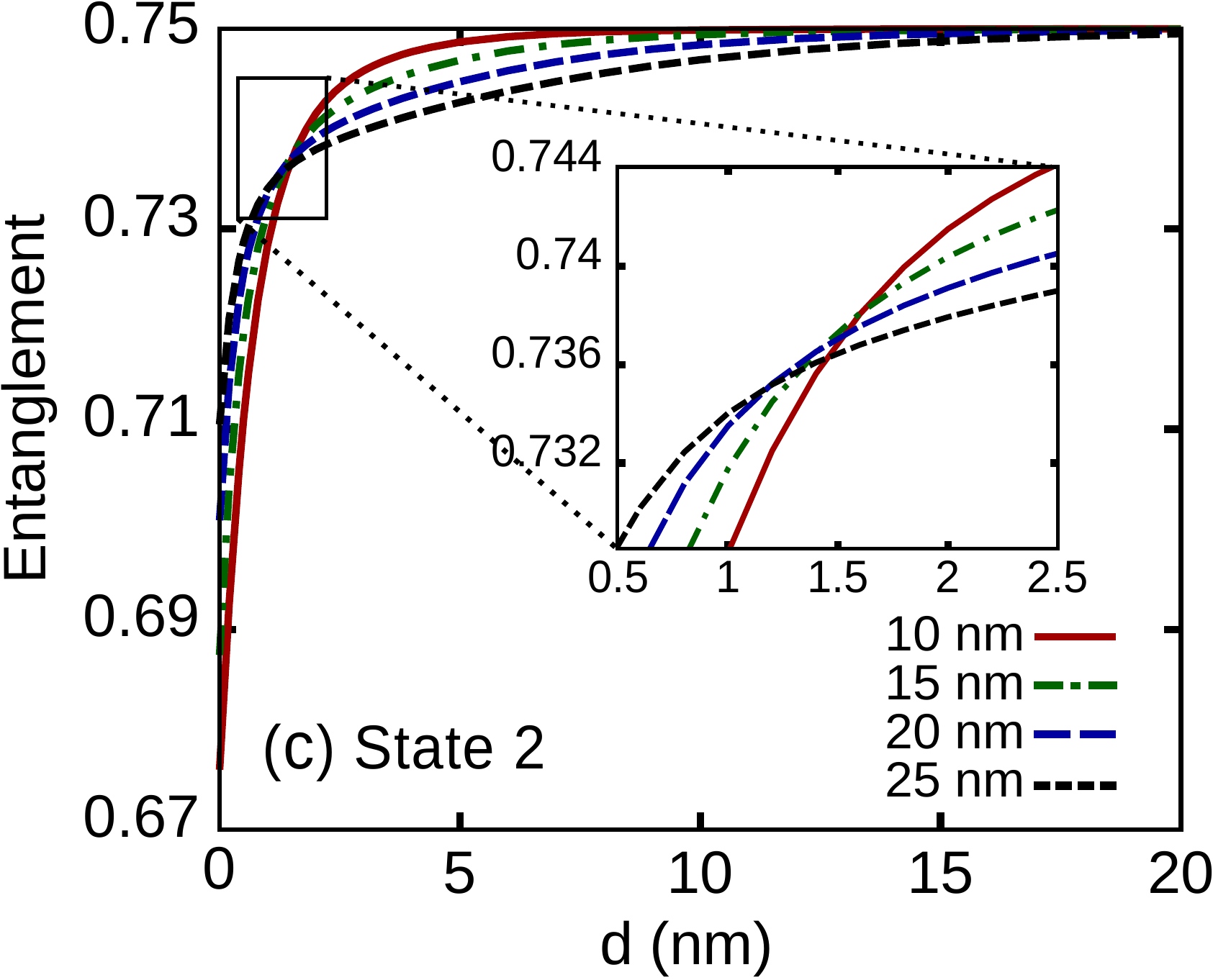}
		%\caption{State 2}
		%\label{fig:sym_ent_vs_distance2}
	\end{subfigure}\\
        
        \vspace*{0.2in}
	\begin{subfigure}[h!]{0.3\textwidth}
		\includegraphics[scale=0.3]{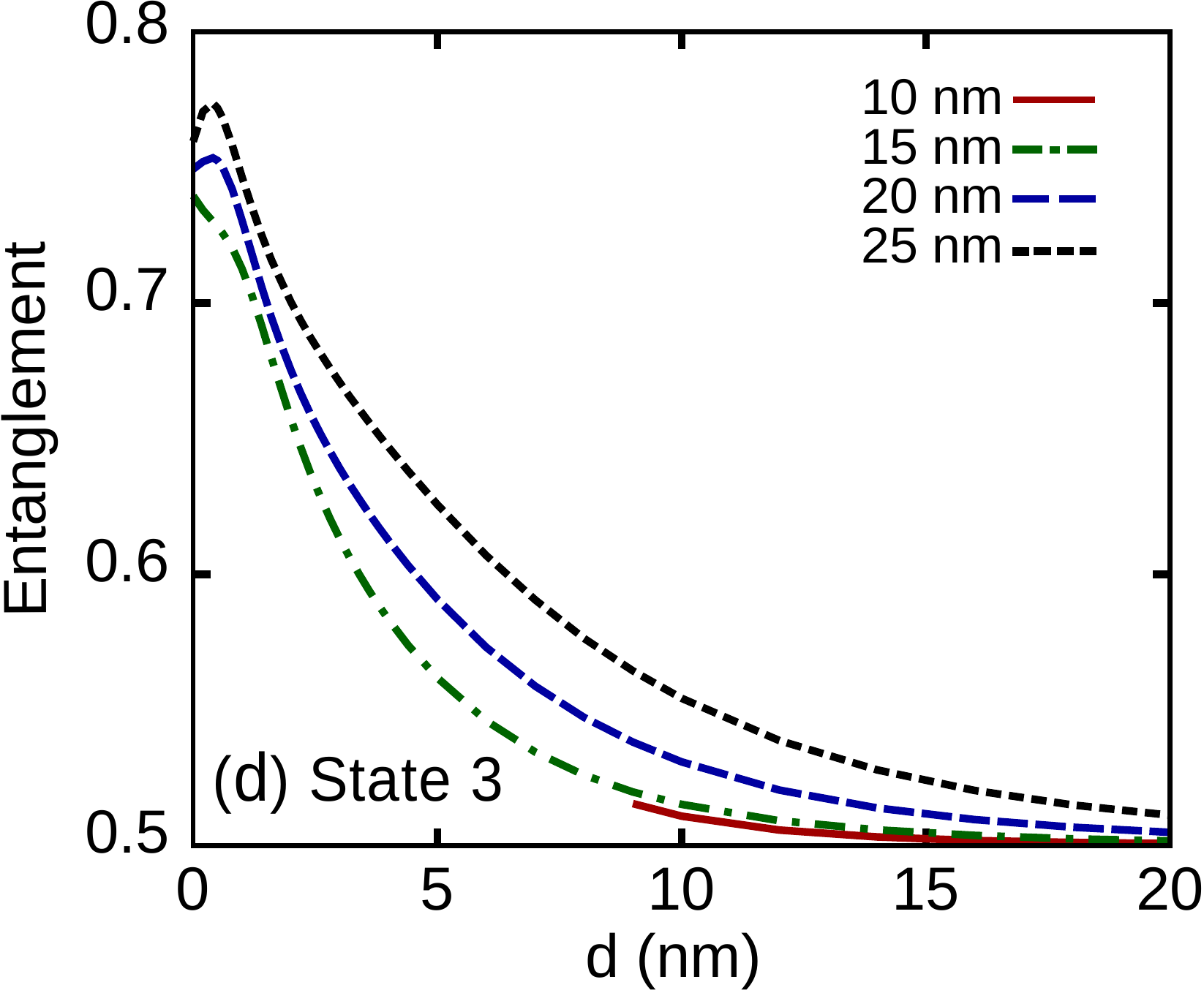}
		%\caption{State 3}
		%\label{fig:sym_ent_vs_distance3}
	\end{subfigure}%
	\begin{subfigure}[h!]{0.3\textwidth}
		%\centering
		\includegraphics[scale=0.3]{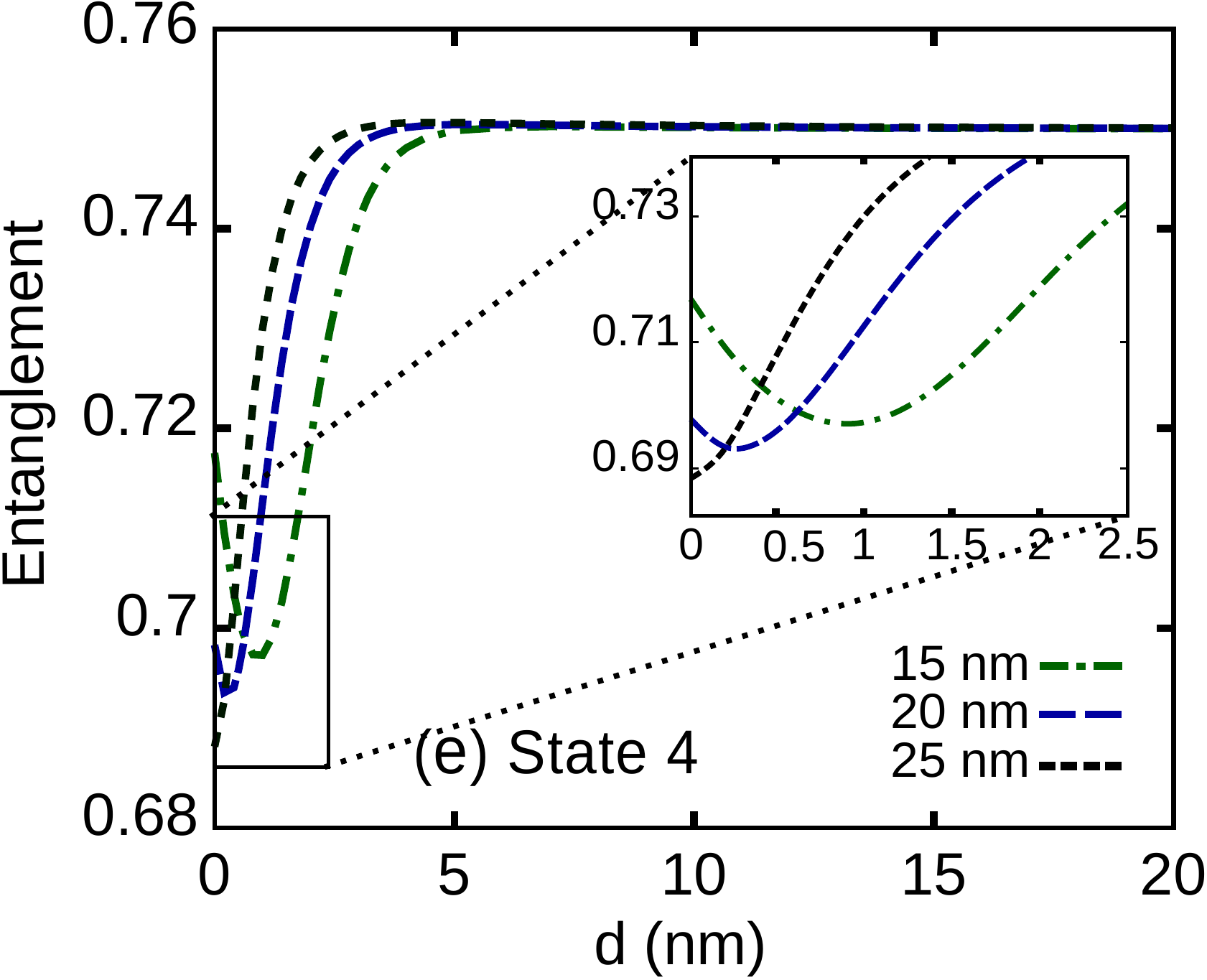}
		%\caption{State 4}
		%\label{fig:sym_ent_vs_distance4}
	\end{subfigure}%
	\begin{subfigure}[h!]{0.3\textwidth}
		%\centering
		\includegraphics[scale=0.3]{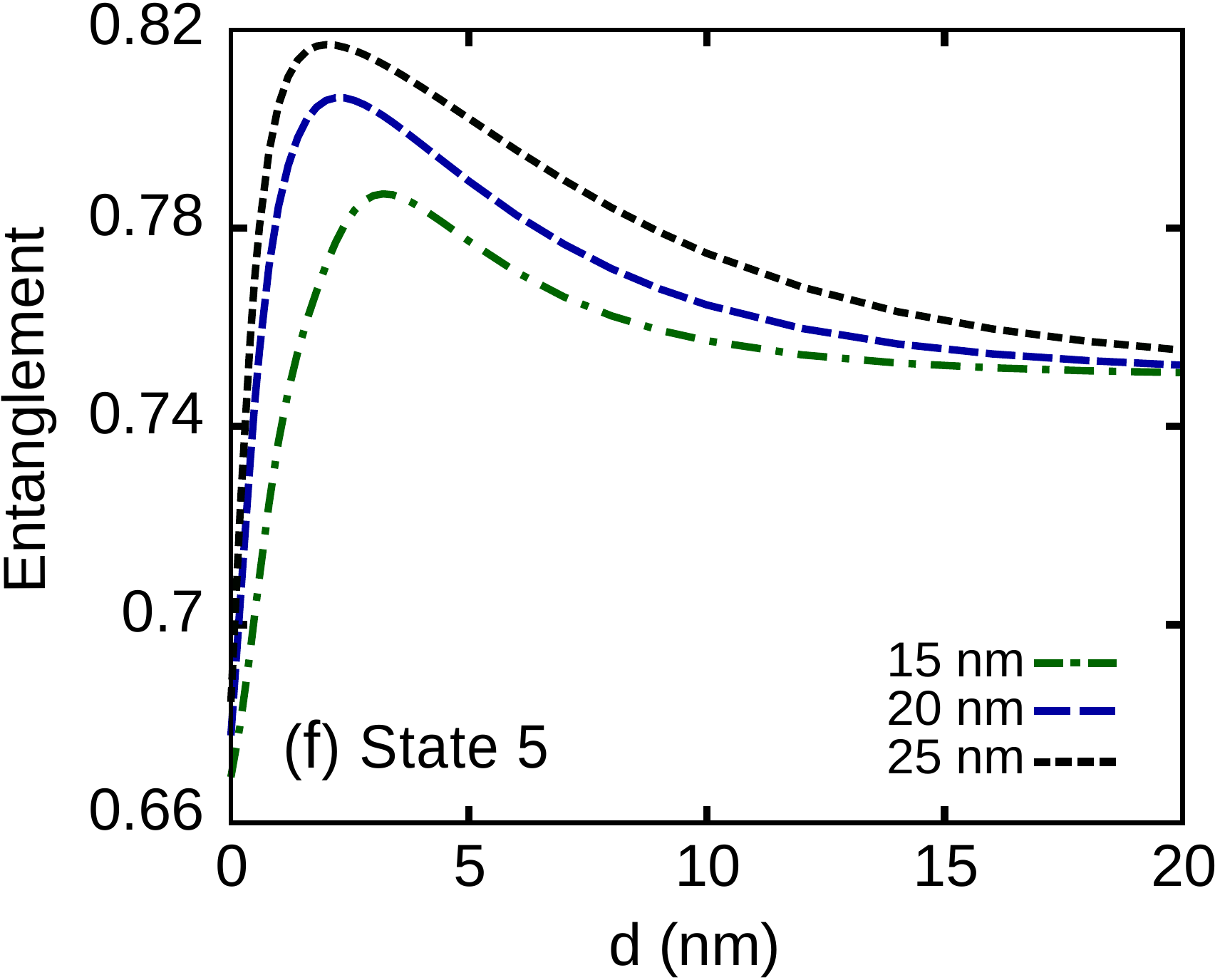}
		%\caption{State 5}
		%\label{fig:sym_ent_vs_distance5}
	\end{subfigure}%
	\caption{\label{fig:sym_ent_vs_distance}Spatial   entanglement
          for  the  first  five  eigenstates of  two  electrons  in  a
          symmetric  double QD  system  is plotted  as  a function  of
          separation distance  $d$, for  four different widths  of the
          QDs:  $10\,  {\rm  nm},15\,  {\rm nm},20\,  {\rm  nm},$  and
          $25\, {\rm nm}$. The  entanglement values saturate to either
          0.5 or 0.75 in the limit $d\rightarrow \infty$.}
\end{figure*}

                                                               In
Figs.~\ref{fig:sym_ent_vs_distance}(a)--\ref{fig:sym_ent_vs_distance}(f),
spatial entanglement of a double dot is plotted as a function of $d$.
We observe  that as $d \rightarrow\infty$,  spatial entanglement ${\cal E}_\ell$
saturates to either  $0.5$ or $0.75$. In fact, in the  asymptotic limit we
see that
\begin{align}[left = {{\cal E}_\ell = \empheqlbrace}]
& 0.5, \quad \alpha=\beta \nonumber\\
& 0.75,\quad \alpha\neq\beta
\end{align}
as                               shown                              in
Figs.\,\ref{fig:sym_ent_vs_distance}(a)--\ref{fig:sym_ent_vs_distance}(f). This
observation can be confirmed by computing analytically ${\cal E}_\ell$
with  the   asymptotic  forms   of  the  wavefunctions   presented  in
Eqs.~(\ref{eq:asymptoticWave1})--(\ref{eq:asymptoticWave4}),   as   in
Appendix \ref{subsec:asymptotic_ent}.

\par One  can also interpret  this result  from a probabilistic  point of
view by  considering the behavior of  the bound states in  a symmetric
system of  two indistinguishable particles. In the  asymptotic limit, each
particle is  located in  a different quantum  dot, unaffected  by the
presence  of the  other particle.  Then, states  $\alpha$ and  $\beta$
describe separated electrons exactly.  For $\alpha=\beta$, there is no
ambiguity  as to  which   single-particle   state   the  particles   are
occupying. However,  due to the indistinguishability  of the electrons
and the QDs being symmetrical, there  is only 0.5 probability of where
each particle is located -- each electron  can either be in QD\,1 or 2,
while the second   electron is then in the other QD.  This 0.5 certainty is
reflected in  the measurement of  the entanglement. On the other  hand, if
$\alpha\neq\beta$,  then there  is an  additional uncertainty;  we not
only have  50\% information on where  a particle is located,  but also
50\%  on what  state the  particle is  occupying in  each dot  (state
$\alpha$  or   $\beta$).   Therefore,  the  amount   of  certainty  is
$0.5\times0.5  =  0.25$;  the  amount  of  entanglement  is  therefore
$1-0.25=0.75$. Note that the  interpretation and the asymptotic values
of entanglement are valid for any type of quantum dot irrespective of
dimensionality and geometry,  as long as the two QDs  are identical to
each other.
The lower bound of the spatial entanglement is observed to be
0.5. This is due to electrons being indistinguishable, and hence the
wavefunction is an eigenfunction of the operator $\hat{\cal  P}$. However,
 if the electrons are distinguished through their spin
directions then the overall entanglement vanishes.

 In         Figs.~\ref{fig:sym_ent_vs_distance}(a)        --
\ref{fig:sym_ent_vs_distance}(c), we see that as $d\rightarrow\infty$,
the entanglement values of the  first three bound states monotonically
reach the saturation values.  The  situation for higher states is more
complex, as  can be  seen in  Figs.~\ref{fig:sym_ent_vs_distance}(d) --
\ref{fig:sym_ent_vs_distance}(f).    For   states   3   and   5,   the
entanglement  reaches a  maximum,  while for  state 4  it  drops to  a
minimum, before approaching the  corresponding asymptotic values.  Such
extrema in the behavior of the  entanglement are a result of two major
competing effects in the system:  ($i$) wavefunction overlapping, and ($ii$)
the  Coulomb  repulsion.   As  the QDs   are  brought  into
proximity,   single-particle  wavefunctions   of  the   two  electrons
localized in these dots can overlap each other. At the same time, the repulsive
Coulomb interaction  between the  electrons becomes stronger  at closer
distances  and  opposes  such   overlap  of  the  wavefunctions.   The
influence  of these  effects  is visible  for  entanglement values  of
excited states  (states 3,  4 and  5), where  a larger  probability of
finding  the  electron  is  distributed outside  the  QDs.   Then  the
wavefunction of one  electron is susceptible to  interactions with the
wavefunction  of the  electron in  the neighboring  QD.  Due  to these
competing effects,  the entanglement  of the system  develops extremal
points, indicating  a switch in  the roles of the two  effects. When the  QDs are
placed  far   apart,  both  the Coulomb interaction  and  the  overlap  of
single-electron   wavefunctions   become    insignificant,   and   the
entanglement reaches a saturation value at the asymptotic limit.

%%%%%%%%%%%%%%%%%%%%%%%%%%%%%%
\section{Asymmetric systems}\label{sec:non_sym} %Sec 5
%%%%%%%%%%%%%%%%%%%%%%%%%%%%%%

\begin{figure*}[ht!] %Fig 5
 \includegraphics[scale=0.33]{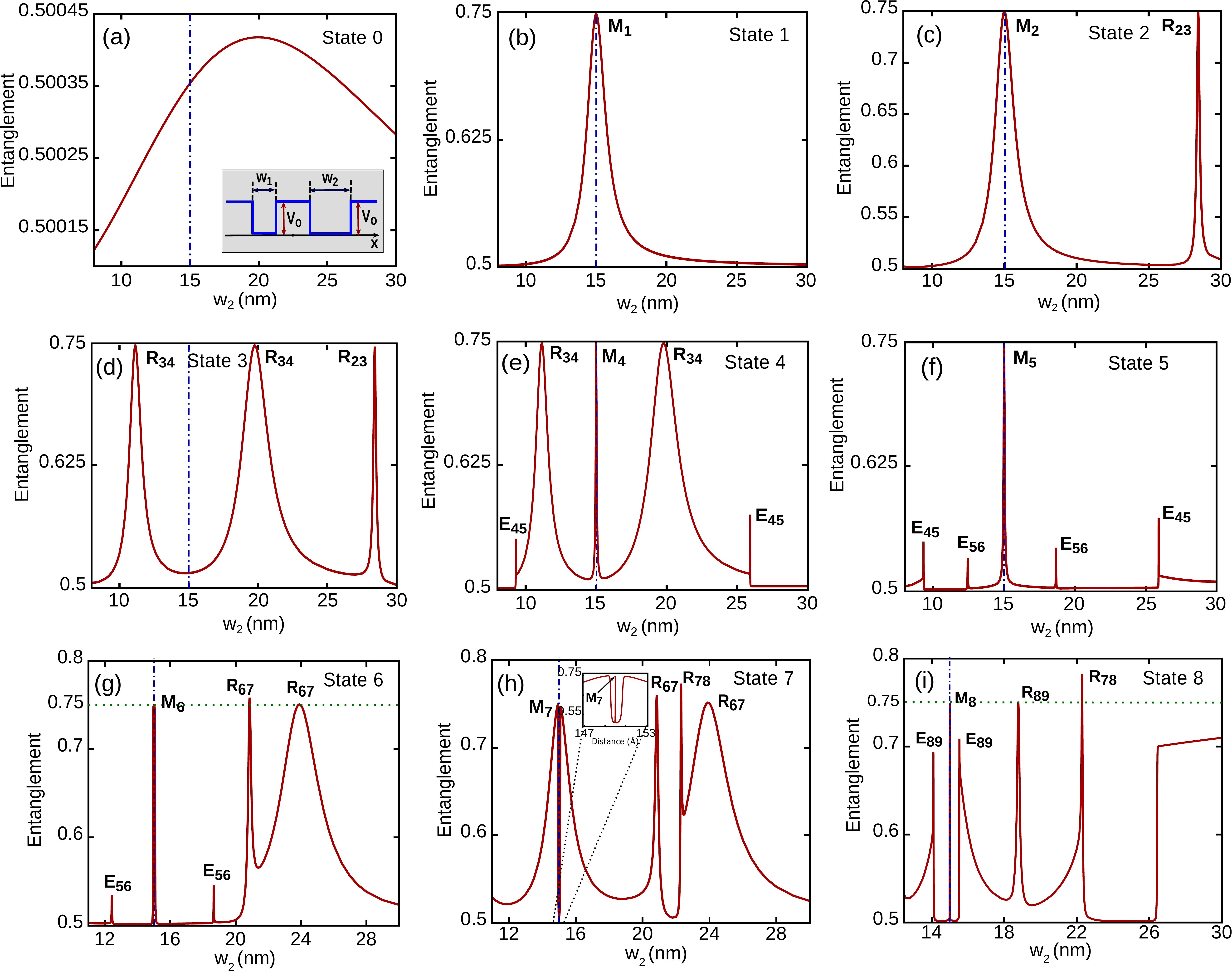}
 \caption{\label{fig:asym_width_ent}  Spatial entanglement  values for
   the first 8 eigenstates of  two electrons in a non-symmetric double
   QD are plotted as  a function of width of the  second QD w$_2$ (see
   inset in  (a)). Width of  the first  QD is w$_1=15\,{\rm  nm}$, and
   kept constant for all calculations.  The distant between the QDs is
   $d=10\,{\rm nm}$.  Here,  the resonance peaks (i) $M_i$  are due to
   the mirror symmetry of the system, (ii) $R_{ij}$ are due to avoided
   level-crossings  (anti-crossings) between  states $i\,$  and $\,j$,
   and (iii) $E_{ij}$ are due to the formation/dissolution of electron
   clusters.}
\end{figure*}
\begin{figure}[bh!] %Fig 6
 \begin{subfigure}[h!]{0.19\textwidth}
 \hspace{-0.6in}\includegraphics[width=1.7in]{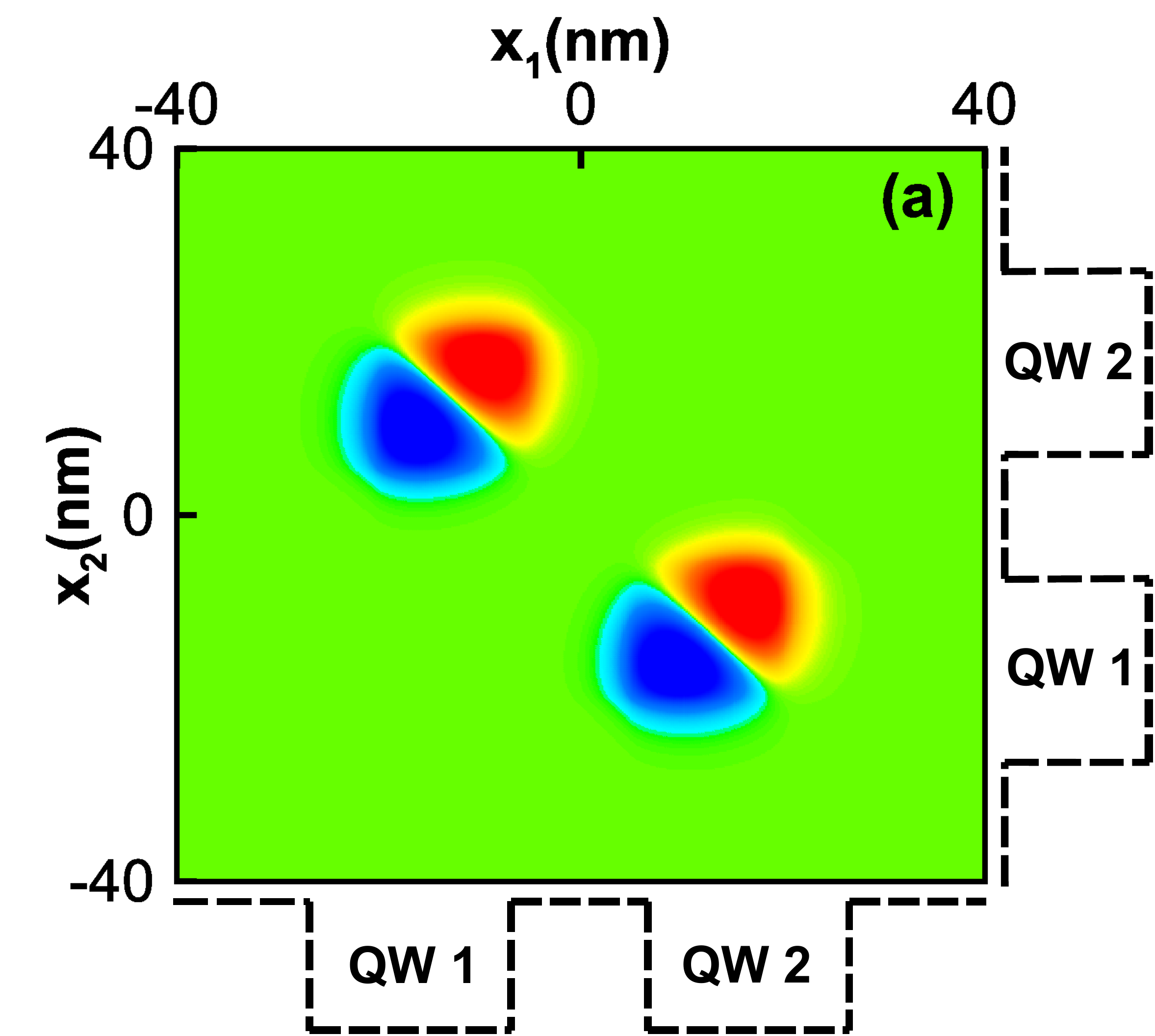}
 \end{subfigure}%
 \begin{subfigure}[h!]{0.19\textwidth}
 \hspace{-0.3in}\includegraphics[width=1.7in]{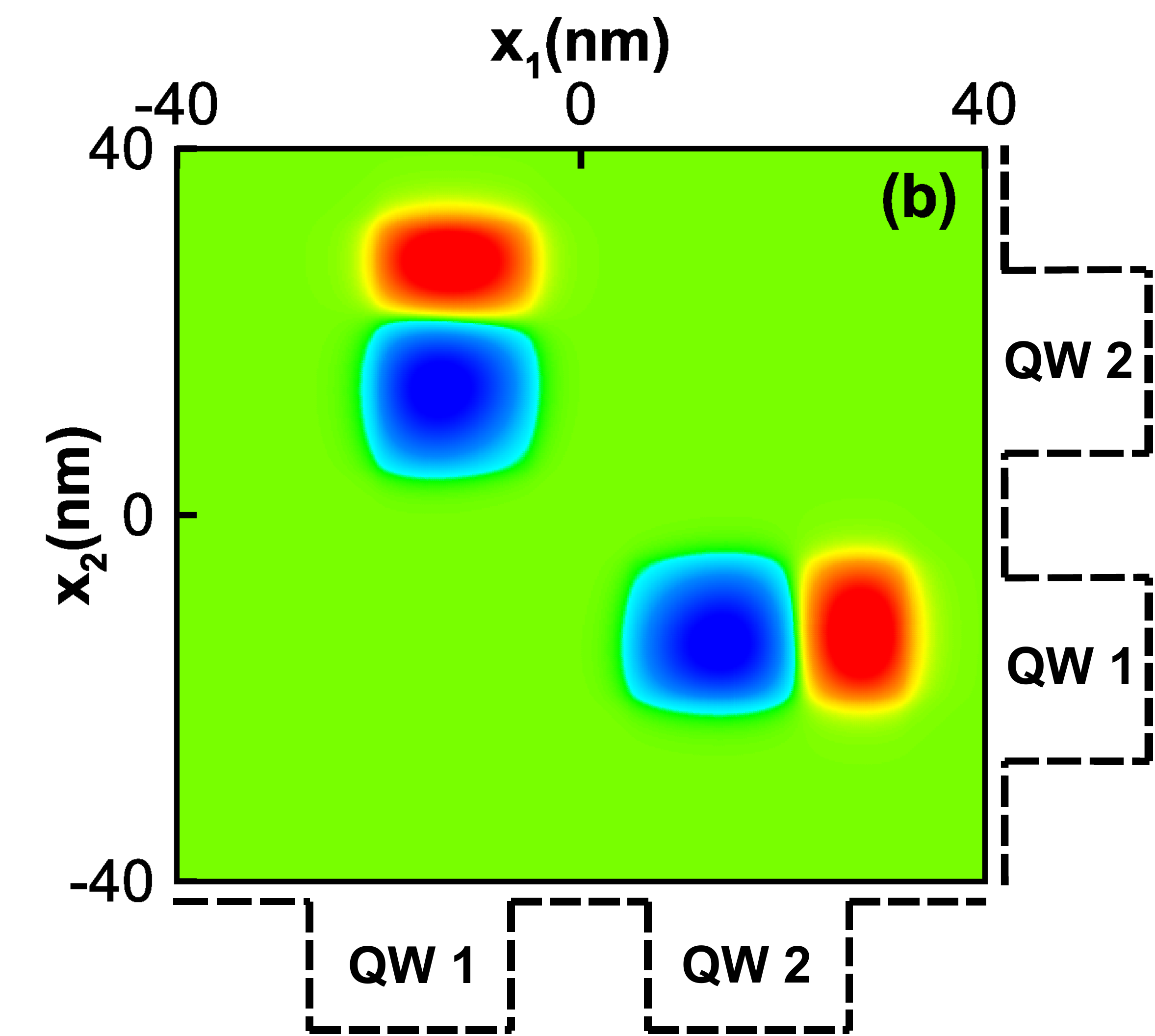}
 \end{subfigure}%
 \caption{\label{fig:asym_width_wave1}  Wavefunctions   of  the  first
   excited  state are  plotted  for: (a)  a  symmetric system  (w$_1=$
   w$_2$) and (b) when the  symmetry is broken (w$_1\neq$ w$_2$). Axes
   $x_1$  and   $x_2$  are  the   coordinates  of  $1^{\rm   st}$  and
   $2^{\rm nd}$ electron.}
\end{figure}

In this section we discuss  how asymmetry in the configuring potential
affects the entanglement. Consider a  system of asymmetric double QDs,
in which the width of the first QD is fixed at $15\,{\rm nm}$, and the
width of the second QD is varied. The distance between the QDs is held
constant at $10\,{\rm  nm}$. The entanglement of the  ground state and
the next eight  excited states as a function of  the second QD's width
w$_2$                 are                  plotted                 in
Figs.~\ref{fig:asym_width_ent}(a)-\ref{fig:asym_width_ent}(i).      In
Fig.~\ref{fig:asym_width_ent}(a), we  see that  there is only  a small
variation  in the  amount of  entanglement  of the  ground state  with
w$_2$. This is because both electrons are in the ground state, and the
change  in w$_2$  will not  lead to  any interaction  with the  higher
energy  levels.  In  other words,  the electron  distribution and  the
entanglement  values are  not  perturbed  significantly with  changing
w$_2$ due to the strong electron confinement within each QD.

\par Effects of  asymmetry are substantially more  significant for the
excited states.  When the  potential has a  mirror symmetry,  we observe
resonances in the  entanglement values for some  excited states labeled
by     M$_1$,     M$_2$,     M$_4$,    M$_5$,     and     M$_6$     in
Fig.~\ref{fig:asym_width_ent}. For example, as soon as the symmetry is
broken either by increasing or  decreasing w$_2$, the entanglement for
the first   excited state  drops rapidly  from $0.75$  to $0.5$
(see Fig.~\ref{fig:asym_width_ent}(b)).  This is due to  QDs no longer
being  identical, hence  the  ambiguity  in assigning  single-particle
states is removed.   This effect can be seen clearly  in the evolution
of the wavefunction of the first excited state as the dot width w$_2$
is varied.   As shown  in  Fig.~\ref{fig:asym_width_wave1}(a), when  the
potential   is   symmetric  \mbox{(w$_1=$w$_2=15\,{\rm   nm}$)},   the
wavefunction  is   an  eigenstate  of  both   the  exchange  operators
$\hat{\mathcal{P}}$   and   $\hat{\mathcal{W}}$    as   discussed   in
Sec.~\ref{sec:energy}.  When the symmetry  is broken, the wavefunction
is   only  an   eigenstate   of   $\hat{\mathcal{P}}$  (see
Fig.~\ref{fig:asym_width_wave1}(b)), and the      asymptotic
representation is now given by
\begin{equation}\label{eq:asymptotic_separable}
\ket{\phi^{\pm}} =
\frac{1}{\sqrt{2}}\big(\ket{1,\alpha}_1\!\ket{2,\beta}_2 \pm
\ket{2,\beta}_1\!\ket{1,\alpha}_2  \big). 
\end{equation} 
\noindent Note that in the above equation,  the index $\alpha$ and
$\beta$ are now specifically assigned to QD 1 and 2, respectively.  

\par Since the states that an electron can have is fixed depending on
which QD it is confined, the ambiguity in choosing between $\alpha$ and
$\beta$ is removed, thereby dropping the entanglement value from $0.75$ towards
$0.5$. Similar behavior (see Figs.~\ref{fig:asym_width_ent}(c),
\ref{fig:asym_width_ent}(e)-\ref{fig:asym_width_ent}(i)) can be seen
in any state whose asymptotic value for the entanglement saturation  is 0.75 in
the case of symmetric QDs.  
 
\subsection{Entanglement resonances due to avoided level-crossings}\label{subsec:resonance} %Sec 5a
\begin{figure*}[ht] %Fig 7
	\begin{subfigure}[h!]{0.3\textwidth}
		%\centering
		\includegraphics[scale=0.25]{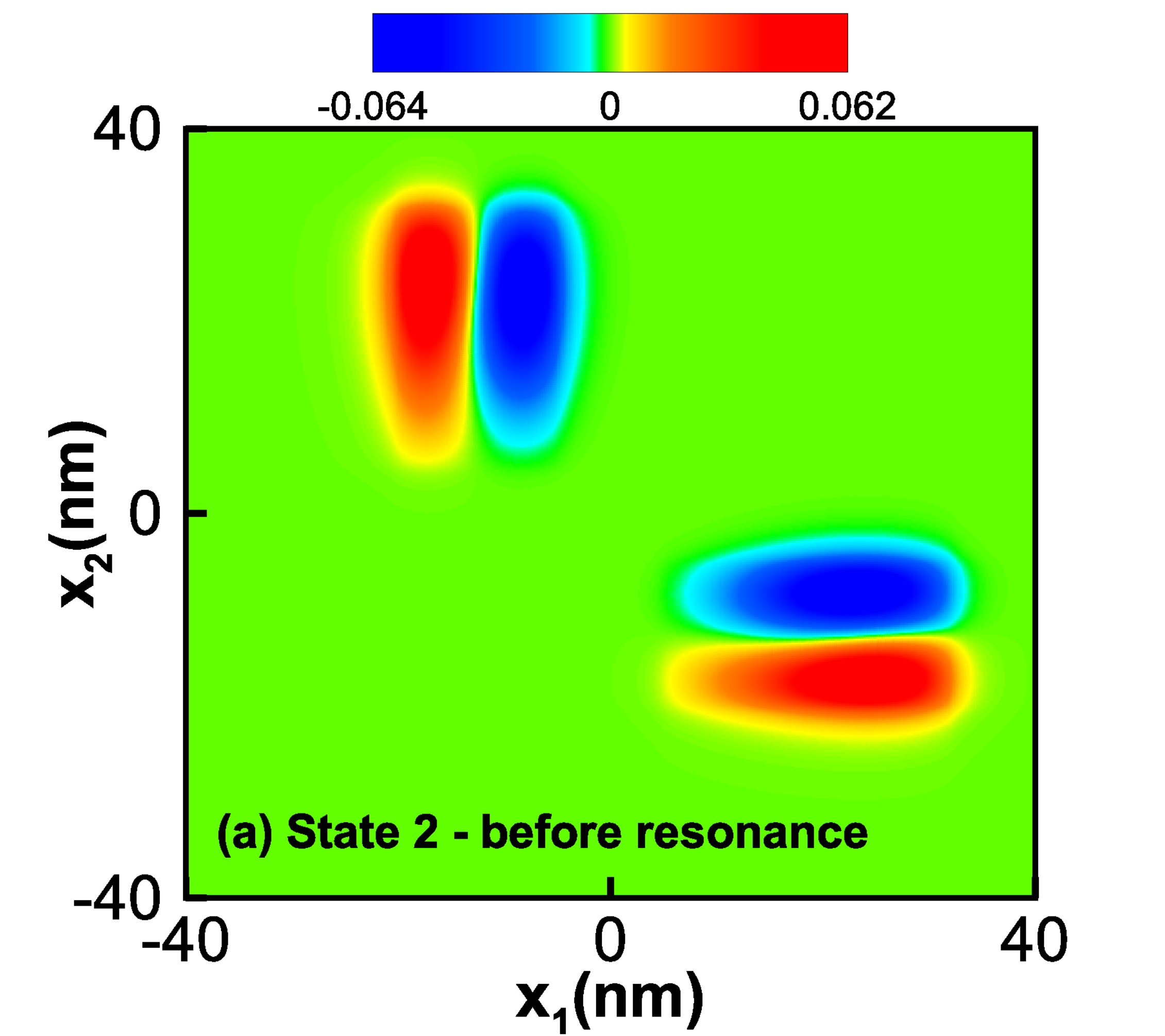}
		%\caption{States 2 - before resonance}
		%\label{fig:asym_width_R1_Peak_wave2_before}
	\end{subfigure}%
	\begin{subfigure}[h!]{0.3\textwidth}
		%\centering
		\includegraphics[scale=0.25]{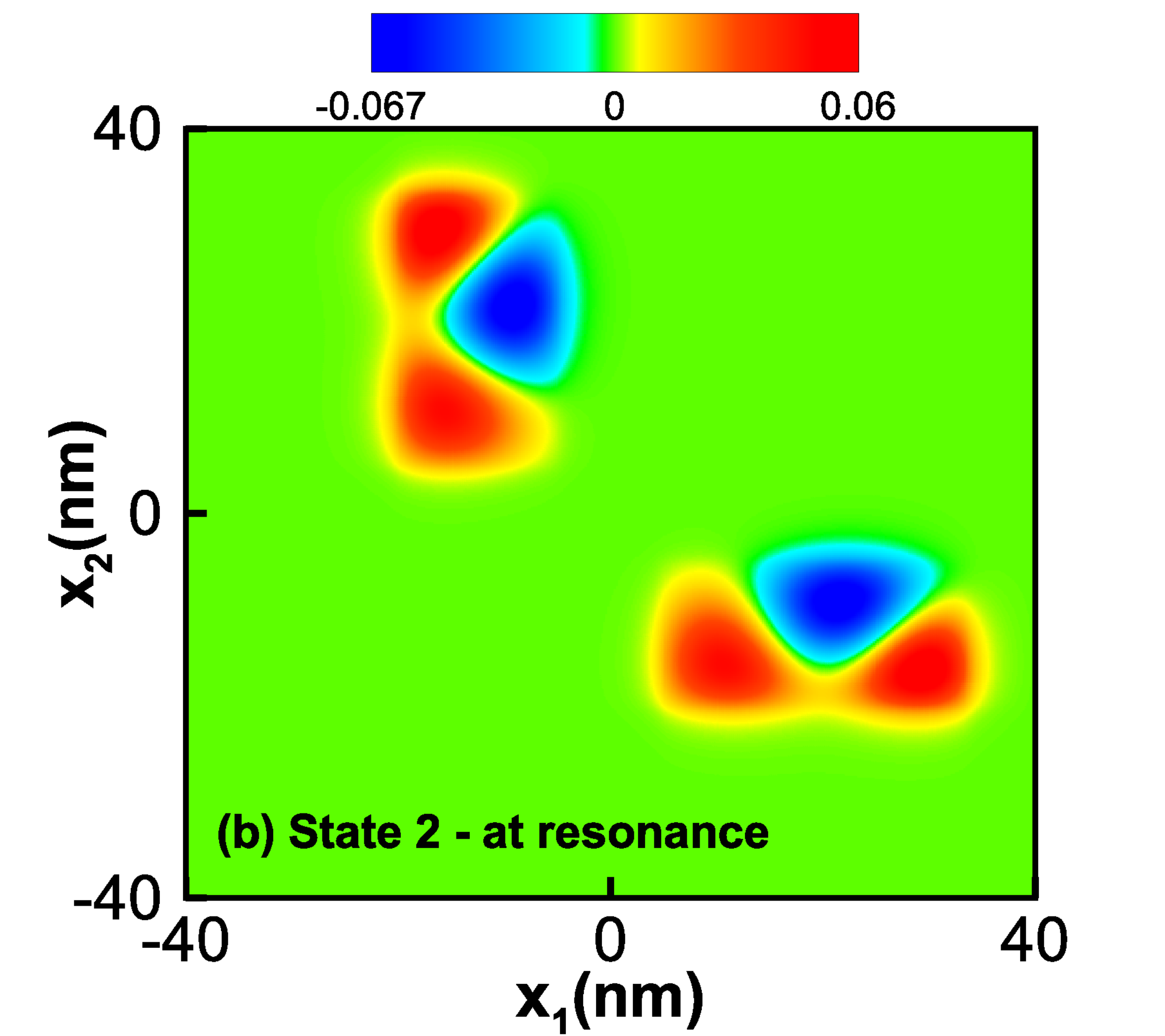}
		%\caption{States 2 - at resonance}
		%\label{fig:asym_width_R1_Peak_wave2_during}
	\end{subfigure}%
	\begin{subfigure}[h!]{0.3\textwidth}
		%\centering
		\includegraphics[scale=0.25]{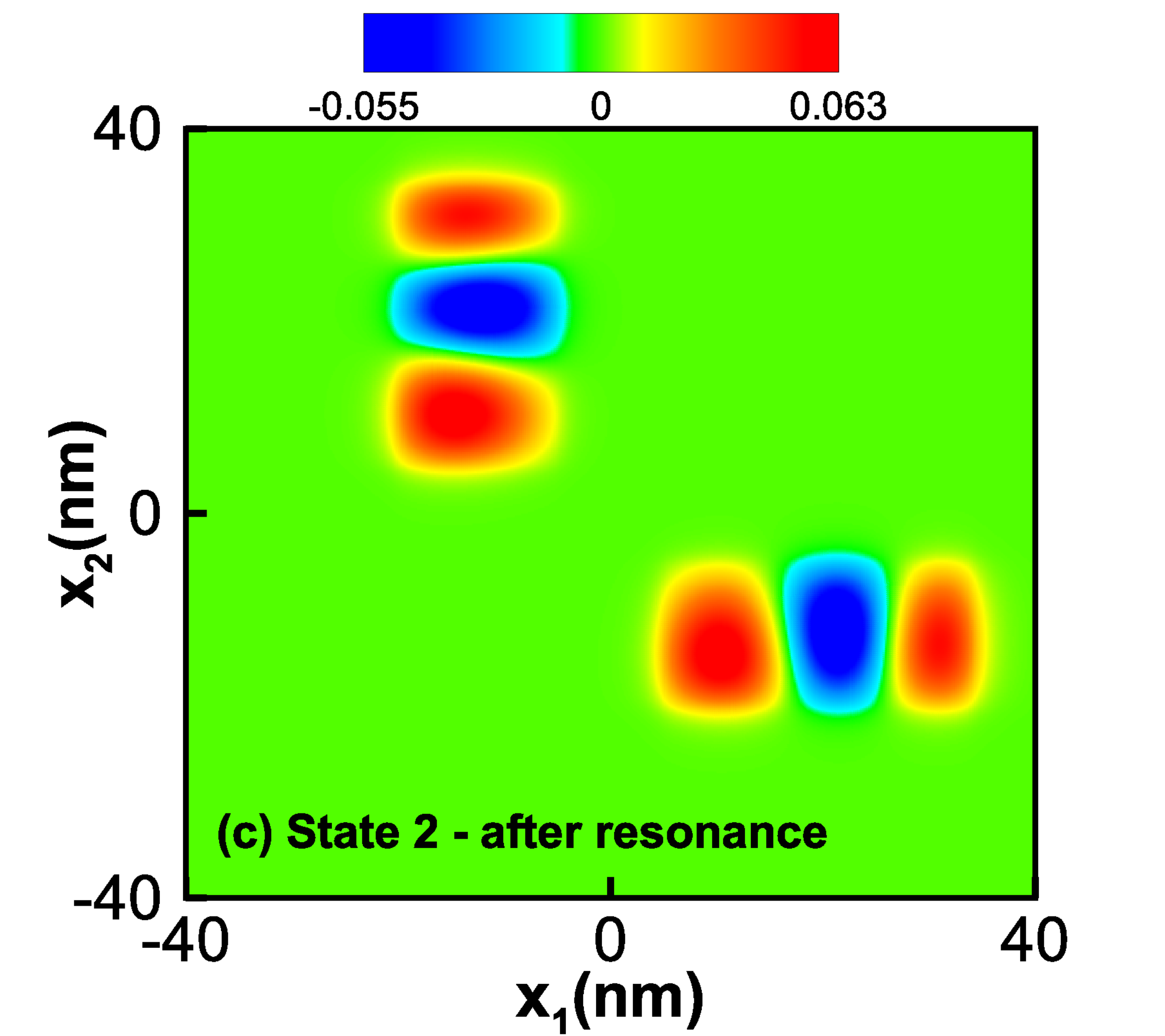}
		%\caption{States 2 - after resonance}
		%\label{fig:asym_width_R1_Peak_wave2_after}
	\end{subfigure}\\
        \vspace*{0.2in}
	\begin{subfigure}[h!]{0.3\textwidth}
		\includegraphics[scale=0.25]{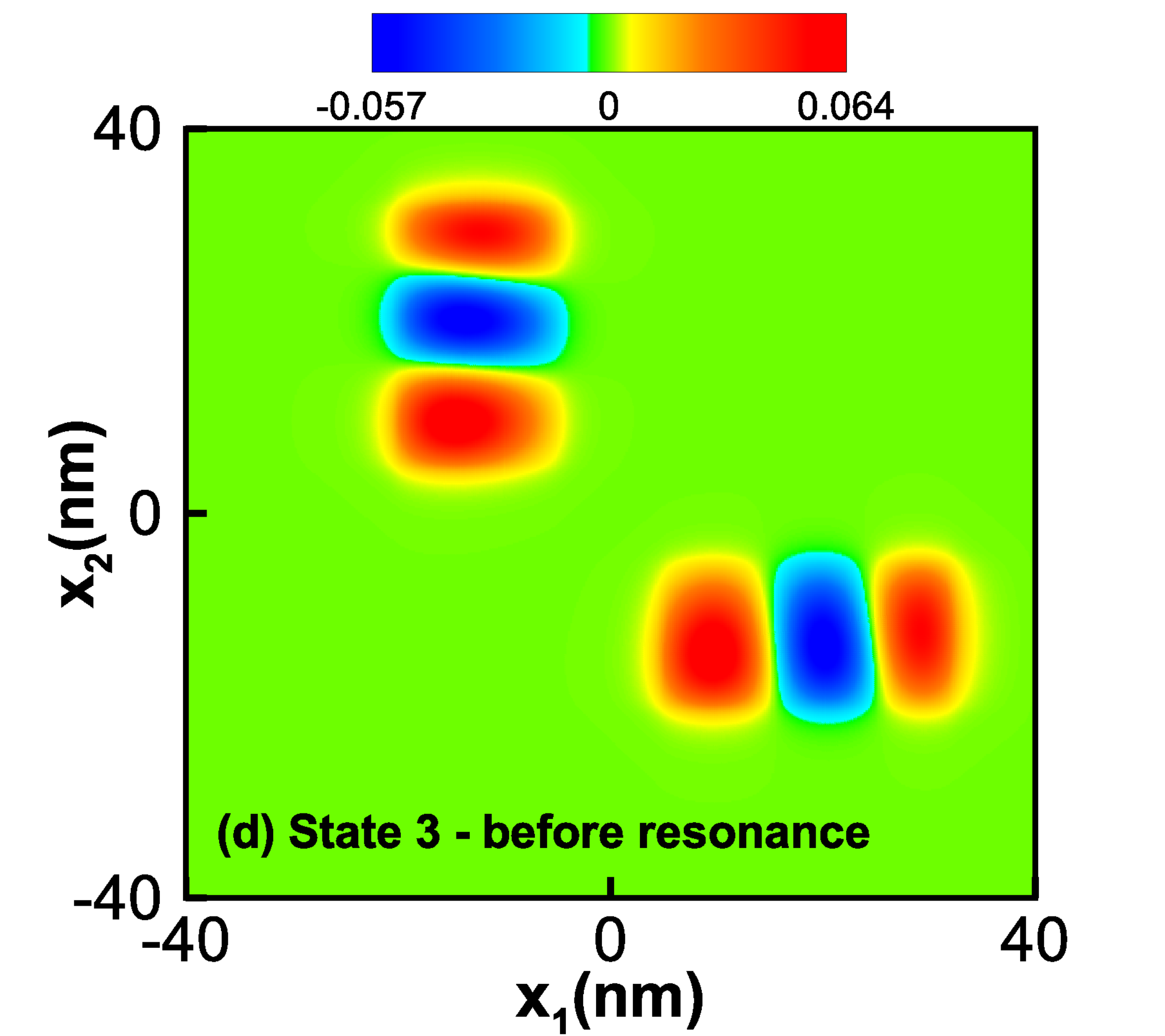}
		%\caption{States 3 - before resonance}
		%\label{fig:asym_width_R1_Peak_wave3_before}
	\end{subfigure}%
	\begin{subfigure}[h!]{0.3\textwidth}
		%\centering
		\includegraphics[scale=0.25]{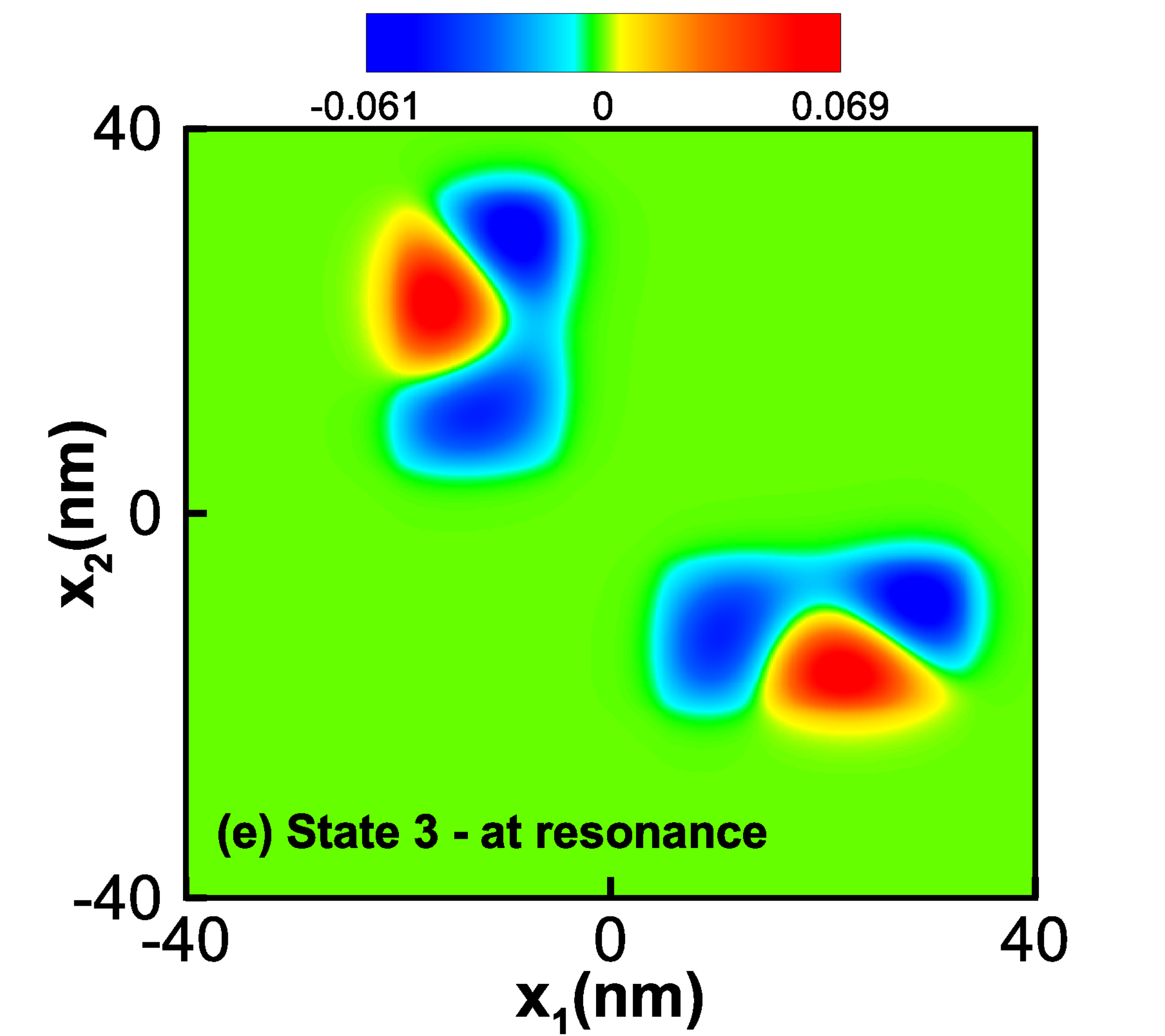}
		%\caption{States 3 - at resonance}
		%\label{fig:asym_width_R1_Peak_wave3_during}
	\end{subfigure}%
	\begin{subfigure}[h!]{0.3\textwidth}
		%\centering
		\includegraphics[scale=0.25]{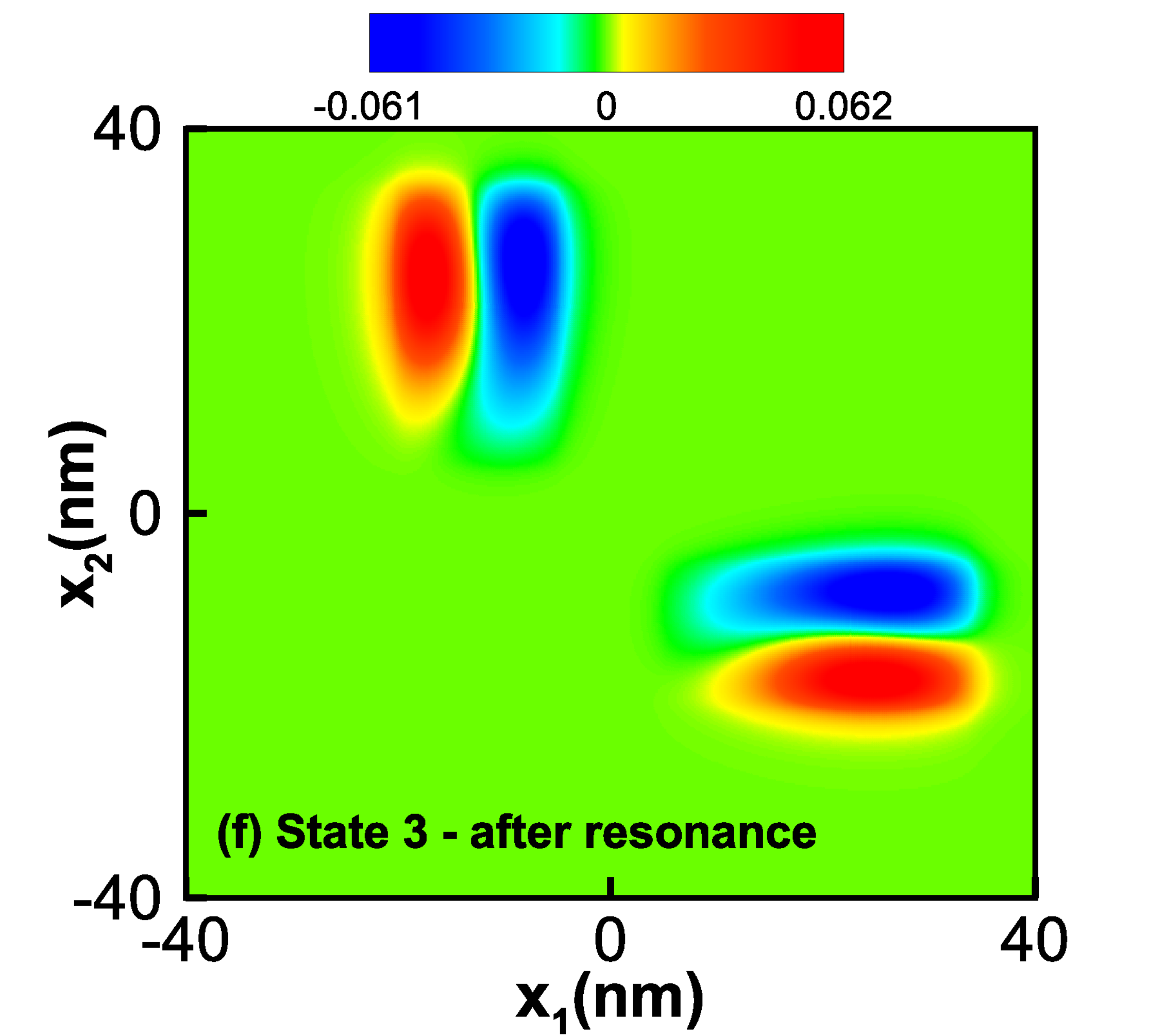}
		%\caption{States 3 - after resonance}
		%\label{fig:asym_width_R1_Peak_wave3_after}
	\end{subfigure}%
	\caption{\label{fig:asym_width_R1_Peak_wave}Wavefunctions of
          the second and third excited states  are plotted before,
          at, and after the resonance R$_{23}$, observed in Fig.~\ref{fig:asym_width_ent}(c).} 
\end{figure*}

\begin{figure*}[th!] %Fig 8
	\includegraphics[width=3.0in]{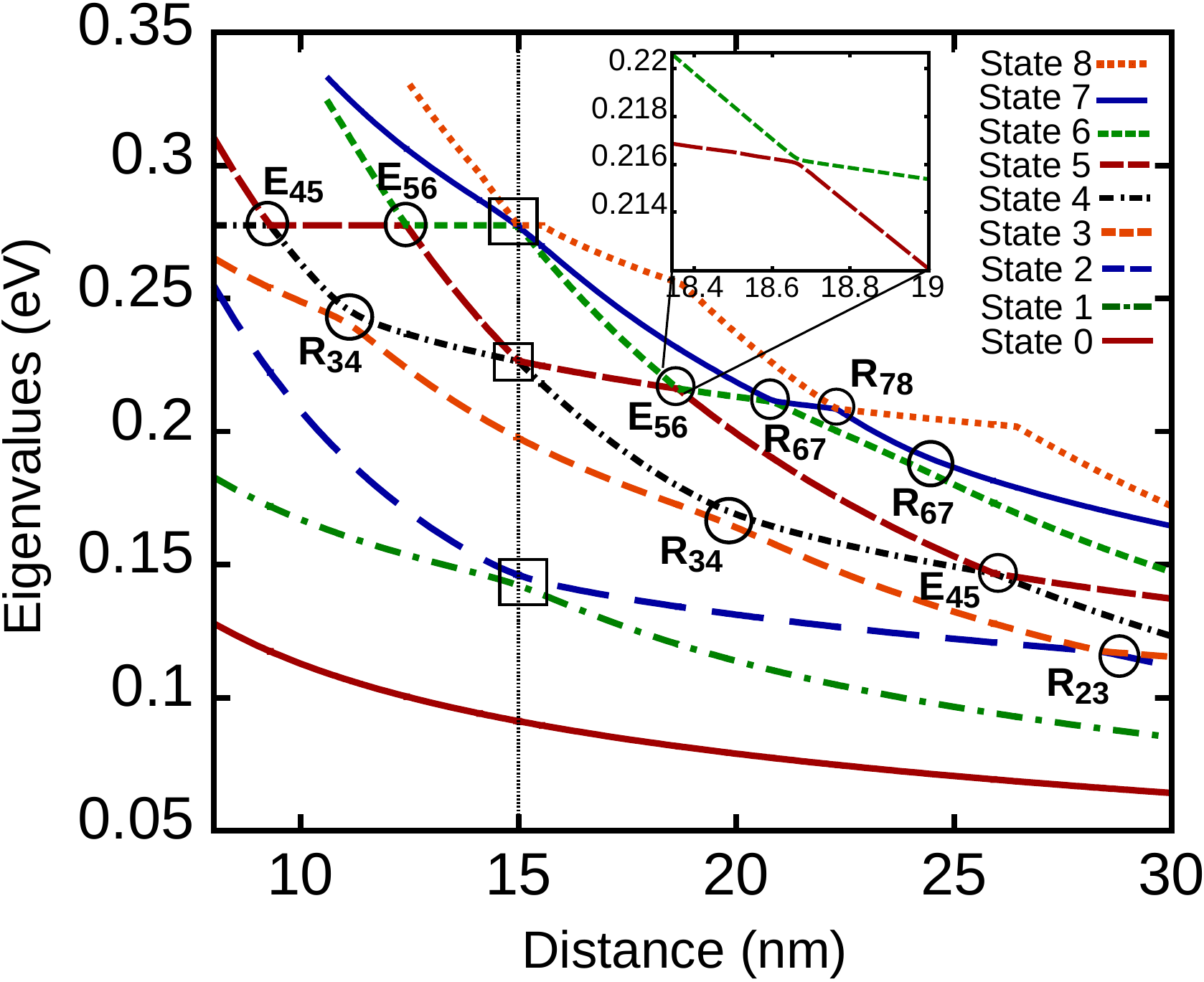}
	\caption{\label{fig:asym_width_eigs} Eigenvalues  of the first
          9 bound states are plotted  as function of w$_2$. Resonances
          corresponding  to  avoided  level-crossings  are  marked  by
          circles,  while  splittings   due  to  symmetry
          breaking  are marked  by squares.  An instance  of avoided
          level crossing is shown in  the inset. $R_{ij}$ and $E_{ij}$
          are  the labels corresponding   to  the   resonance  peaks   observed  in
          Fig.~\ref{fig:asym_width_ent}.}
\end{figure*} 

In  Fig.~\ref{fig:asym_width_ent},  we  observe  several  entanglement
resonances  even when  w$_1$\!$\neq\,$w$_2$. These  resonances are  due to
avoided  level-crossings of  the eigenstates.  For  example, in
Fig.~\ref{fig:asym_width_ent}(c) and  (d), we observe  an entanglement
resonance  labeled  as  R$_{23}$  at  w$_2=28.43\,{\rm  nm}$  for  the
$2^{\rm nd}$ and $3^{\rm rd}$ excited state, respectively. This is due
to the interaction  between adjacent states $2$ and  $3$ which undergo
an   avoided  level-crossing   at  R$_{23}$.   After  the   resonance,
wavefunctions  for   these  two   adjacent  states  evolve   into  one
another.  In Fig.~\ref{fig:asym_width_R1_Peak_wave},  we have  plotted
the wavefunction for state $2$ and $3$ before, at, and after the resonance
R$_{23}$.   We   observe   that    the   wavefunction   of   state   2
(Fig.~\ref{fig:asym_width_R1_Peak_wave}(a)) transforms into that of state
3 (Fig.~\ref{fig:asym_width_R1_Peak_wave}(c)) after the resonance, and
vice versa.  Around R$_{23}$, wavefunctions  of both these  states are
heavily  deformed  due to  maximal  interaction  between these  adjacent
states,  as  seen  in  Figs.~\ref{fig:asym_width_R1_Peak_wave}(b)  and
\ref{fig:asym_width_R1_Peak_wave}(e). Such  inter-mixing results  in a
high level of entanglement. Similar mechanisms will explain rest of the
resonances  labeled  as  R$_{ij}$,  which  indicate  the  interaction
between the  excited states $i$  and $j$. For instance,  the resonance
R$_{34}$ shown  in Fig.~\ref{fig:asym_width_ent}(d) and (e)  is due to
avoided level crossing  between the excited states $3$  and $4$. Also,
since asymmetry  in the system  can be  created both by  increasing or
decreasing w$_2$,  similar resonances are  observed on either  side of
the symmetric width \mbox{w$_2=15\,{\rm nm}$}.

Eigenvalues for the first nine bound  states are plotted as a function
of  w$_2$ in  Fig.~\ref{fig:asym_width_eigs}. Avoided  level-crossings
are seen  at the  entanglement resonance  positions R$_{ij}$  that are
observed  in  Fig.~\ref{fig:asym_width_ent}. We  note that  around
w$_1$=w$_2$, the  spectrum is almost  doubly degenerate due  to the
mirror   symmetry  of   the  system.   This  is   distinct  from   the
avoided-crossings observed in the asymmetric systems.

\subsection{Electron cluster formation} \label{subsec:double_occupancy} %Sec 5b

\begin{figure}[h!] %Fig 9
	\begin{subfigure}[h!]{0.17\textwidth}
          \hspace{-0.6in}%
          \includegraphics[width=1.7in]{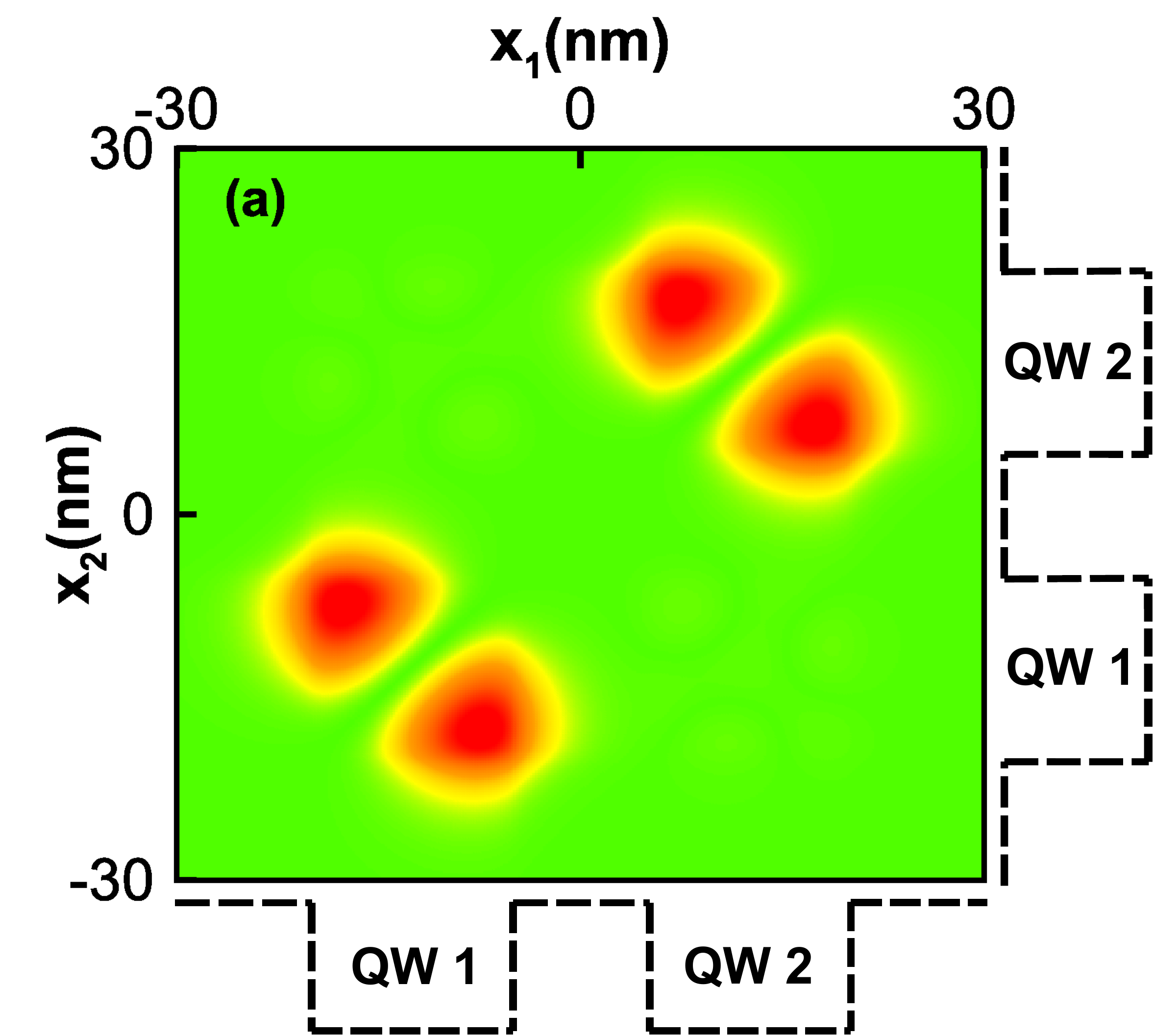}
	\end{subfigure}%
	\begin{subfigure}[h!]{0.17\textwidth}
          \hspace{-0.3in}%
          \includegraphics[width=1.7in]{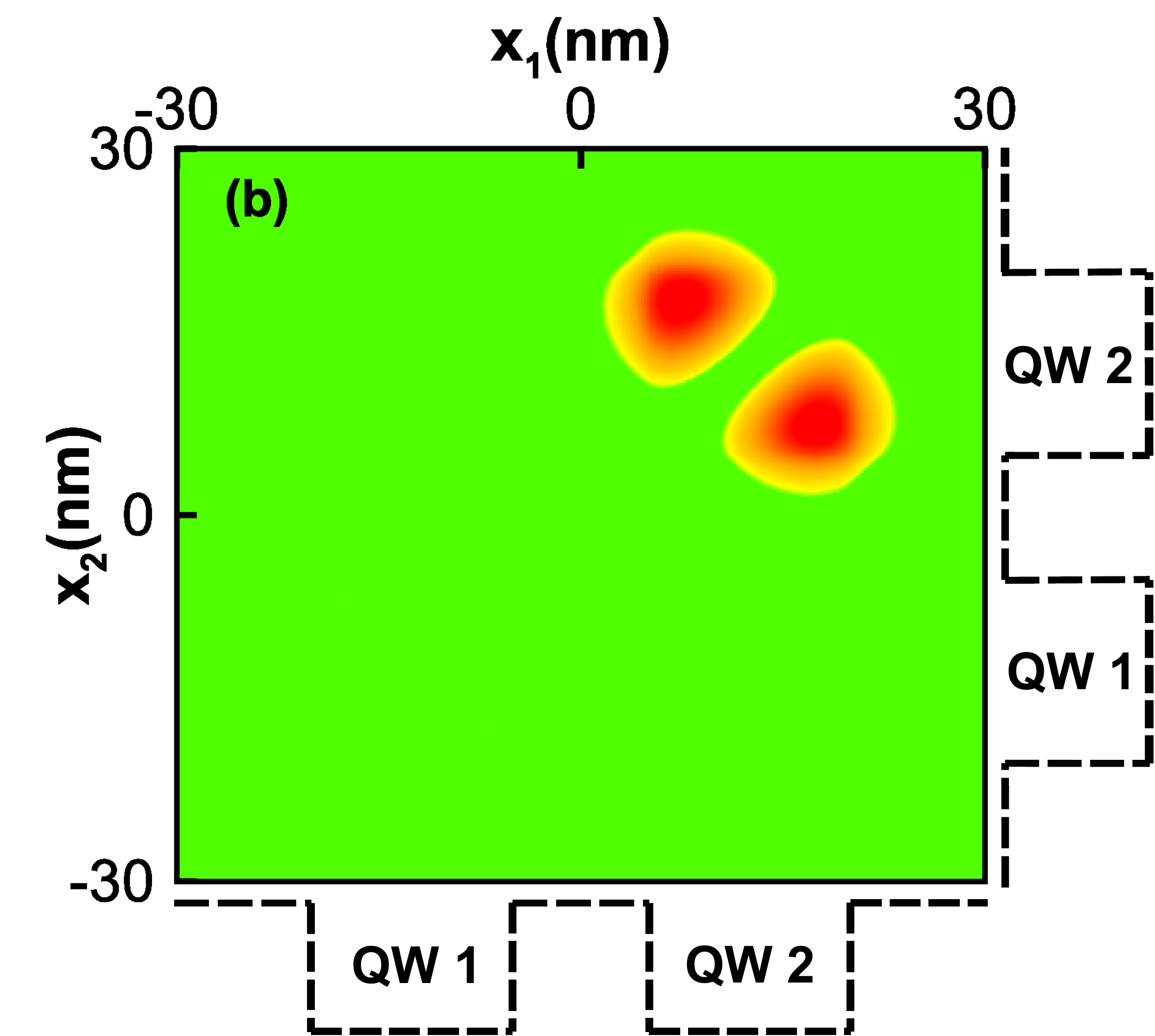}
	\end{subfigure}%
 \caption{\label{fig:2electroncluster}  Wavefunctions of  the 7$^{th}$
   excited state when (a) the QDs are identical, \mbox{${\rm w}_1={\rm w}_2=150$} nm,
   and (b) when  symmetry is broken with  w$_1=150$ nm, \mbox{w$_2=150.03$
   nm} are plotted.}
\end{figure}

\begin{figure*}[ht]  %Fig 10
	\begin{subfigure}[h!]{0.3\textwidth}
		%\centering
		\includegraphics[scale=0.25]{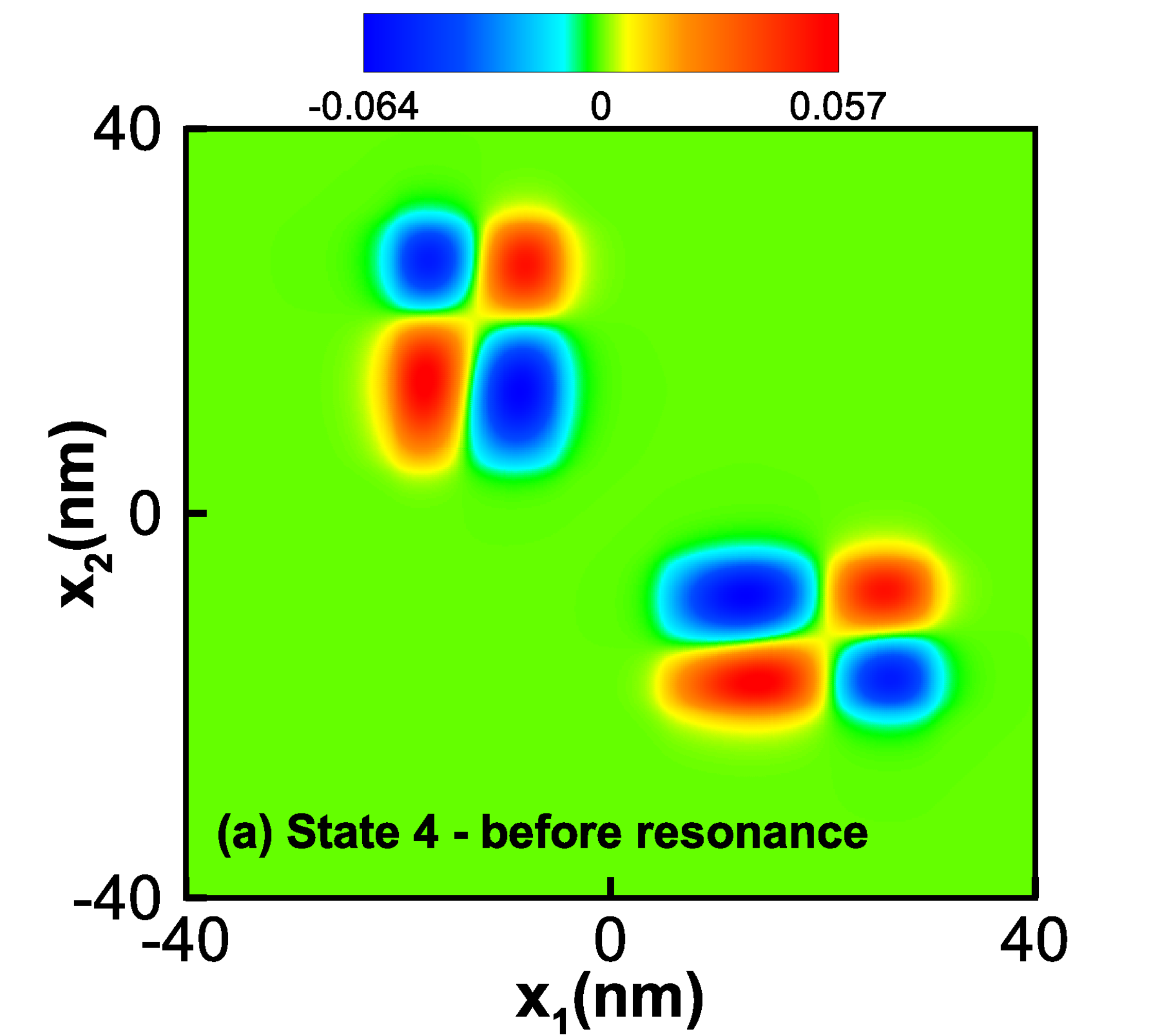}
		%\caption{States 4 - before resonance }
		%\label{fig:asym_width_R4_Peak_wave4_before}
	\end{subfigure}%
	\begin{subfigure}[h!]{0.3\textwidth}
		%\centering
		\includegraphics[scale=0.25]{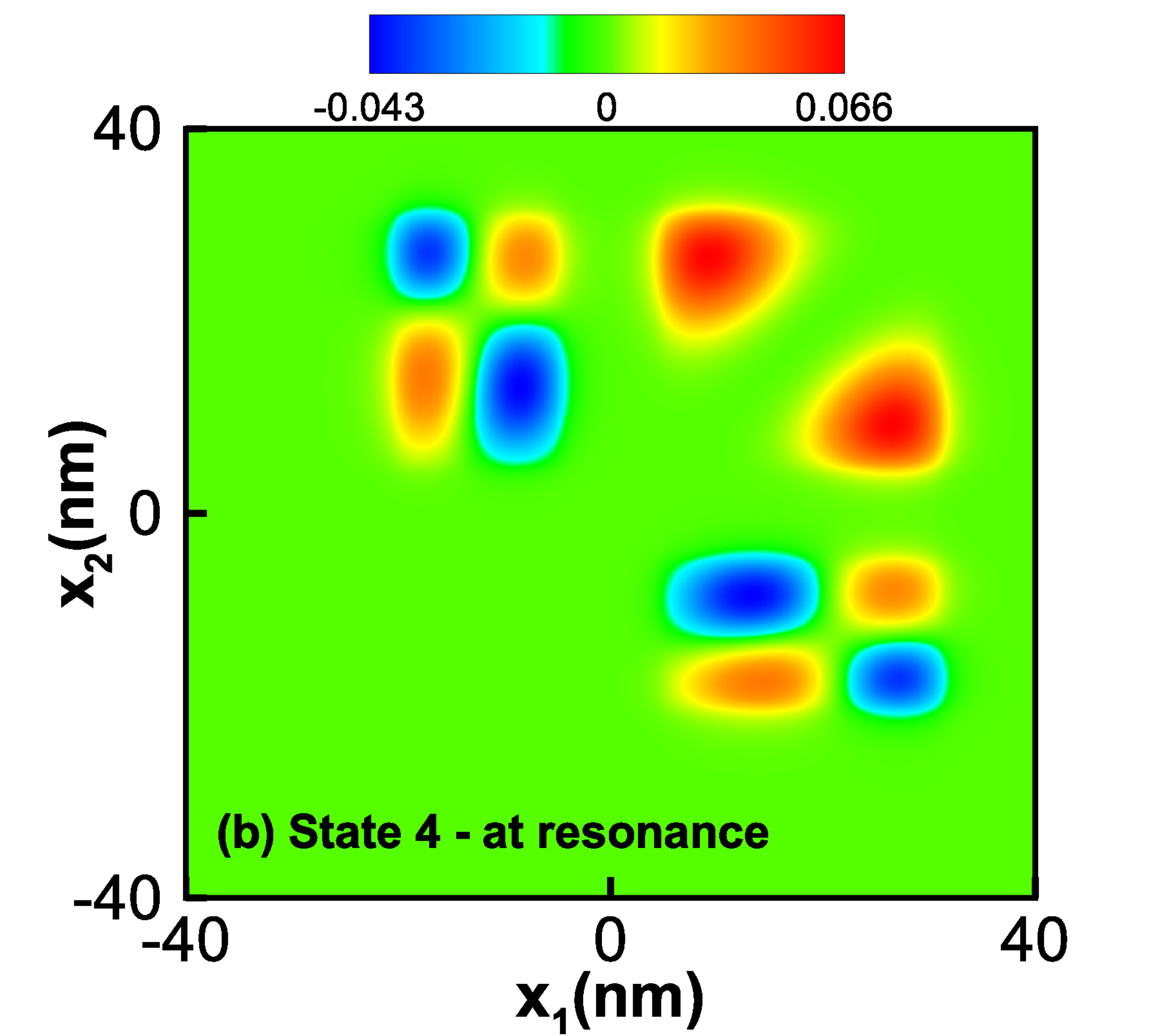}
		%\caption{States 4 - at resonance}
		%\label{fig:asym_width_R4_Peak_wave4_during}
	\end{subfigure}%
	\begin{subfigure}[h!]{0.3\textwidth}
		%\centering
		\includegraphics[scale=0.25]{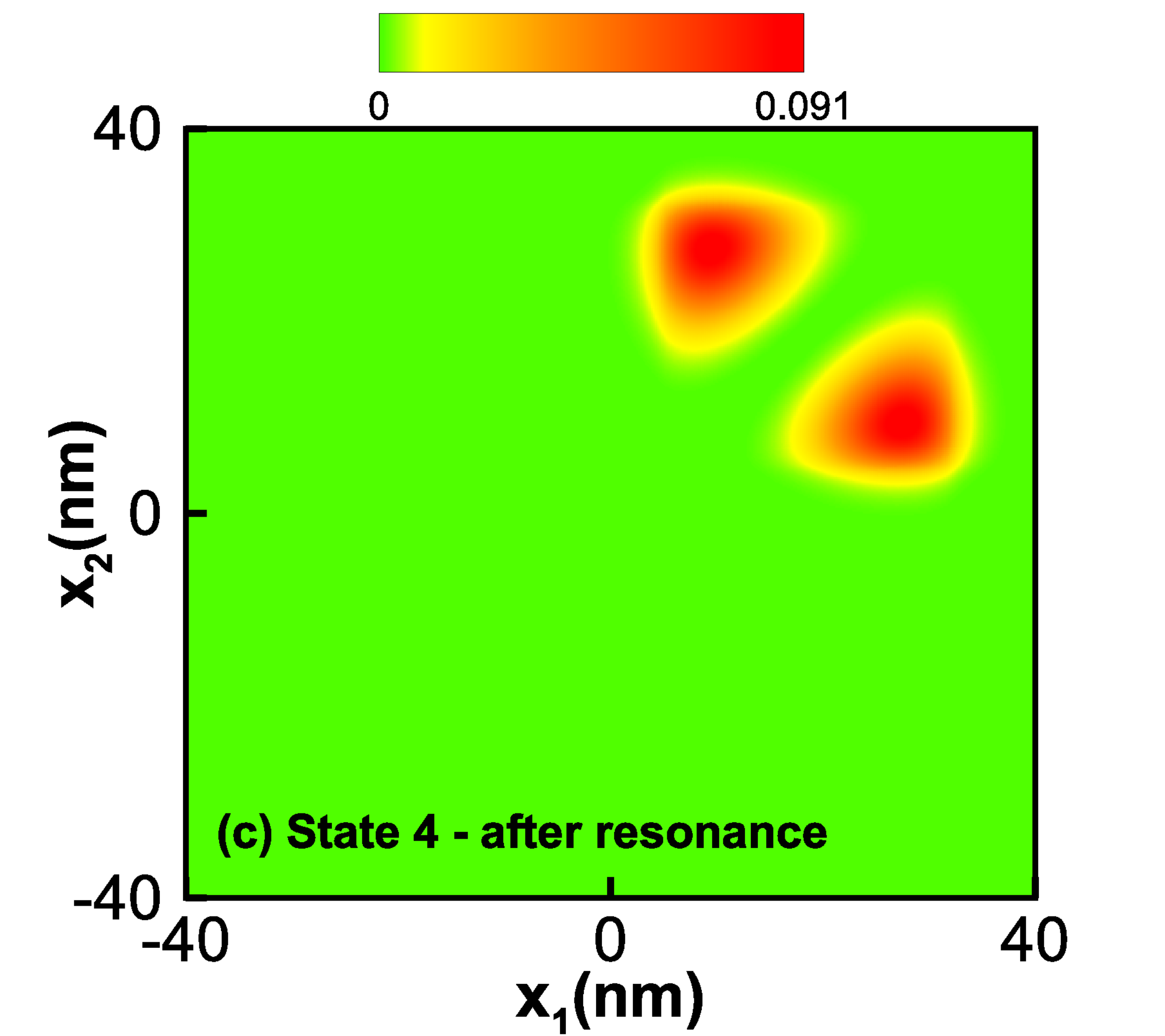}
		%\caption{States 4 - after resonance}
		%\label{fig:asym_width_R4_Peak_wave4_after}
	\end{subfigure}\\

        \vspace*{0.2in}
	\begin{subfigure}[h!]{0.3\textwidth}
		\includegraphics[scale=0.25]{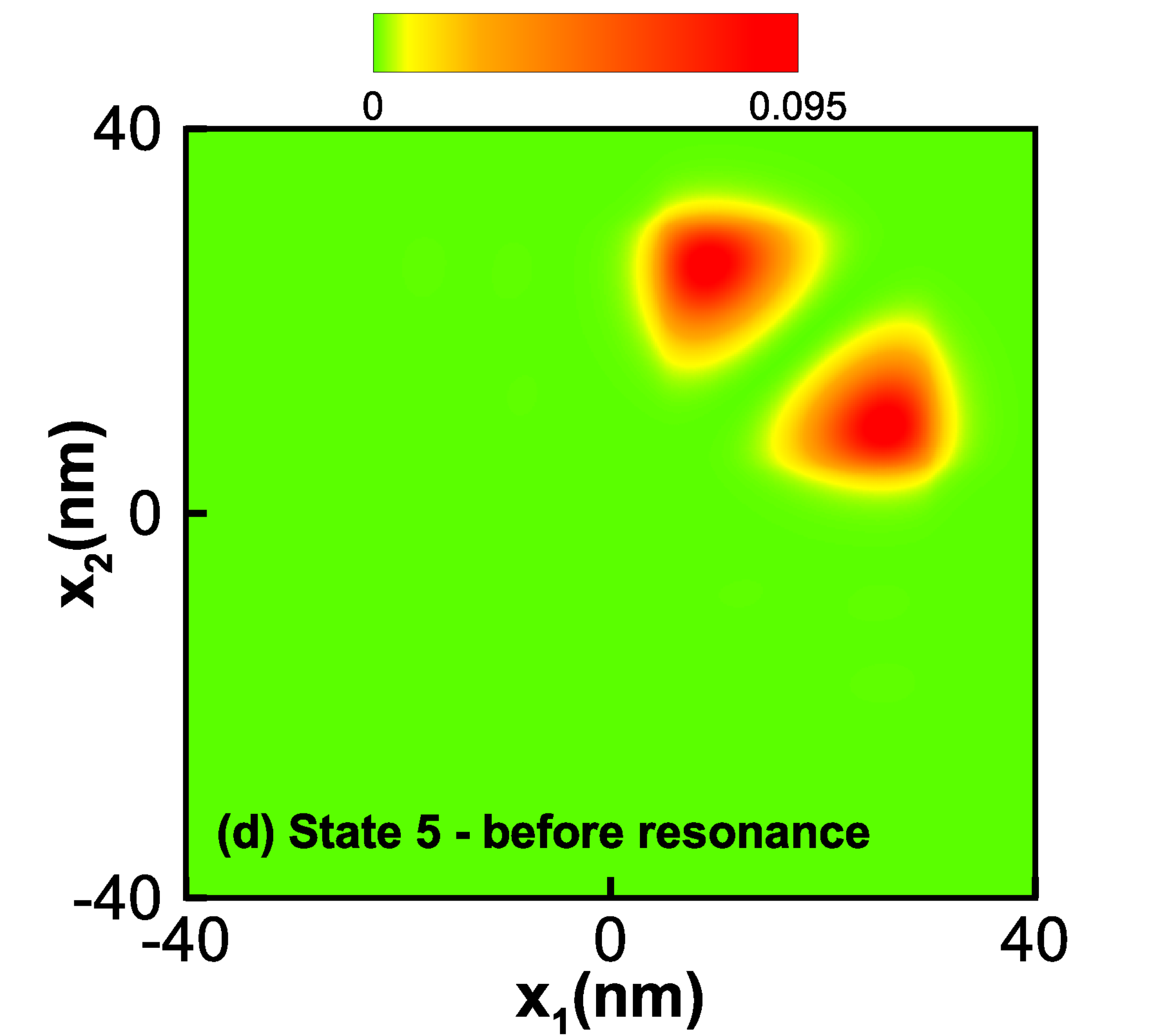}
		%\caption{States 5 - before resonance}
		%\label{fig:asym_width_R4_Peak_wave5_before}
	\end{subfigure}%
	\begin{subfigure}[h!]{0.3\textwidth}
		%\centering
		\includegraphics[scale=0.25]{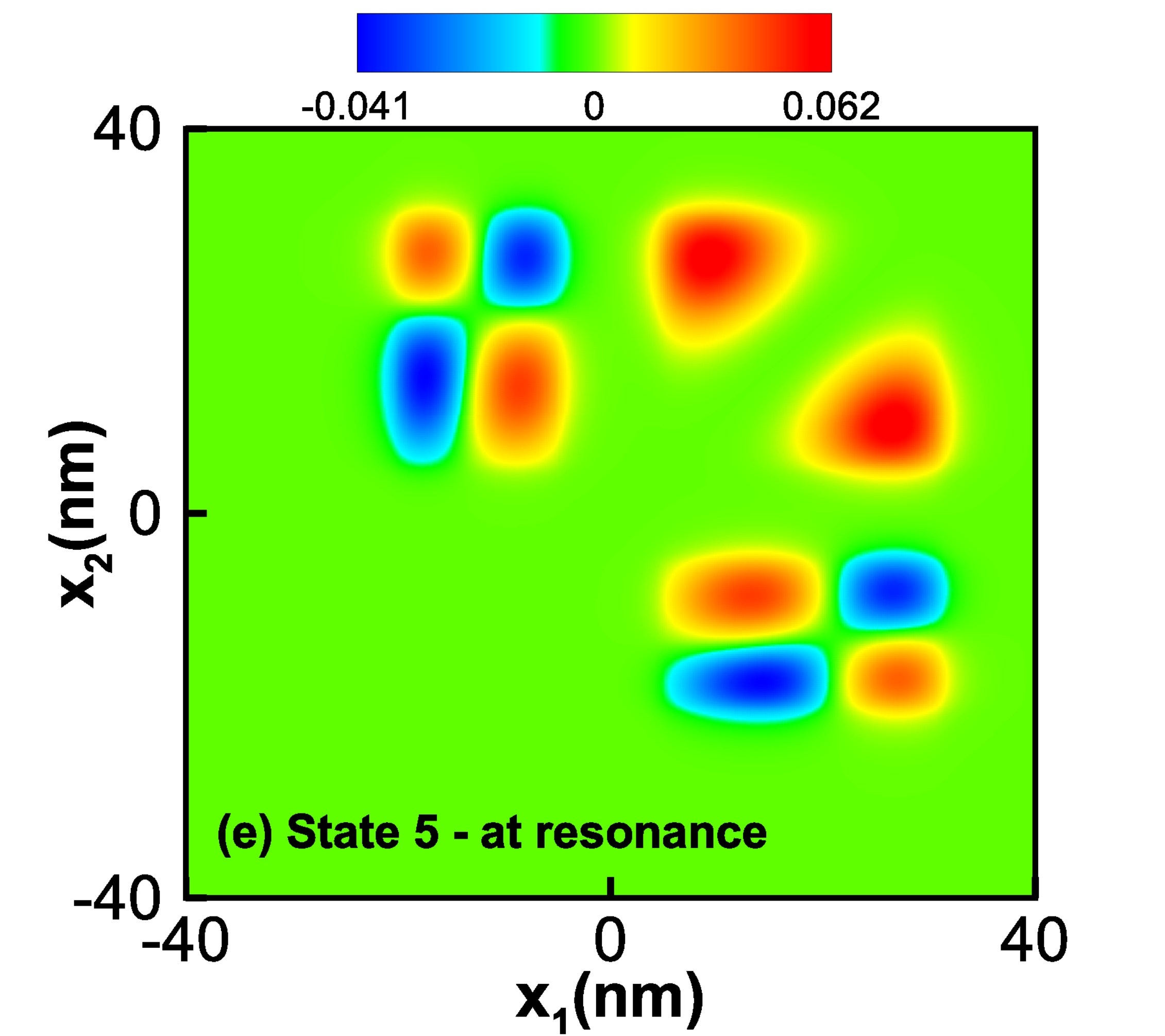}
		%\caption{States 5 - at resonance}
		%\label{fig:asym_width_R4_Peak_wave5_during}
	\end{subfigure}%
	\begin{subfigure}[h!]{0.3\textwidth}
		%\centering
		\includegraphics[scale=0.25]{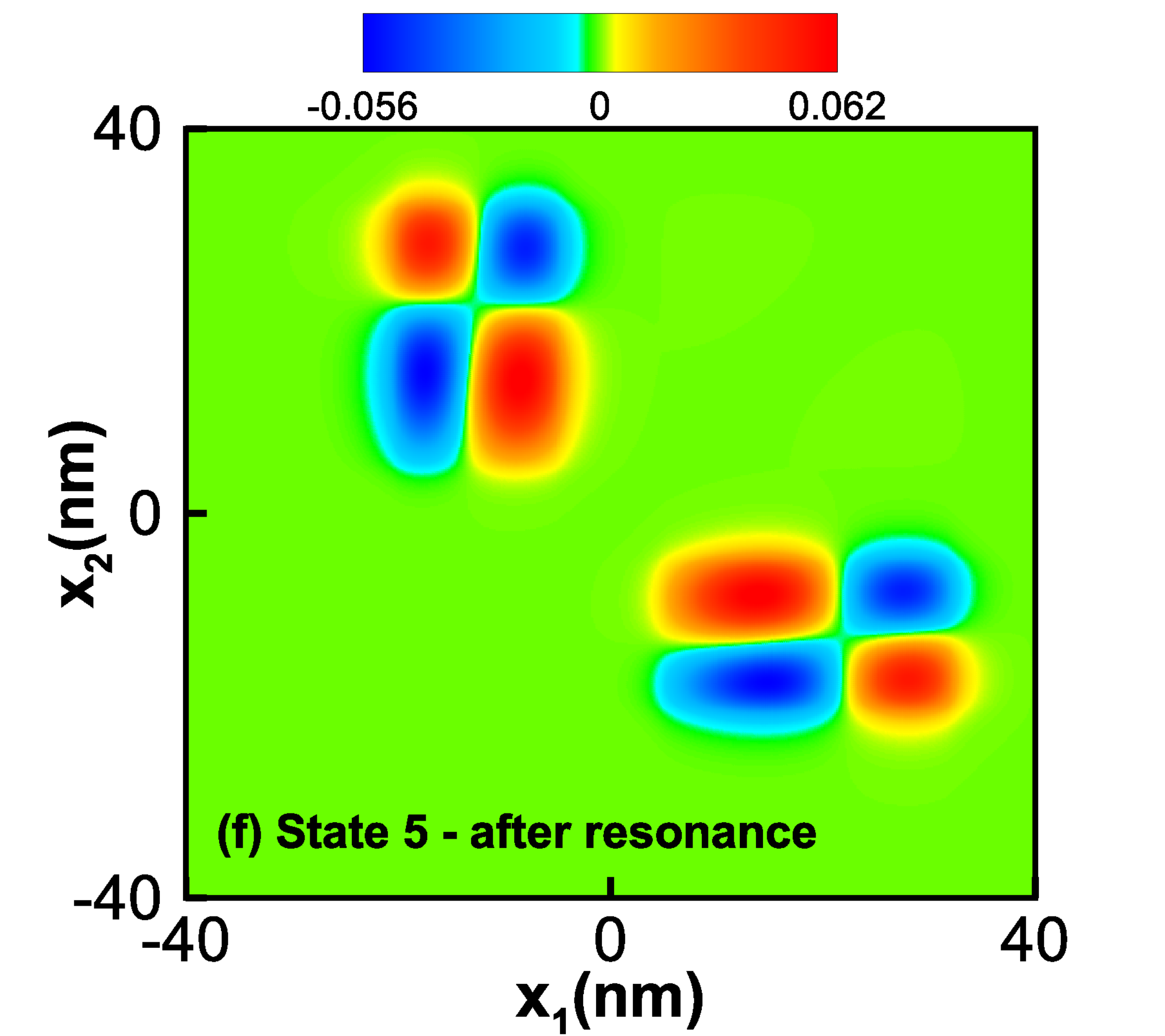}
		%\caption{States 5 - after resonance}
		%\label{fig:asym_width_R4_Peak_wave5_after}
	\end{subfigure}%
	\caption{\label{fig:asym_width_R4_Peak_wave}  Wavefunctions of
          the forth  and fifth excited  states are plotted before,
          at, and after the resonance E$_{45}$, observed in Fig.~\ref{fig:asym_width_ent}.}
\end{figure*}
A special case of avoided  level-crossing occurs when accompanied with
the  formation/dissolution  of electron  clusters, and  are  observed  as
additional maxima in the entanglement values. We can tune the width,
separation distance,  and the potential depth  of the QDs so  that the
two electrons  are in  the same quantum  dot, forming  a ``cluster.''
Such two-electron clusters can also be formed at higher excited states
of double QDs.  In Fig.~\ref{fig:2electroncluster}(a) we have shown an
occurrence of a two-electron cluster in the 7$^{\rm th}$ excited state
for a system of  symmetric double QDs. We see that  the cluster has an
equal probability distribution across both  QDs due to mirror symmetry
of the  potential.  However,  in case of  asymmetric QDs,  this mirror
symmetry is lost. As  seen in Fig.~\ref{fig:2electroncluster}(b), the
cluster will occupy only QD\,2 out of the two QDs.

To elaborate further,  we note that the $x_1$-  ($x_2$-) axis represents
the  position of the  first  (second) electron.  Following the  potential
pattern of the QDs  in Fig.~\ref{fig:2electroncluster}(a), we see that
both electrons have probability distribution in either of QDs. This is
in contrast  with, for example  Fig.~\ref{fig:asym_width_wave1}, where
we see  that if the  first electron is in the  first  QD,  then the
second  electron   will  be   in  the second   QD,   and  vice
versa.  Formation of  such two-electron  clusters leads  to additional
resonances  in   the  spatial   entanglement.   These   resonances  are
classified into two  categories: (1) resonance due  to mirror symmetry
of the  potential, (2)  local maxima due  to the evolution of two
separate electrons into a two-electron cluster, and vice versa.

\subsubsection{Resonance due  to mirror symmetry of  the potential}
We have observed  that the $7^{\rm  th}$ and $8^{\rm th}$  excited states
have an electron cluster occupancy. For    example, this  is seen for the 7$^{\rm th}$
state   in    Fig.~\ref{fig:2electroncluster}.    In
Figs.~\ref{fig:asym_width_ent}(h) and  \ref{fig:asym_width_ent}(i), we
have plotted the entanglement values for these states as a function of
the second dot width w$_2$. We  see that breaking the mirror symmetry
of  the  system  while  going  away  from  the  symmetric  dot  width
w$_2=15\,$nm, the entanglement value drops from $0.75$ to $0.5$. These
resonances     are    labeled     as    $M_7$,     and    $M_8$     in
Figs.~\ref{fig:asym_width_ent}(h)   and   \ref{fig:asym_width_ent}(i),
respectively.  This can  be explained  as follows:  when the  electron
cluster  is in  a  single QD  for the  case  of w$_1\neq\,$w$_2$,  the
entanglement value ${\cal E}_\ell$ is  close to $0.5$.
   %    This phenomenon
% is similar to the one in Fig.~\ref{fig:sym_ent_vs_distance}(a), where,
% in the limit $d\rightarrow 0$, electrons are essentially confined in a
% single QD, and ${\cal E}_\ell$ is close to 0.5.
For w$_1=\,$w$_2$, the probability of finding the 
electron  cluster is  distributes  equally  in each  QD,  and  this  will
contribute an  additional 50\% uncertainty in  the particle occupation
within either dot. Hence, we  observe an increase in the entanglement
value to $0.75$.

\subsubsection{Transition between two separate
  electrons to a two-electron cluster}
In the  case of  asymmetric double  QDs, the  electron cluster  can be
formed  at  lower  energies.   Transition between  a  state  with  two
separate electrons  to a two-electron  cluster can be  deduced through
the  occurrence of  local resonance  maxima in  the entanglement.   In
Figs.~\ref{fig:asym_width_ent}(e),   \ref{fig:asym_width_ent}(f),  and
\ref{fig:asym_width_ent}(g), such  resonances are labeled  as E$_{45}$
and E$_{56}$, which indicates  the avoided-level crossings between the
neighboring   states  4   and  5,   and  5  and  6,   respectively.  In
Fig.~\ref{fig:asym_width_eigs}, we have  shown these crossings between
the eigenvalues.

Wavefunctions  for the  excited states  4 and  5 around  the resonance
E$_{45}$                 are                plotted                 in
Figs.~\ref{fig:asym_width_R4_Peak_wave}(a)-~\ref{fig:asym_width_R4_Peak_wave}(f). After 
the resonance E$_{45}$, the  two-electron cluster occupancy is favored
over the single electron localization  within each QD.  Hence it occurs
at  a lower  energy.   Therefore,  the sharp  peak  at E$_{45}$
 occurs  along  with the  formation/dissolution  of  electron
clusters as displayed  in Fig.~\ref{fig:asym_width_R4_Peak_wave}.  For
example, in  Fig.~\ref{fig:asym_width_ent}(e), state 4 has 
two separate electrons in each QDs before the resonance E$_{45}$.
Whereas, after  the resonance,  the state 4  will have  a two-electron
cluster occupancy. Analogous mechanisms explain the resonance E$_{56}$
in  Fig.~\ref{fig:asym_width_ent}(f),  where  we see  that  after  the
resonance the state 5 has a two-electron cluster occupancy.

\section{Resonances in entanglement with applied
  external electric  fields}\label{sec:Efield}%Sec 6
\begin{figure*}[ht]  %Fig 11
	\begin{subfigure}[h!]{0.3\textwidth}
		%\centering
		\includegraphics[scale=0.3]{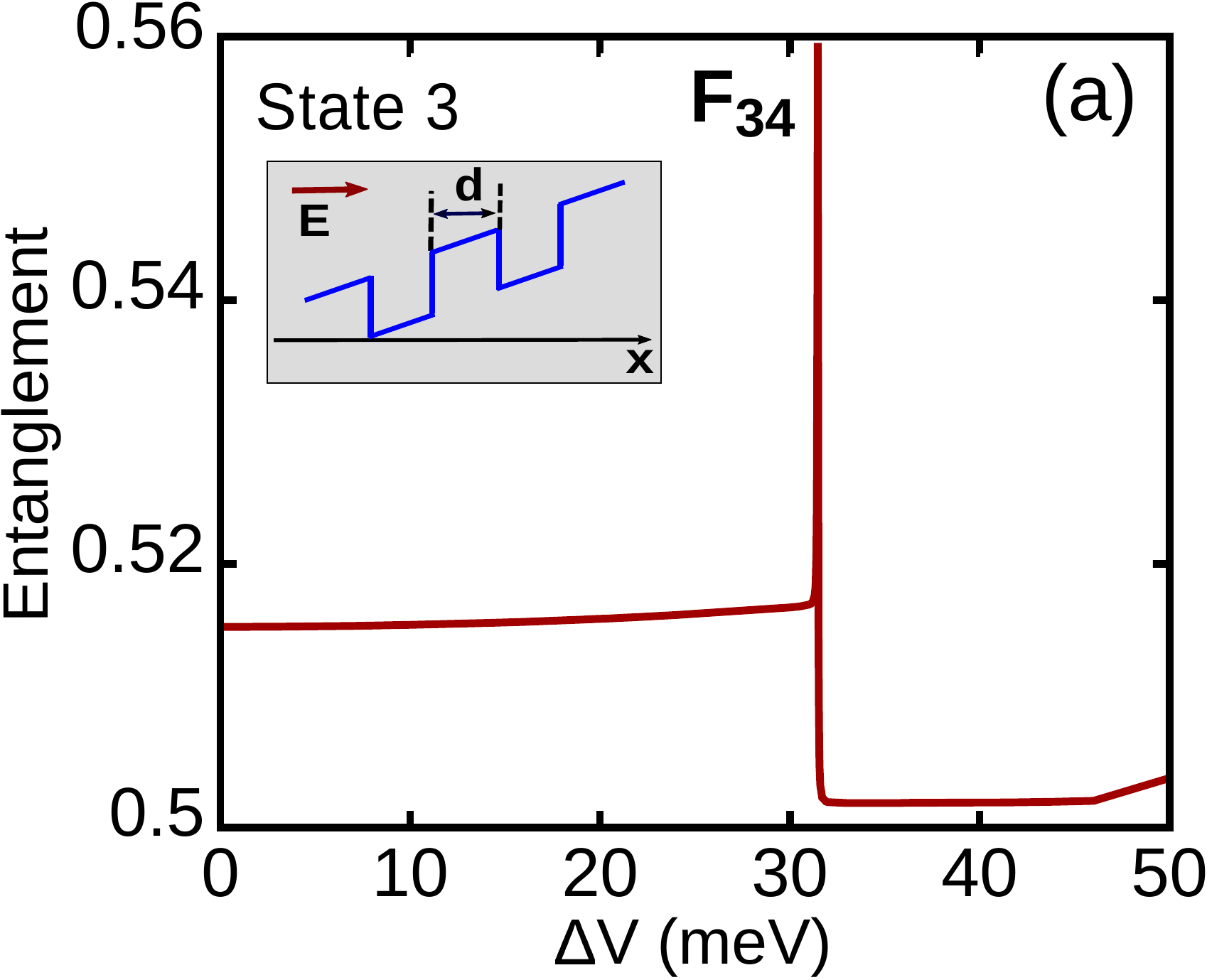}
		%\caption{States 3}
		%\label{fig:state3_efield_ent}
	\end{subfigure}%
	\begin{subfigure}[h!]{0.3\textwidth}
		%\centering
		\includegraphics[scale=0.3]{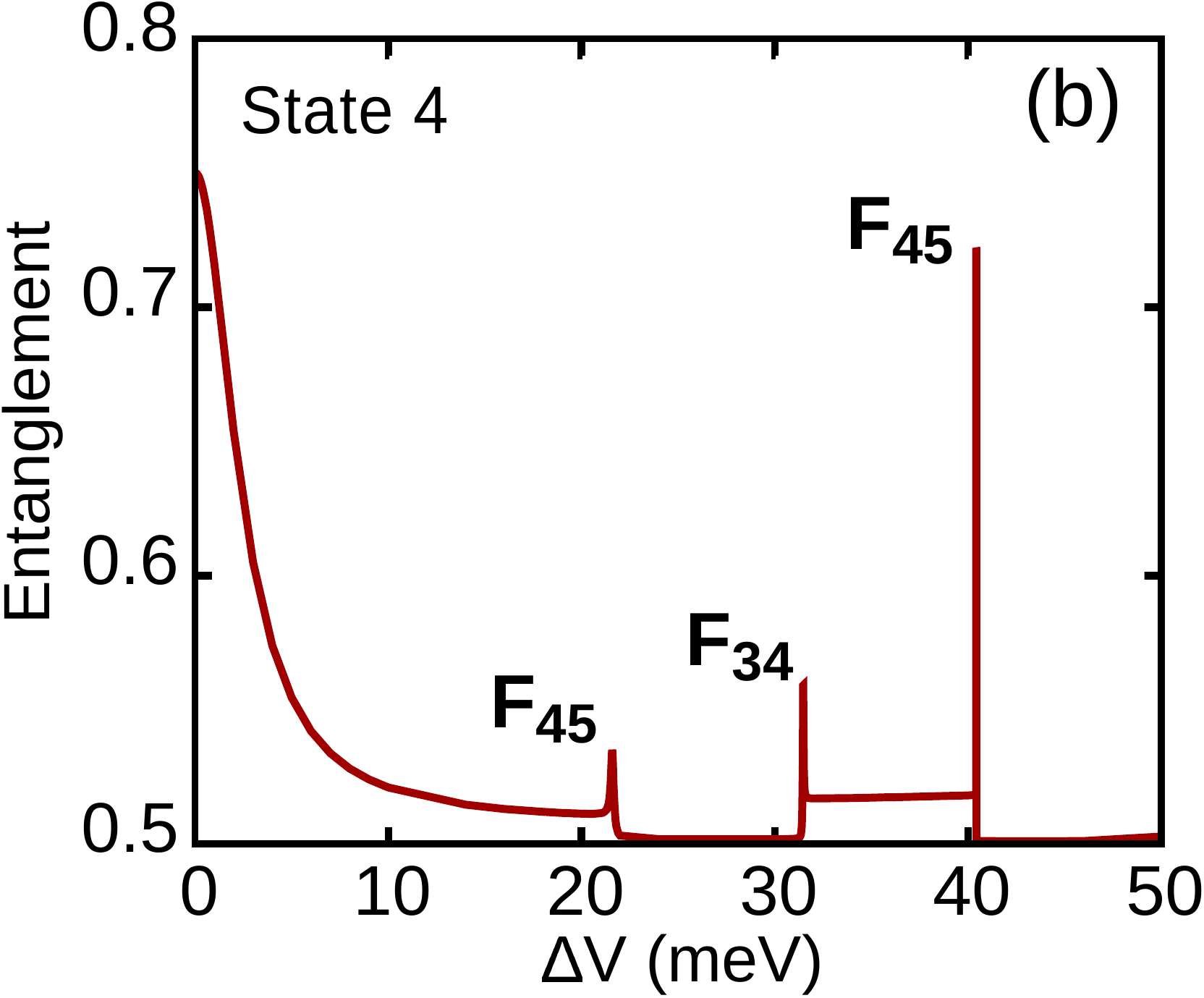}
		%\caption{States 4}
		%\label{fig:state4_efield_ent}
	\end{subfigure}%
	\begin{subfigure}[h!]{0.3\textwidth}
		%\centering
		\includegraphics[scale=0.3]{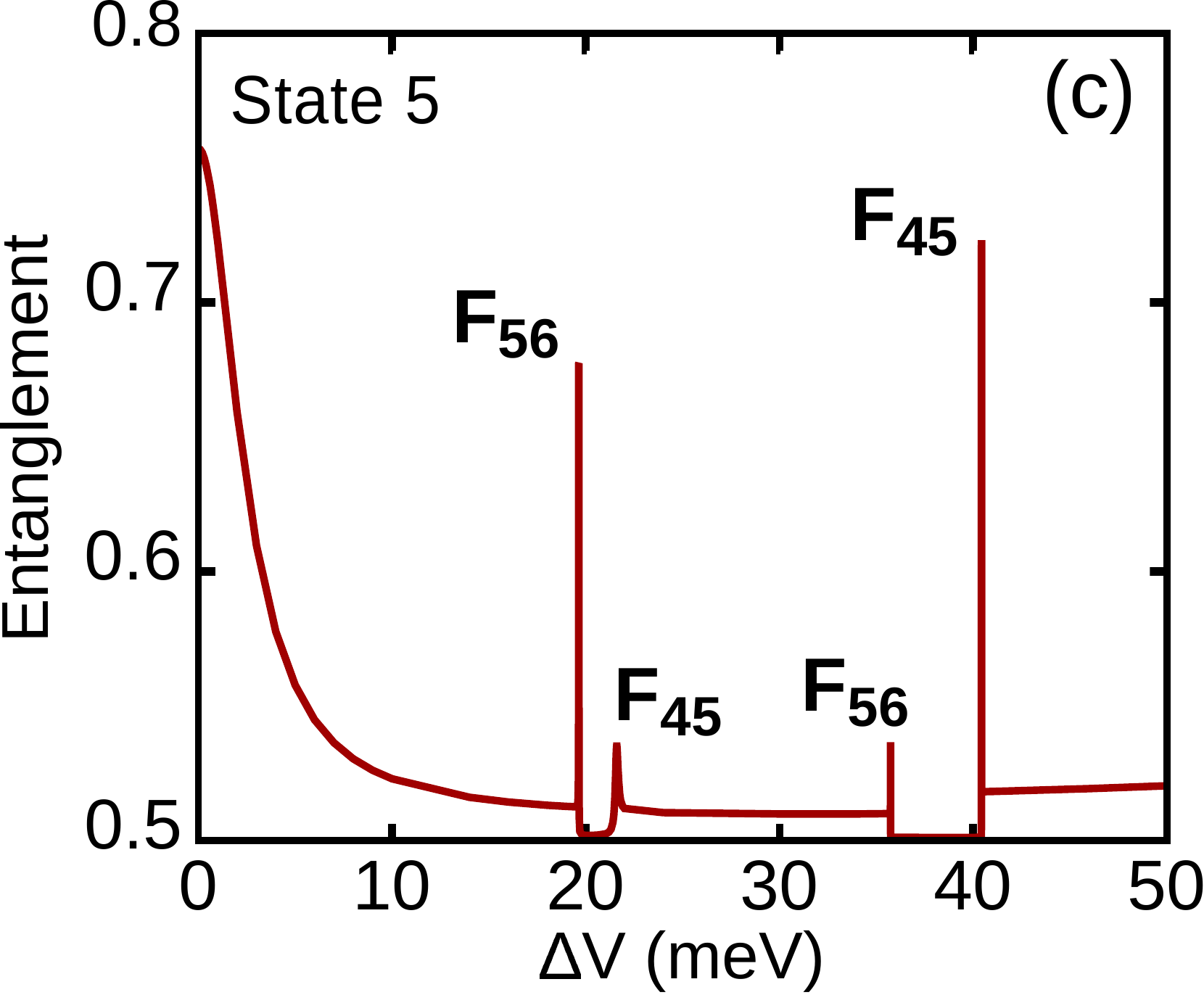}
		%\caption{States 5}
		%\label{fig:state5_efield_ent}
	\end{subfigure}
        \caption{\label{fig:efield_ent}   Spatial    entanglement   of
          excited states 3, 4, and 5  for two electrons in a symmetric
          double QDs  is plotted  as a  function of  applied potential
          $\Delta V$.  The distance between  QDs are kept  constant at
          $10\,{\rm         nm}$,         and        the         width
          \mbox{${\rm  w}_1={\rm w}_2=15\,{\rm  nm}$}. Here,  the peak
          $F_{ij}$  corresponds to  an avoided  level-crossing between
          the excited states $i$ and $j$.}
\end{figure*}

%%%%%%%%%%%%%%%%%%%%%%%%%%%%%% 
In this section, we study the entanglement properties in symmetric
double QDs with an applied constant electric field {\bf E}. The
additional term in the Hamiltonian is given by 
\begin{equation}
H'(x) = |e|Ex.
\end{equation}
The ``ramp"  potential breaks the  mirror symmetry of the  system. The
entanglement of the two electrons as a  function of $\Delta\! V=|e|Ed$ is shown
for  the  3$^{rd}$,   $4^{\rm  th}$,   and  $5^{\rm   th}$  states   in
Figs.~\ref{fig:efield_ent}(a)-\ref{fig:efield_ent}(c).

\par  Although  the mechanism  for  creating  asymmetry is  different,
phenomena  similar to  the case  of the  asymmetric QDs,  discussed in
Sec.~\ref{sec:non_sym}, are  observed. Spatial Entanglement  in states
with $\alpha\!\neq\!\beta$ decreases rapidly with the applied electric
field  (see  Figs.~\ref{fig:efield_ent}(b),  \ref{fig:efield_ent}(c)),
and resonant behavior associated with avoided level-crossings are also
present                                                           (see
Figs.~\ref{fig:efield_ent}(a)-\ref{fig:efield_ent}(c)). 
Electron cluster  formation occurs  with the applied  field as  the QD
potential  profile is  tilted.   In Figs.~\ref{fig:efield_ent}(a)  and
\ref{fig:efield_ent}(b), the  sharp resonance peak F$_{34}$  is formed
due to the avoided level-crossing  between the state with two separate
electrons (state  3), and  the state that  corresponds to  an electron
cluster (state 4).  A similar mechanism explains the  formation of the
entanglement resonance  F$_{45}$ in  Figs.~\ref{fig:efield_ent}(b) and
\ref{fig:efield_ent}(c),  which  is  formed due  to  multiple  avoided
level-crossings between the states 4 and 5.

\par Since  one can vary  the magnitude  of the field  with reasonable
ease in  experiments, the  occupancy of  electrons in  each QD  can be
controlled and varied effectively  by external electric fields without
having to  re-fabricate the width  of QDs. Moreover,  since resonances
associated with transitions  from a double to a  single occupancy (and
vice  versa)  are  extremely  sharp, entanglement  measurement  is  an
excellent indicator for the electron cluster formation/dissolution.

\par
\begin{figure}[h]  %Fig 12
	\includegraphics[width=2.5in]{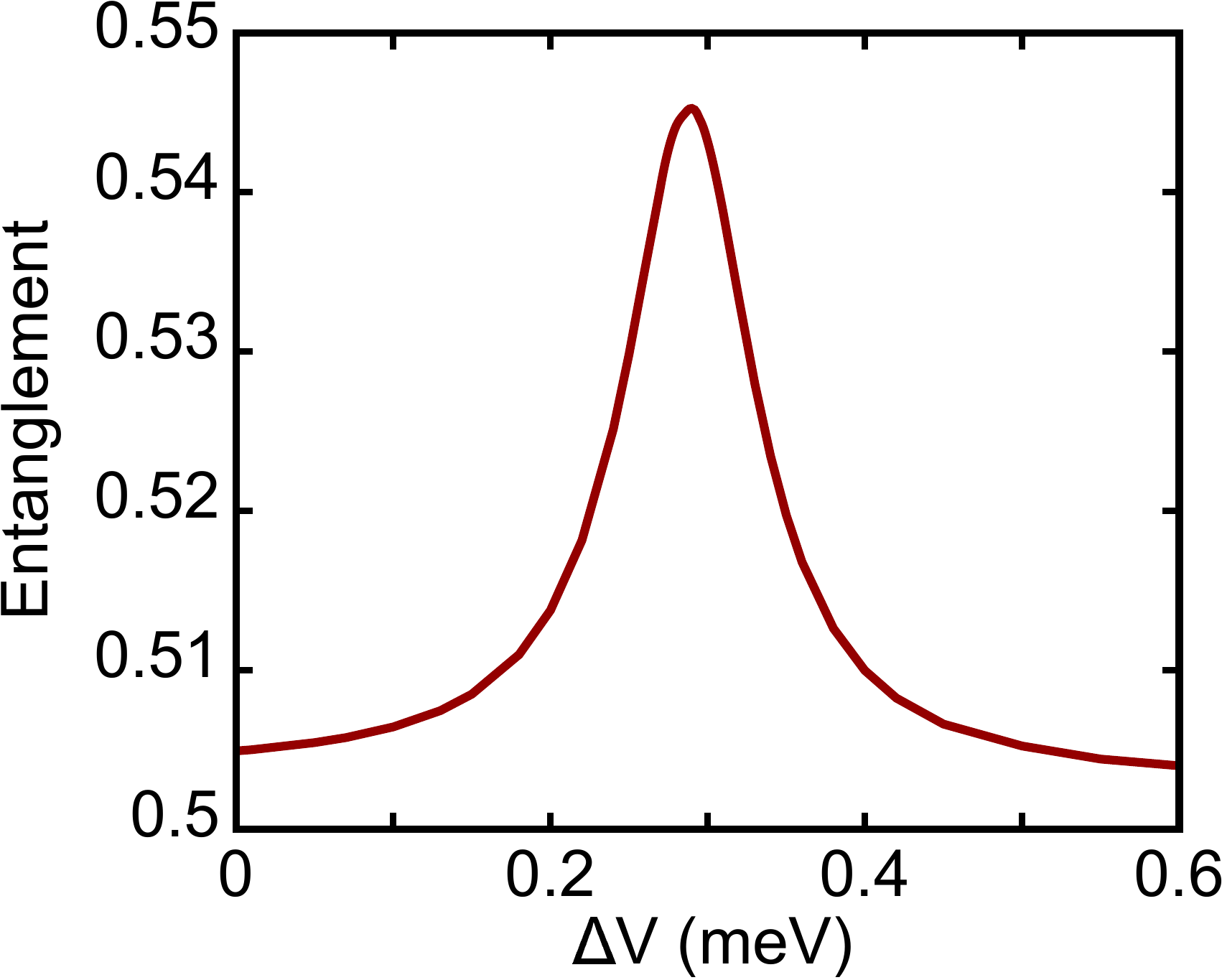}
        \caption{Spatial entanglement of state  5 for two electrons in
          an asymmetric double QDs is plotted as a function of applied
          potential $\Delta  V$. The dot widths are  \mbox{w$_1=15\,{\rm nm}$},
          \mbox{w$_2=18.6\,{\rm  nm}$},  and the  separation  distance is
          \mbox{$d=10\,$nm}.}
	\label{fig:Efield_nonSymWidth}
\end{figure}

  Sensitivity  of   the  system  to  small   changes  in  symmetry
facilitates  the use  of very  small electric  fields to  initiate the
formation of electron clusters, given  that the system is already near
the resonance due to the  inherent asymmetry in the potential profile.
This can be done by designing  two QDs having different widths in such
a way  that the system  is close to a  resonance, and a  weak electric
field can then  initiate this transition. The process  can be reversed
efficiently  by reversing  the direction  of the  electric field.   To
illustrate    this,    consider     the    resonance    E$_{56}$    in
Fig.~\ref{fig:asym_width_ent}(f). The  width w$_1$ of the  first QD is
$15\,{\rm nm}$,  and w$_2=18.6\,{\rm  nm}$.  Now instead  of achieving
the resonance by  slowly increasing w$_2$, we have  introduced a small
E-field.   The change  in entanglement  of the  system by  varying the
magnitude of {\bf E} is shown in Fig.~\ref{fig:Efield_nonSymWidth}. As
can   be   noted  from   the   range   of   the  $x$-axis   label   in
Fig.~\ref{fig:Efield_nonSymWidth},  due to  the built-in  asymmetry of
the system, the required electric  field to reach resonance, and hence
to create an electron cluster, is substantially reduced.

%%%%%%%%%%%%%%%%%%%%%%%%%%%%%%
\section{Concluding Remarks}\label{sec:conclusions}
%%%%%%%%%%%%%%%%%%%%%%%%%%%%%%
We  have  developed  a   variational  formalism  for  calculating  the
coordinate-space      representation     for      the     few-particle
wavefunctions. Finite-element discretization based on the principle of
stationary action provides an  accurate eigen-spectrum for any complex
geometry. Singularities arising while  evaluating the Coulomb integral
are circumvented using multiple  Gauss quadrature of different orders.
Our  scheme  can be  easily  extended  to study  few-electron  quantum
confinements  in  higher  dimensions  for  any  complicated  potential
distribution.

We obtained the solutions for two-electrons in double QDs, and studied
the  spatial  entanglement  properties   of  their  wavefunctions.  We
investigated  the  dependence  of  the energy  spectrum  and  the  spatial
entanglement  on various  geometrical  parameters.   We derived  exact
asymptotic  wavefunctions  to  explain the  degeneracy  spectrum,  and
universal saturation values for the spatial entanglement as separation
distance $d\rightarrow\,\infty$.   These saturation values  were found
to   be  $0.5$ for  quantum  numbers   $\alpha=\beta$,  and   $0.75$  for
$\alpha\neq \beta$.

We observed resonances  in the entanglement values for  the first time
in both  symmetric and  asymmetric QDs. These  entanglement resonances
are found as a consequence of  (a) breaking the mirror symmetry of the
potential, (b) avoided level-crossings  between the excited states, or
(c) due  to the interaction between  states supporting single-electron
and the two-electron  clusters. We note that  the spatial entanglement
value is  a good indicator for  the formation and dissolution  of such
electron clusters.

Further, we showed  that a precise  tuning of
the  entanglement   values  is  feasible  with   an  applied  electric
field. Since  the magnitude of  the external  field can be  tuned readily
in  experiments, the  occupancy of  electrons in  each QD  can be
controlled  and varied  effectively.   Our results  dictate that  when
several QDs are in proximity, spatial correlation between electrons in
the system becomes significant.  The  ability to tune the entanglement
values with  external parameters unveils  new avenues for  forming and
manipulating quantum bits.

In this article, we have focused on the theory of spatial entanglement
in  QDs. Preliminary  work on  quantum  wires  and  the two  electron
spatial entanglement in them shows very similar trends. These studies
of  spatial entanglement  will  be reported  separately  in the  near
future.

%%%%%%%%%%%%%%%%%%%%%%%%%%%%%%
\section{Acknowledgments}
%%%%%%%%%%%%%%%%%%%%%%%%%%%%%%
We  thank  P.   K.   Aravind for  valuable  discussions.   DNP  thanks
Worcester  Polytechnic  Institute  for summer  undergraduate  research
fellowships.  Part  of the calculations presented  here were performed
using computational  resources supported by the  Academic and Research
Computing Group at WPI.

%%%%%%%%%%%%%%%%%%%%%%%%%%%%%%
\appendix
\section{Entanglement in double QDs in the asymptotic
  limit}\label{subsec:asymptotic_ent} 
In  this Appendix,  we show  the spatial  entanglement, ${\cal E}_\ell$,  for a
symmetric double QD as  $d\rightarrow\infty$. In the asymptotic limit,
the  two-particle   states  are  represented by  one of  the four
wavefunctions                                                       in
Eqs.~(\ref{eq:asymptoticWave1})--(\ref{eq:asymptoticWave4}).  Here  we
calculate ${\cal E}_\ell$ for  Eq.~(\ref{eq:asymptoticWave1}), but similar results
can be  derived for the other  three representations. The  reduced density
matrix for  $\ket{\psi^{+}_{\rm s}}$ in  Eq.~(\ref{eq:asymptoticWave1}) is
given by
\begin{eqnarray}
\rho_1 &=&
\frac{1}{4(1+\delta_{\alpha\beta})}\int\!\!\braket{\mathbf{r_2}|{\psi^{+}_{\rm
           s}}}
\!\!\braket{{\psi^{+}_{\rm s}}|\mathbf{r_2}}\rm{d}\mathbf{r_2} 
         \nonumber\\ 
&=&\hspace{-0.02in} \frac{1}{4(1+\delta_{\alpha\beta})}\int\!\! \Big(\!
                        \braket{\mathbf{r_2}|2,\beta}_2\!\ket{1,\alpha}_1
                        +
                        \braket{\mathbf{r_2}|2,\alpha}_2\!\ket{1,\beta}_1
                        \nonumber\\ 
&& \hspace{-0.3in}+ \braket{\mathbf{r_2}|1,\beta}_2\!\ket{2,\alpha}_1
+\braket{\mathbf{r_2}|1,\alpha}_2\!\ket{2,\beta}_1 \!\Big) \nonumber\\
&&\hspace{-0.3in}\times \Big(\! \braket{2,\beta|\mathbf{r_2}}_2\!\bra{1,\alpha}_1+
\braket{2,\alpha|\mathbf{r_2}}_2\!\bra{1,\beta}_1 \nonumber\\
&&\hspace{-0.3in}+\braket{1,\beta|\mathbf{r_2}}_2\!\bra{2,\alpha}_1 +
 \braket{1,\alpha|\mathbf{r_2}}_2\!\bra{2,\beta}_1
                                                                \!\Big)\rm{d}\mathbf{r_2}. 
\end{eqnarray}
Then
\begin{eqnarray}
\rho_1&=&\frac{1}{4(1+\delta_{\alpha\beta})}\bigg[\ket{1,\alpha}\!\!\bra{1,\alpha}
   +
   \ket{1,\beta}\!\!\bra{1,\beta} + \ket{2,\alpha}\!\!\bra{2,\beta}\nonumber\\
&&\hspace{0.3in}+ \ket{2,\beta}\!\!\bra{2,\beta}  
+ \delta_{\alpha\beta}\Big( \ket{1,\alpha}\!\!\bra{1,\beta} +
\ket{1,\beta}\!\!
\bra{1,\alpha} \nonumber\\
&&\hspace{0.3in}+ \ket{2,\alpha}\!\!\bra{2,\beta} + \ket{2,\beta}\!\!
\bra{2,\alpha} \Big)\bigg].
\end{eqnarray}
In the above expression, the  indices are dropped on the bras and
kets, since all of them
correspond  to  the first  electron. The  trace
$\rm{Tr}(\rho_1^2)$ gives
\begin{eqnarray}\label{lintrace}
\rm{Tr}(\rho_1^2)
&=&\int\!\!|\braket{\mathbf{r'_1}|\rho_1|\mathbf{r_1}}|^2\rm{d}\mathbf{r'_1}\rm{d}\mathbf{r_1} 
\nonumber\\  
&&\hspace{-0.3in}=\frac{1}{16(1+\delta_{\alpha\beta})^2}\!\!\int\!\bigg|\!
\braket{\mathbf{r'_1}|1,\alpha}\!\!\braket{1,\alpha|\mathbf{r_1}}\!+\!
\braket{\mathbf{r'_1}|1,\beta}\!\!\braket{1,\beta|\mathbf{r_1}} \nonumber \\ 
&&\hspace{-0.3in} + \braket{\mathbf{r'_1}|2,\alpha}\!\!\braket{2,\alpha|\mathbf{r_1}}
+ \braket{\mathbf{r'_1}|2,\beta}\!\!\braket{2,\beta|\mathbf{r_1}} \\
&&\hspace{-0.3in} + \delta_{\alpha\beta}\Big(\!
\braket{\mathbf{r'_1}|1,\alpha}\!\!\braket{1,\beta|\mathbf{r_1}}
+\braket{\mathbf{r'_1}|1,\beta}\!\!\braket{1,\alpha|\mathbf{r_1}} \nonumber\\ 
&&\hspace{-0.3in} +\braket{\mathbf{r'_1}|2,\alpha}\!\!\braket{2,\beta|\mathbf{r_1}}
\!\!+\!\!\braket{\mathbf{r'_1}|2,\beta}\!\!\braket{2,\alpha|\mathbf{r_1}}\!\Big)
\bigg|^2\rm{d}\mathbf{r'_1}\rm{d}\mathbf{r_1}.\nonumber
\end{eqnarray}
\begin{figure}[th!]  % Fig 13 
	%\begin{subfigure}[h!]{0.3\textwidth}
		%\centering
		\includegraphics[scale=0.3]{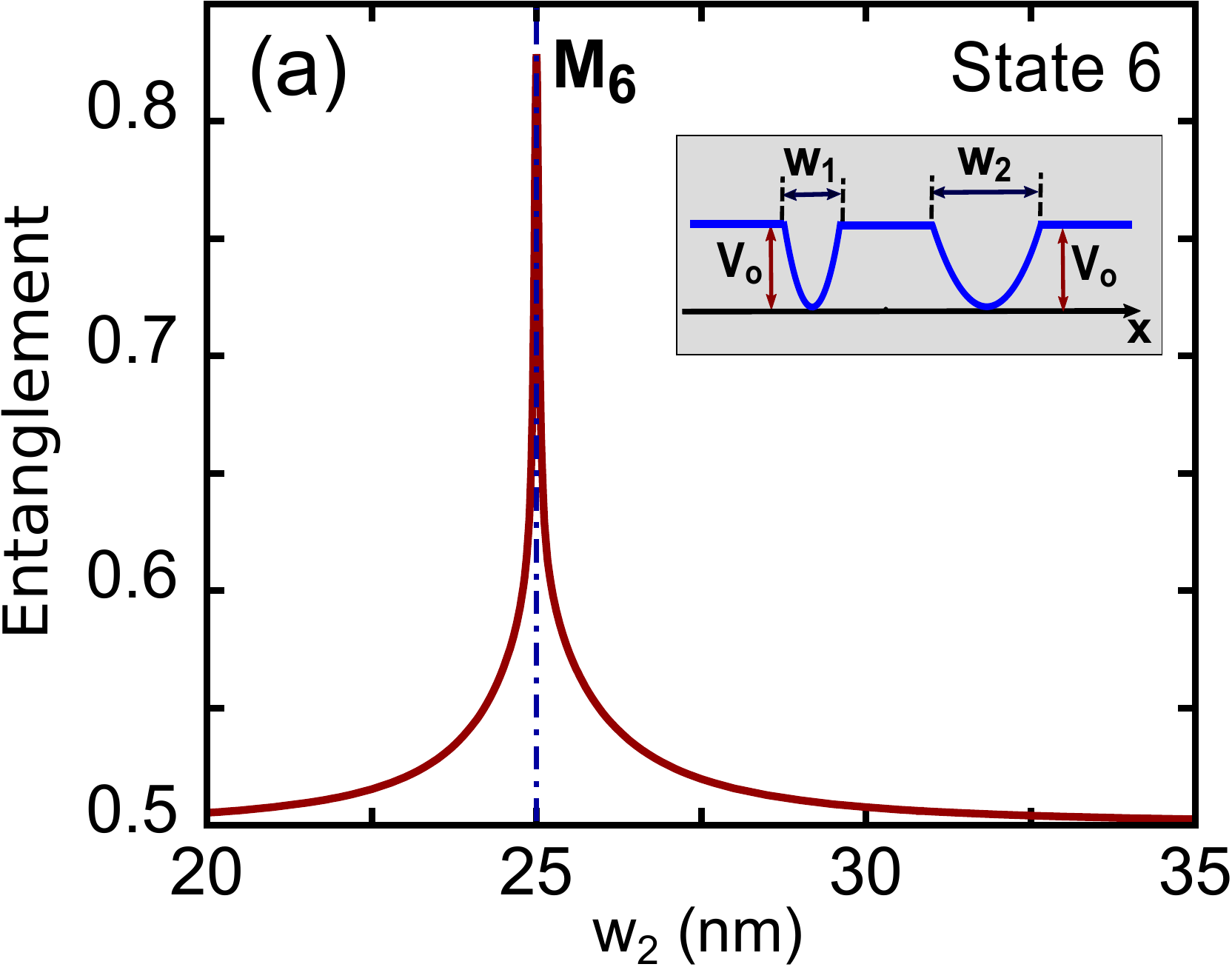}
		%\caption{States 6}
		%\label{fig:state6_parabolic_nonsym_ent}
                % \end{subfigure}%

                \vspace{0.18in}

%	\begin{subfigure}[h!]{0.3\textwidth}
		%\centering
		\includegraphics[scale=0.3]{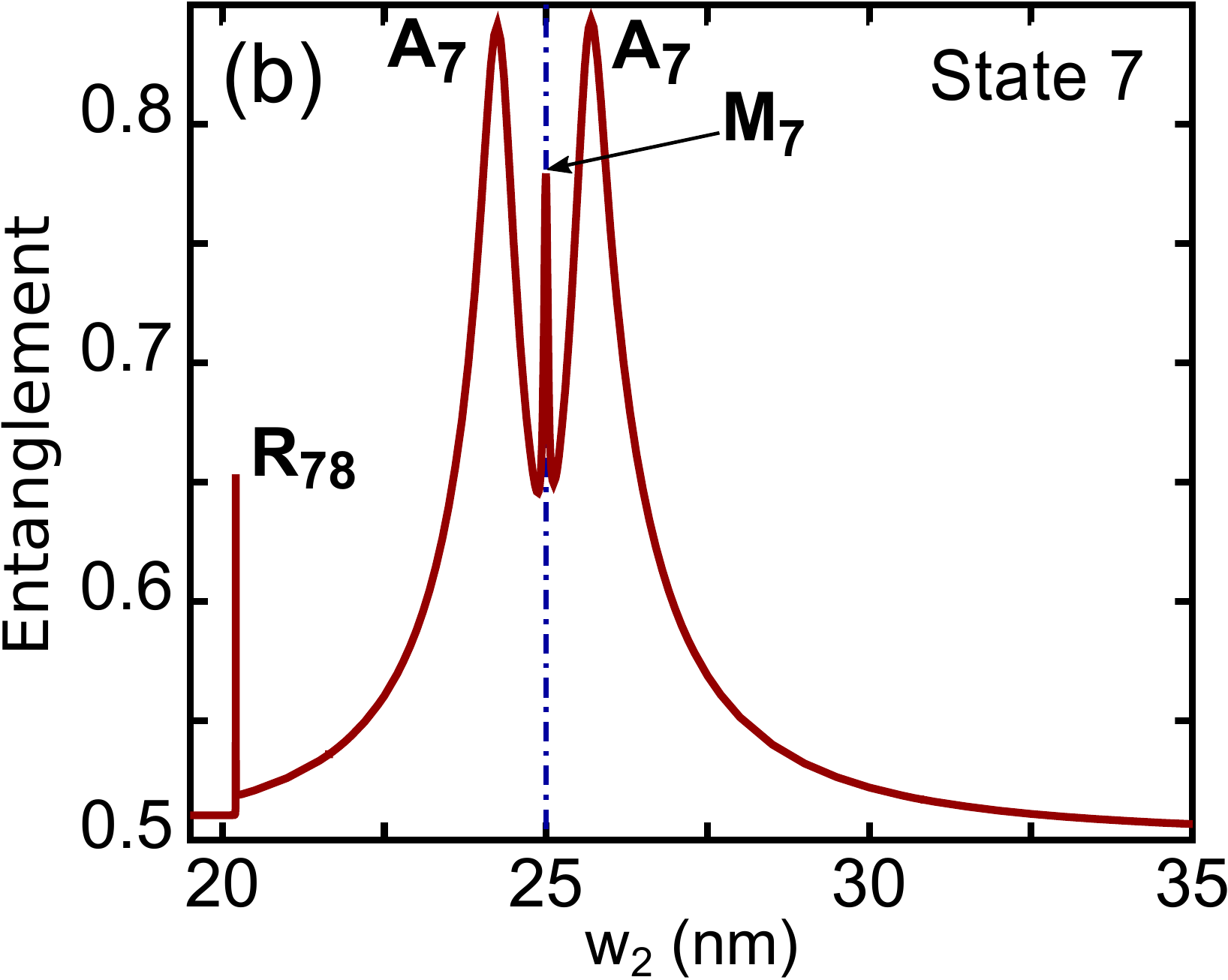}
		%\caption{States 7}
		%\label{fig:state7_parabolic_nonsym_ent}
                % \end{subfigure}%

          \vspace{0.18in}

%	\begin{subfigure}[h!]{0.3\textwidth}
		%\centering
		\includegraphics[scale=0.3]{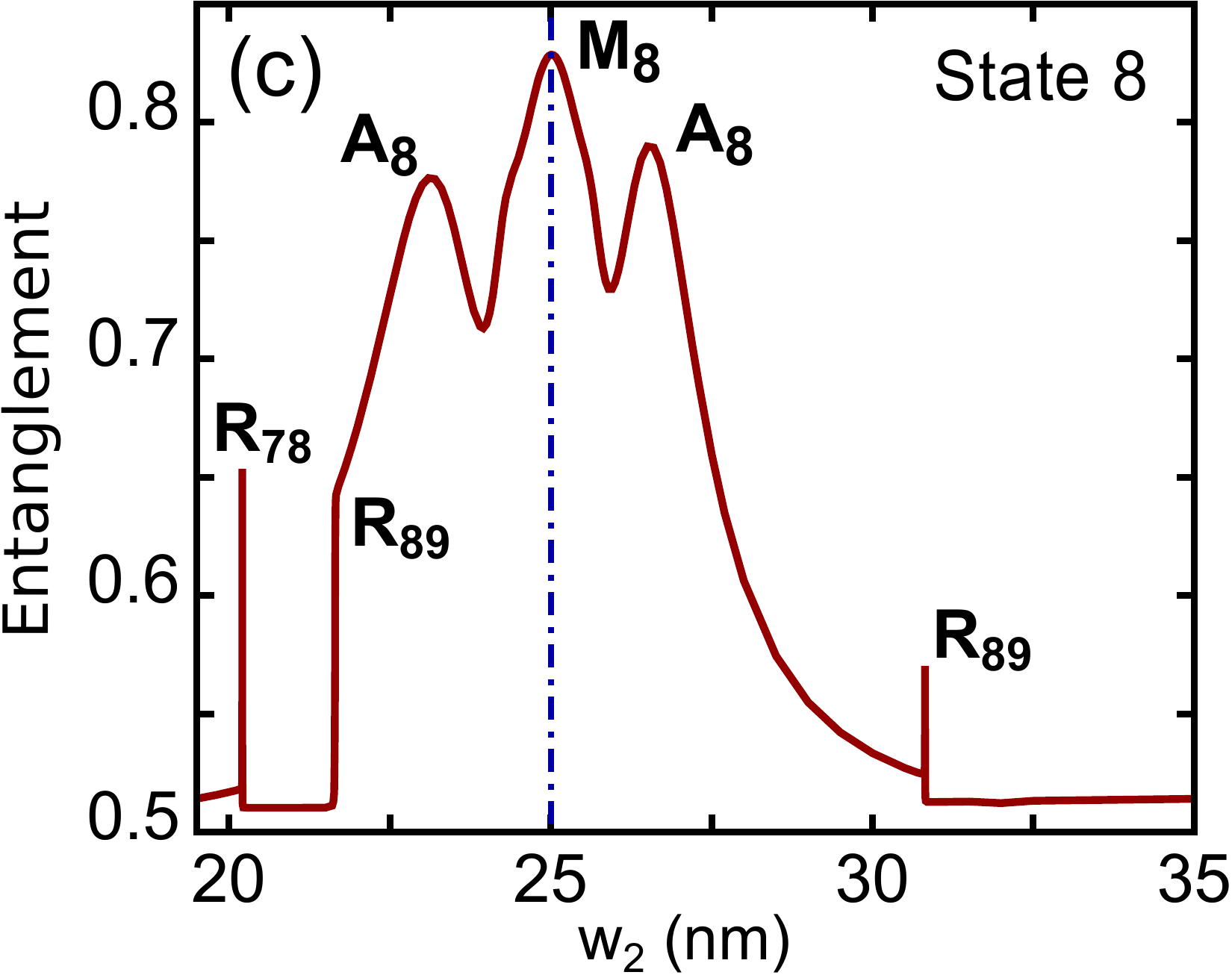}
		%\caption{States 8}
		%\label{fig:state8_parabolic_nonsym_ent}
%	\end{subfigure}
	\caption{\label{fig:parabolic_nonsym_ent}Spatial  entanglement
          for  the  excited  states  6,  7,  and  8  are  plotted  for
          asymmetric double parabolic QDs as a function of w$_2$.  The
          width w$_1$  is kept constant  at $25\,{\rm nm}$.  Here, the
          resonance peaks (i) $M_i$ are  due to the mirror symmetry of
          the system, (ii) $R_{ij}$ are due to avoided level-crossings
          between states  $i\,$ and $\,j$,  and (iii) $A_{i}$  are due
          transition between different mode numbers. }
\end{figure}
Note that we have the relation
\begin{equation}
\braket{i,\alpha|j,\beta} = \delta_{ij}\delta_{\alpha\beta},
\end{equation}  
since single-particle states $\alpha$ and $\beta$ are orthonormal, and
bound-state   wavefunctions   of   one   QD   do   not   overlap with  the
other in the asymptotic limit.  Therefore,  with  some  rearrangements  the  expression  in
Eq.~(\ref{lintrace}) is given by
\begin{align}[left = {\rm{Tr}(\rho_1^2) = \empheqlbrace}]
& 0.5,\quad \alpha=\beta \nonumber\\
& 0.25,\quad \alpha\neq\beta
\end{align}
Hence the amount of entanglement is given  by
\begin{align}[left = {{\cal E}_\ell = 1 - \rm{Tr}(\rho_1^2) = \empheqlbrace}]
& 0.5,\quad \alpha=\beta \nonumber\\
& 0.75,\quad \alpha\neq\beta,
\end{align}
which  is  consistent  with the  qualitative  probabilistic  arguments
presented in Sec.~\ref{subsec:sym}.

\section{Parabolic finite quantum dots} \label{sec:parabolic}
\begin{figure}[th!] %Fig 14
	\includegraphics[width=2.5in]{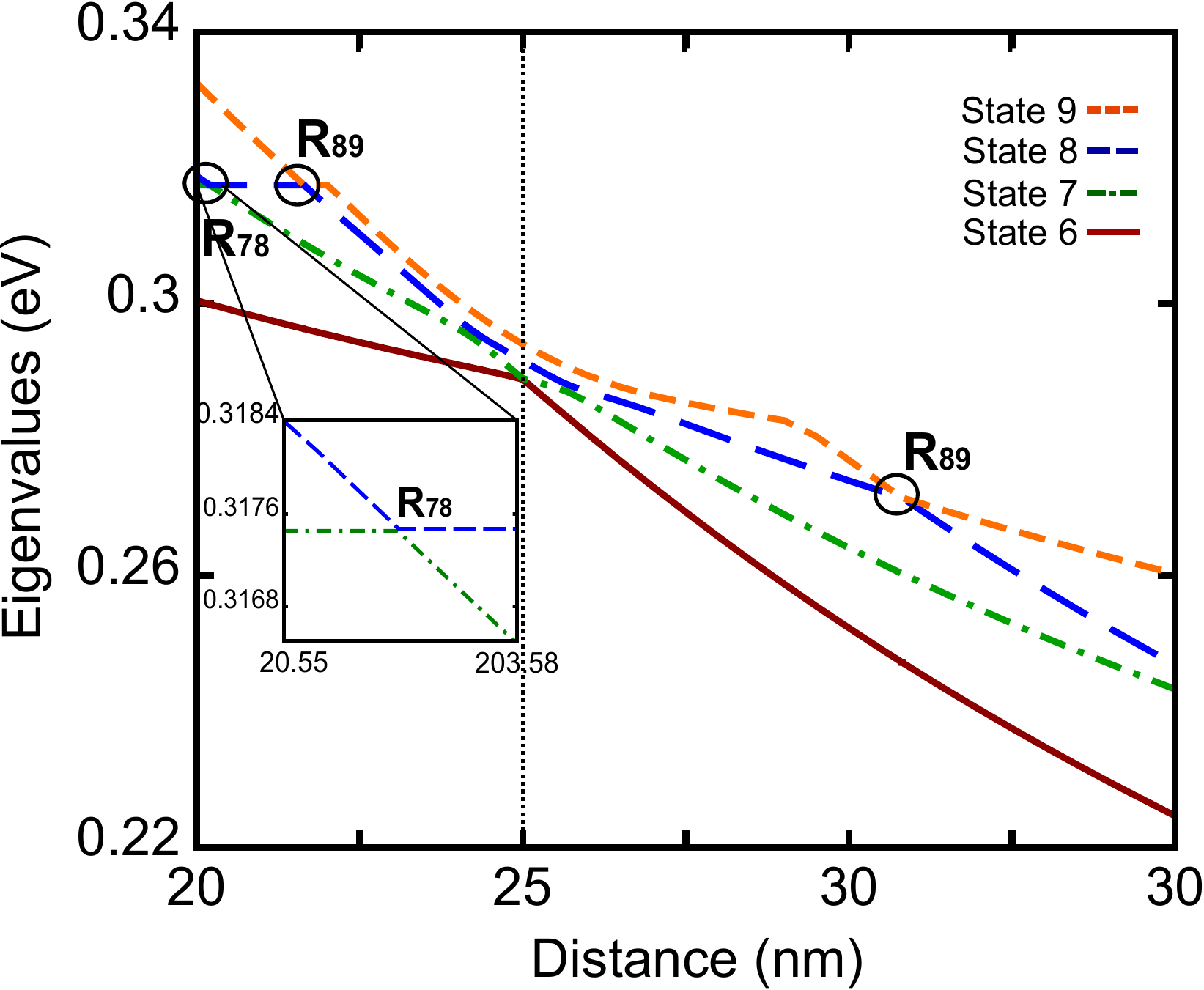}
	\caption{\label{fig:parabolic_eigen}  Eigenenergies   for  two
		electrons in double parabolic QDs  are plotted as a function
		of width of the second dot,  w$_2$. The first dot width is given by
		w$_1=25\,$nm, and dot depths  are $276\,$meV. Avoided level
		crossings are  observed at R$_{78}$ and  R$_{89}$, and their
		resonance       peaks       can       be       seen       in
		Fig.~\ref{fig:parabolic_nonsym_ent}.}
\end{figure}

%%%%%%%%%%%%%%%%%%%%%%%%%%%%%%
In this  Appendix, we investigate  the spatial entanglement  in double
parabolic QDs.   The depth  of each QD  is chosen to  be 276  meV, the
width w$_1$  is fixed at  $25\,{\rm nm}$, and  the width w$_2$  of the
second QD is  varied.  The spatial entanglements for  a selected number
of         excited         states         are         shown         in
Figs.~\ref{fig:parabolic_nonsym_ent}(a)-\ref{fig:parabolic_nonsym_ent}(c).
Avoided  level-crossings  are  observed   at  the  resonance  positions
R$_{78}$        and        R$_{89}$,         as        shown        in
Figs.~\ref{fig:parabolic_nonsym_ent}                               and
~\ref{fig:parabolic_eigen}. Resonances  due to breaking of  the mirror
symmetry  of  the potential  are  observed  at M$_{6}$,  M$_{7}$,  and
M$_{8}$,     located      at     w$_2=25\,$nm     as      shown     in
Fig.~\ref{fig:parabolic_nonsym_ent}.

However, unlike in rectangular finite  dots, here in parabolic dots there
are many additional  entanglement maxima that are  not associated with
any crossing  of states. These  maxima are  labeled by A$_i$  with $i$
being the  state index,  such as A$_7$  and A$_8$, as  can be  seen in
Fig.~\ref{fig:parabolic_nonsym_ent}. In  double parabolic QDs,  if the
Coulomb      interaction      is     absent,      the      eigenenergy
\mbox{$E  \sim  (n\,\omega_1+m\,\omega_2)$},  where $(n,m)$  are  mode
numbers,  and  $\omega_1,\omega_2$  are angular  frequencies.  We  can
achieve the  same eigenenergy  through different  choices of  mode numbers
$(n,m)$.  Around  these  resonances   we  observe  transitions  between
different choices of mode numbers $(n,m)$, and their intermixing leads
to maxima  in the  entanglement. For example,  at resonance  $A_7$ the
transition occurs between modes $(0,3)$ and $(1,2)$.
\begin{figure}[th]  % Fig 15
	\includegraphics[width=3.0in]{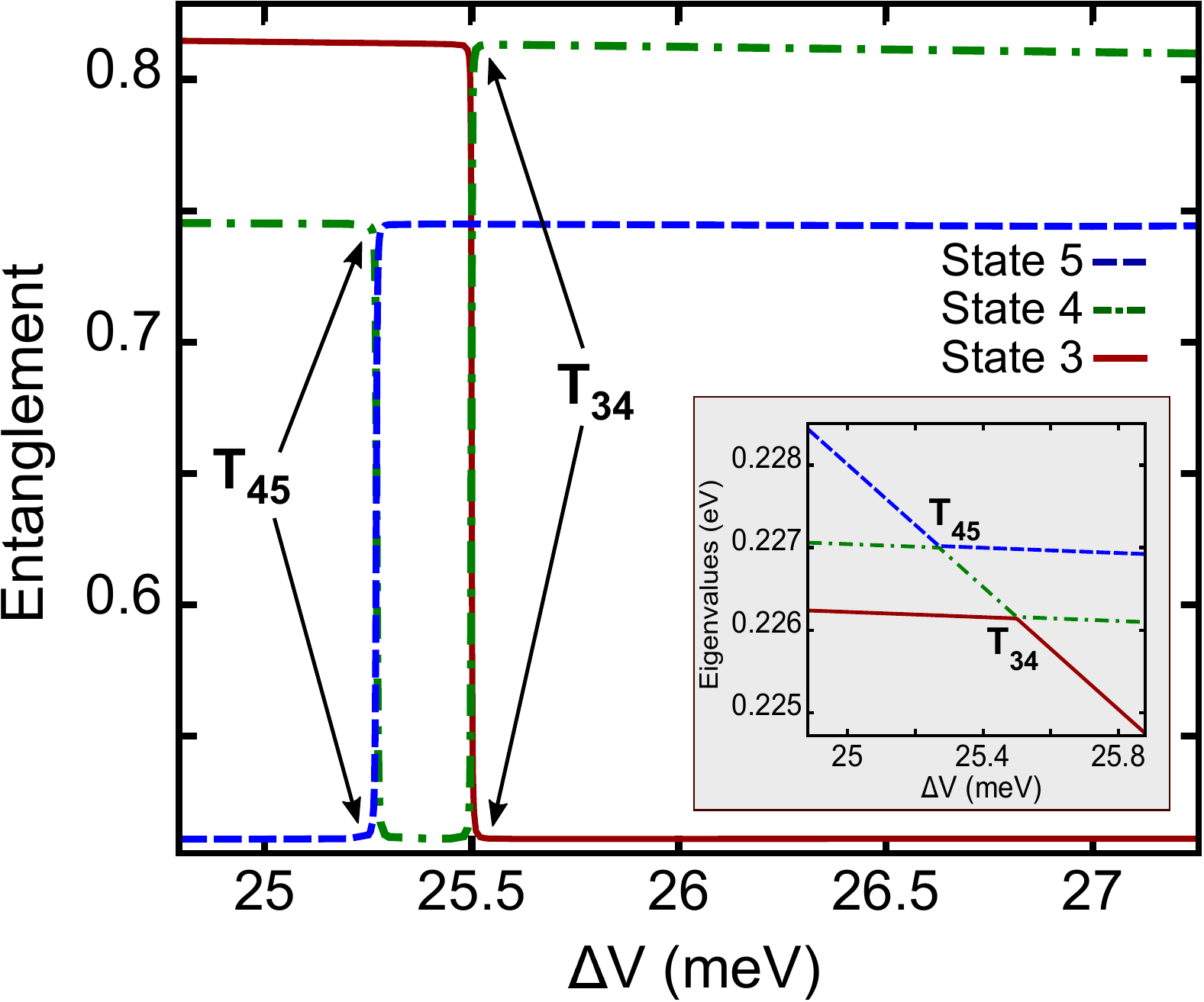}
	\caption{Spatial entanglement of excited states in  symmetric
		double  parabolic QDs  in an  external electric  field is
		plotted  as  a  function  of  the applied potential  $\Delta  V$.  The
		step-wise  transitions in  the entanglement  are due  to the
		avoided level crossing between the adjacent states.}
	\label{fig:Efield_parabolic}
\end{figure}

In Fig.~\ref{fig:Efield_parabolic},  we plot the entanglement values
for selected states in a  symmetric  double  parabolic dot  as  a
function  of applied potential $\Delta V$.  We observe several avoided
level-crossings (see inset in Fig.~\ref{fig:Efield_parabolic}) that
are manifested through a stepwise behavior in the entanglement values.
These   stepwise transitions are   labeled    as    T$_{ij}$. The
labeling T$_{ij}$ represents the interactions between states $i$ and
$j$. For example, before resonance T$_{45}$, state 4 has an
entanglement value that is close to $0.5$, and the state 5 has value close to
$0.75$. Whereas, after the avoided level-crossing between state 4 and
5, their entanglement values are interchanged to have a stepwise
behavior in the case of parabolic QDs. These  characteristics may find
applications in designing quantum bits, where the entanglements can be
tuned through an external electric field. 

%\clearpage

%%%%%%%%%%%%%%%%%%%%%%%%%%%%%%
\end{document}